\documentclass[12pt,english]{article}
\pdfoutput=1
\usepackage[T1]{fontenc}
\usepackage{graphicx,array}
\usepackage{cite}
\usepackage{color}
\usepackage{bm}
\usepackage{dsfont}
\usepackage{latexsym}
\usepackage{amsthm}
\usepackage{amsmath}
\usepackage{stackengine}
\usepackage[titletoc]{appendix}
\usepackage{enumitem}
\usepackage{amssymb}
\usepackage[unicode=true,
 bookmarks=true,bookmarksnumbered=false,bookmarksopen=false,
 breaklinks=false,pdfborder={0 0 1},backref=false,colorlinks=true]
 {hyperref}
\hypersetup{pdftitle={Spinning Sum Rules for the Dimension-Six SMEFT},
 pdfauthor={Grant N. Remmen, Nicholas L. Rodd},
 citecolor=blue,linkcolor=black,urlcolor=black}
\usepackage{breakurl}
\usepackage[hang,flushmargin]{footmisc}

\newcolumntype{L}[1]{>{\raggedright\let\newline\\\arraybackslash\hspace{0pt}}m{#1}}
\newcolumntype{C}[1]{>{\centering\let\newline\\\arraybackslash\hspace{0pt}}m{#1}}

\def\simgt{\mathrel{\lower2.5pt\vbox{\lineskip=0pt\baselineskip=0pt
           \hbox{$>$}\hbox{$\sim$}}}}
\def\simlt{\mathrel{\lower2.5pt\vbox{\lineskip=0pt\baselineskip=0pt
           \hbox{$<$}\hbox{$\sim$}}}}

\newcommand{\be}{\begin{equation}}
\newcommand{\ee}{\end{equation}}
\newcommand{\bea}{\begin{equation}\begin{aligned}}
\newcommand{\eea}{\end{aligned}\end{equation}}
\newcommand{\Eq}[1]{Eq.~(\ref{#1})}
\newcommand{\Sec}[1]{Sec.~\ref{#1}}

\newcommand{\Tab}[1]{Table~\ref{#1}}

\newcommand{\Refc}[1]{Ref.~\cite{#1}}
\newcommand{\Refs}[1]{Refs.~\cite{#1}}

\newcommand{\overleftrightarrowalt}[1]{\overset{\text{\tiny$\bm\leftrightarrow$}}{#1}}
\newcommand{\tW}{\theta_{\scriptscriptstyle W}}
\newcommand{\mS}{m_{\scriptscriptstyle S}}
\newcommand{\mV}{m_{\scriptscriptstyle V}}
\newcommand{\mW}{m_{\scriptscriptstyle W}}

\newcommand{\ket}[1]{\left| #1 \right\rangle}

\newcommand*\oline[1]{%
  \vbox{%
    \hrule height 0.5pt
    \kern0.68ex
    \hbox{%
      \kern-0.1em
      \ifmmode#1\else\ensuremath{#1}\fi
      \kern-0.1em
    }
  }
}

\definecolor{nicered}{rgb}{0.7,0.1,0.1}
\definecolor{nicegreen}{rgb}{0.1,0.5,0.1}
\hypersetup{
 colorlinks,citecolor=black,linkcolor=black,urlcolor=black}
\usepackage{xcolor}

\begin{document}

\hfill CERN-TH-2022-105

\interfootnotelinepenalty=10000
\baselineskip=18pt
\hfill

\vspace{1.36cm}
\thispagestyle{empty}
\begin{center}
{\LARGE \bf
Spinning Sum Rules for the \\\vspace{1.5mm} Dimension-Six SMEFT
}\\
\bigskip\vspace{1cm}{
{\large Grant N. Remmen${}^{a,b}$ and Nicholas L. Rodd${}^{c}$}
} \\[7mm]
{
\it ${}^a$Kavli Institute for Theoretical Physics, \\[-1mm]
University of California, Santa Barbara, CA 93106, USA\\[1.5mm]
${}^b$Department of Physics, \\[-1mm]
University of California, Santa Barbara, CA 93106, USA \\[1.5 mm]
${}^c$Theoretical Physics Department, \\[-1mm]
CERN, 1 Esplanade des Particules, CH-1211 Geneva 23, Switzerland}
\let\thefootnote\relax\footnote{e-mail: 
\url{remmen@kitp.ucsb.edu}, \url{nrodd@cern.ch}}
 \end{center}

\bigskip
\centerline{\large\bf Abstract}
\begin{quote} \small
We construct new dispersive sum rules for the effective field theory of the standard model at mass dimension six.
These {\it spinning sum rules} encode information about the spin of UV states: the sign of the IR Wilson coefficients carries a memory of the dominant spin in the UV completion.
The sum rules are constructed for operators containing scalars and fermions, although we consider the dimension-six SMEFT exhaustively, outlining why equivalent relations do not hold for the remaining operators.
As with any dimension-six dispersive argument, our conclusions are contingent on the absence of potential poles at infinity---so-called boundary terms---and we discuss in detail where these are expected to appear.
There are a number of phenomenological applications of spinning sum rules, and as an example we explore the connection to the Peskin-Takeuchi parameters and, more generally, the set of oblique parameters in universal theories.
\end{quote}
	
\setcounter{footnote}{0}

\newpage
\tableofcontents
\newpage

\section{Introduction}\label{sec:intro}

The Standard Model Effective Field Theory (SMEFT) has emerged as the central organizing framework in the search for new physics at the energy and precision frontiers.
Assuming only that the scale where new states appear is beyond experimental energies, Wilsonian effective field theory (EFT) codifies the possible infrared (IR) relics of the ultraviolet (UV).
But the SMEFT parameter space is vast.
At mass dimension six, where the first deviations from the standard model (SM) are expected to appear, the SMEFT basis requires 84 operators, even when considering just a single generation of fermions~\cite{Grzadkowski:2010es}, and 3,045 operators for three generations.\footnote{Here we follow the convention in \Refc{Henning:2015alf}, where operators and their hermitian conjugates are counted separately.}
Within the context of specific completions, the full space of EFT operators collapses to reveal a set of predictive correlations between the Wilson coefficients and observables.
In the fully UV-agnostic Wilsonian picture, however, the parameter space remains unstructured.

Yet a middle ground exists between these two pictures: general assumptions about the UV impose a detailed topography on the EFT landscape.
The restrictions arise as the higher-dimension operators in an EFT introduce interactions that can mediate scatterings that manifestly violate unitarity, locality, or causality for generic choices of the operator coefficients, which can be revealed through dispersive arguments.
These constraints were first discovered in the context of chiral-perturbation theory~\cite{Pham:1985cr,Ananthanarayan:1994hf,Pennington:1994kc} and then outlined more generally in \Refc{Adams:2006sv}.
Given the primacy of EFT to the language of modern physics, this ``EFT consistency'' program has been pursued in many directions, a non-exhaustive list of which includes the chiral Lagrangian~\cite{Distler:2006if,Jenkins:2006ia,Vecchi:2007na,Manohar:2008tc,Wang:2020jxr}, electroweak effective theory~\cite{Bellazzini:2018paj,Zhang:2018shp,Remmen:2019cyz,Bi:2019phv}, various scalar theories~\cite{Jenkins:2006ia,Nicolis:2009qm,Dvali:2012zc,deRham:2017imi,Chandrasekaran:2018qmx,Herrero-Valea:2019hde,Li:2021cjv,Remmen:2021zmc}---including recent work on the EFThedron~\cite{Arkani-Hamed:2020blm} and other powerful new techniques~\cite{Bellazzini:2020cot,Tolley:2020gtv,Caron-Huot:2020cmc,Bellazzini:2021oaj,Caron-Huot:2021rmr}---along with the $a$-theorem~\cite{Komargodski:2011vj,Elvang:2012st}, fermionic theories~\cite{Adams:2008hp,Bellazzini:2016xrt,Bellazzini:2017bkb,Remmen:2020vts,Remmen:2020uze}, massive higher-spin particles and massive gravity~\cite{Cheung:2016yqr,deRham:2017xox,Camanho:2014apa,Camanho:2016opx,Bellazzini:2017fep,Bonifacio:2018vzv,Bonifacio:2016wcb,deRham:2017zjm,Hinterbichler:2017qyt,deRham:2018qqo,Bellazzini:2019bzh,Alberte:2019xfh,Alberte:2019zhd,Wang:2020xlt}, nonlinear electromagnetism~\cite{Jenkins:2006ia,Adams:2006sv,Remmen:2019cyz,Alberte:2020bdz,Gorghetto:2021luj}, cosmological settings~\cite{Nicolis:2009qm,Baumann:2015nta,Baumann:2019ghk,Grall:2021xxm,deRham:2021fpu,Grall:2020tqc,Melville:2021lst}, quantum corrections to the Einstein equations~\cite{Bellazzini:2015cra,Cheung:2016wjt,Gruzinov:2006ie,Guerrieri:2021ivu,deRham:2021bll,Caron-Huot:2022ugt,Chiang:2022jep}, and Einstein-Maxwell theory and its connections to the Weak Gravity Conjecture~\cite{Cheung:2014ega,Bellazzini:2019xts,Cheung:2019cwi,Cheung:2018cwt,Charles:2019qqt,deRham:2018dqm,Andriolo:2020lul,Alberte:2020jsk,Arkani-Hamed:2021ajd}.
A focus of recent investigations has been on the application of the technology of analyticity and unitarity to systematically constrain the Wilson coefficients of the full EFT of the SM itself~\cite{Low:2009di,Falkowski:2012vh,Bellazzini:2014waa,Ema:2018jgc,Remmen:2019cyz,Remmen:2020vts,Bellazzini:2018paj,Zhang:2018shp,Bi:2019phv,Remmen:2020uze,LianTao,Zhang:2020jyn,Fuks:2020ujk,Yamashita:2020gtt,Gu:2020ldn,Trott:2020ebl,Bonnefoy:2020yee,Davighi:2021osh,Zhang:2021eeo,Li:2022tcz,Chala:2021wpj,Azatov:2021ygj}. This is the direction we pursue in this work.

The sharpest realization of these ideas arises at dimension eight.
Consider, for instance, the SMEFT operator ${\cal O}_{e}^{(8)} = - c^{(8)}_{mnpq} \partial_{\mu} (\bar{e}_m \gamma_{\nu} e_n) \partial^{\mu} (\bar{e}_p \gamma^{\nu} e_q)$, with $e=e_R$ the right-handed electron field in the unbroken SMEFT---a singlet under color and weak isospin---and $m,n,p,q$ flavor indices.
The traditional Wilsonian picture is that any value of the coefficients is allowed, so long as $|c_{mnpq}| \lesssim 1$, to ensure the EFT power counting is not violated.
Nevertheless, if the UV theory that generates ${\cal O}_{e}^{(8)}$ is unitary, local, and causal, then the Wilson coefficients can be shown to satisfy a dispersion relation that considerably limits their allowed structure.
Specifically, as shown in \Refc{Remmen:2020vts}, consistency of two-to-two scattering of pure electron-type states requires $c_{1111} > 0$, whereas elastic scattering of two arbitrary flavor superpositions, $\alpha$ and $\beta$, reveals that
\be
c^{(8)}_{mnpq} \alpha_m \beta_n \bar{\beta}_p \bar{\alpha}_q > 0 \hspace{0.3cm} \forall \; \alpha;\beta.
\label{eq:dim8fermionbound}
\ee
The boundary defined by this relation contains considerable structure.
The inequality also imports physical consequences.
At the level of this operator both flavor and CP violation are strictly bounded by the size of the equivalent interactions that conserve the effects; correlations between operators and observables have emerged from bedrock UV tenets.
However, this is just a single operator, and considerable work has gone into providing a more detailed map of the consistent boundary of the dimension-eight SMEFT~\cite{Remmen:2019cyz,Remmen:2020vts,Bellazzini:2018paj,Zhang:2018shp,Bi:2019phv,Zhang:2020jyn,Fuks:2020ujk,Yamashita:2020gtt,Gu:2020ldn,Trott:2020ebl,Bonnefoy:2020yee,Davighi:2021osh,Zhang:2021eeo,Li:2022tcz} (the full basis of which has now been enumerated~\cite{Murphy:2020rsh,Li:2020gnx}).

Progress at dimension six has been less rapid and is the focus of the present work.
As at higher dimensions, if the EFT emerged from a unitary, local, and causal completion, then the Wilson coefficients can be shown to satisfy a dispersion relation.
The fundamental challenge is that as the forward amplitude now scales linearly with Mandelstam $s$, as opposed to quadratically as at dimension eight, the interpretation of the resulting dispersive arguments is less straightforward.
We will review the obstructions later in this work.
For the moment we simply note that while at dimension eight the forward dispersion relation involves a sum of integrated cross sections, and is therefore strictly positive, at dimension six it involves differences of integrated cross sections and also a sign-indefinite contribution from a pole at infinity.
Conventionally, the dispersive relations at dimension six are referred to as sum rules.

While there are challenges at dimension six, the phenomenological importance of the SMEFT at this order mandates the determination of the structure imposed by a consistent UV.
This question was considered in the seminal work of Low, Rattazzi, and Vichi~\cite{Low:2009di}, in the context of operators that modify observable Higgs properties at the LHC, and has been taken up in a number of additional works (see, for example, \Refs{Falkowski:2012vh,Bellazzini:2014waa,Ema:2018jgc,Azatov:2021ygj}), including a recent determination in \Refc{LianTao} of the full set of sum rules that can be established by studying the forward amplitude (obtained by taking Mandelstam $t$ to zero).
Sum rules derived from the forward amplitude only constrain the coefficient of $s$.
However, two-to-two scattering amplitudes also depend on $t$, and constraints on these beyond-forward contributions can also be established, as shown in \Refs{Nicolis:2009qm,deRham:2017avq}.
For example, \Refc{Davighi:2021osh} recently demonstrated that the set of possible bounds can be considerably extended with beyond-forward dispersion relations.

Combining forward and non-forward dispersion relations, \Refc{Remmen:2020uze} demonstrated that a novel class of sum rules can be derived if one assumes that the UV theory sufficiently improves the scaling of the forward amplitude at large $s$ in order to avoid poles at infinity. 
While intuitively one may expect UV completions to reduce the momentum scaling of an amplitude---in many instances they do---we emphasize that the assumption is far less fundamental than causality, unitarity, and locality; indeed, tree-level UV completions that involve massive states exchanged in the $t$ channel can violate the conditions for poles at infinity to vanish, and we will discuss when we expect obstructions in detail. 
Regardless, if the UV completion sufficiently improves the asymptotic scaling of the amplitude, then the result in \Eq{eq:dim8fermionbound} can effectively be replaced by a novel sum rule that implies that the Wilson coefficients carry a memory of the UV state from which they were dominantly generated, rather than direct positivity constraints.
For instance, the lower-dimension analogue of ${\cal O}_{e}^{(8)}$ is ${\cal O}_{e}^{(6)} = -c^{(6)}_{mnpq} (\bar{e}_m \gamma_{\mu} e_n) (\bar{e}_p \gamma^{\mu} e_q)$, and as shown in \Refc{Remmen:2020uze}, it satisfies the following inequality:\footnote{Importantly, note that as in \Refc{Remmen:2020uze}, we use the mostly-plus metric convention.
As ${\cal O}_{e}^{(6)}$ contains the Lorentz contraction $\sim g_{\mu \nu} \gamma^{\mu} \gamma^{\nu}$, changing signature would invert the sign of the result in \Eq{eq:dim6fermionbound}.
The same is not true of \Eq{eq:dim8fermionbound}: as ${\cal O}_{e}^{(8)}$ contains two Lorentz contractions, the sign of the bound is independent of signature.
We emphasize more generally that the signature must be considered carefully.}
\be
c^{(6)}_{mnpq} \alpha_m \beta_n \bar{\beta}_q \bar{\alpha}_q\;
\begin{array}{l} > 0~\text{scalars~dominate~the~UV},\\ < 0~\text{vectors~dominate~the~UV}, \end{array}
\label{eq:dim6fermionbound}
\ee
so that the sign of the Wilson coefficient encodes the spin of the UV completion, giving rise to {\it spinning sum rules}.\footnote{A similar phenomenon appears in the ``low-spin dominance'' bounds on gravitational EFTs in Ref.~\cite{Bern:2021ppb}.}
Further, as long as either sign dominates, one still expects that CP and flavor violation are bounded from above by the corresponding conservation processes.
Beyond the additional assumption invoked, this result relied on the fermion spin in that it made use of the Jacob-Wick expansion~\cite{Jacob:1959at}, where amplitudes are expanded in the total angular momentum (cf. the Legendre partial-wave expansion, which is performed in orbital angular momentum).

Nevertheless, \Refc{Remmen:2020uze} only applied this new approach to fermions.
In the present work, we derive spinning sum rules for the full dimension-six SMEFT, with a view to adding as much character to the space as possible.
Again, the spirit of our work is not to simply derive clear-cut positivity bounds, but rather to consider any restrictions on combinations of the SMEFT Wilson coefficients dictated by our assumptions.
We will show that the use of the Jacob-Wick expansion is not critical, and therefore spinning sum rules can be established for the purely scalar operators involving the Higgs.
In this way, we will be able to explain several features of the results in \Refc{Low:2009di}, in particular, why specific UV completions generated specific signs.
More generally, we find bounds on the four-Higgs and four-fermion operators with sign dependent on whether the UV is dominated by states of low or high spin.
For the remaining operators, while spinning sum rules cannot be established, we show that the dispersion relations are still informative of the UV; in particular, a connection emerges between the form of the UV amplitude and the presence of the poles at infinity, also referred to as boundary terms.
For Higgs-fermion scattering, if the boundary terms vanish, then the UV must be dominated by a fermionic mode, whereas for Higgs-gauge boson or quartic gauge boson operators a stronger statement is possible: if the boundary terms vanish, then the UV amplitudes mediating interactions between these states must be identically zero.
While the main application of our results is to the dimension-six SMEFT, we also demonstrate their utility by establishing a connection to the electroweak oblique parameters ($S$, $T$, $W$, $Y$, $Z$) describing universal theories.
As these parameters can be connected to SMEFT operators, we show how we can establish effective spinning sum rules for a number of the observables in universal theories.

We will organize our analysis and discussion as follows.
In Sec.~\ref{sec:dispersion} we introduce and develop our formalism for deriving our spinning sum rules.
After summarizing the relevant basis of SMEFT operators in Sec.~\ref{sec:basis}, we systematically apply these bounds to various sectors of SM scattering processes in Sec.~\ref{sec:sumrules}, obtaining our main results.
In Sec.~\ref{sec:boundary}, we perform an extended investigation of where poles at infinity are expected to appear by considering a wide range of UV completions and demonstrate that, at tree level, boundary terms are associated with UV states exchanged in the $t$ channel.
Finally, we apply our results to the electroweak oblique parameters in Sec.~\ref{sec:oblique} and conclude in Sec.~\ref{sec:conclusions}.

\section{Dispersion Relations at Dimension Six}\label{sec:dispersion}

In this section, we construct the dispersion relations that we will use to derive spinning sum rules for dimension-six operators in the SMEFT.
These sum rules will allow us to write the Wilson coefficients of these operators as sums over spins of states in the UV completion.\footnote{Wilson coefficients depend on the scale at which they are evaluated.
Throughout, we will always assume the coefficients are evaluated at the matching scale, where the new physics is integrated out, or at a lower scale only if the theory is sufficiently weakly coupled that the running can be neglected.
For a discussion of how the running can impact the types of relations considered in this work, see, for example, \Refs{Manohar:2008tc,Trott:2020ebl,Bellazzini:2020cot,Arkani-Hamed:2020blm,Chala:2021pll,DasBakshi:2022mwk}.}
In \Sec{sec:sumrules}, we will be able to use these relationships to associate the signs of certain Wilson coefficients with the spin of the UV state.
Our methods generalize those developed in \Refc{Remmen:2020uze} (which considered exclusively four-fermion operators) to states of arbitrary spin.

As already discussed, the standard methods of establishing positivity bounds on EFT coefficients of quartic operators (see, e.g., \Refc{Adams:2006sv}) do not immediately extend to terms of mass dimension six.
This can be seen by straightforward power counting: two-to-two EFT amplitudes---functions of $s$ and $t$, denoted ${\cal A}(s,t)$---mediated by dimension-six operators will go like $p^2$, where $p$ is some generic external momentum.
This scaling has the result that applying a dispersion relation to the forward, $t \to 0$, amplitude to extract the $\propto s$ part results in a mixed-sign relation:\footnote{The integral over $s$ in the second line is associated with the discontinuity (due to poles and branch cuts) of the forward amplitude in the complex $s$ plane.
If the theory allows for loops of massless states to be exchanged in the scattering, then the discontinuity will run to $s=0$ on the real axis, as we show on the second line.
Formally, this would imply that the contour $\gamma$ cannot be defined, but this obstruction can always be removed by introducing an IR regulator for the massless states, as discussed in~\Refc{Adams:2006sv}.
Accordingly, technically the integral over $s$ here and in later similar expressions should be viewed as starting at the scale associated with the IR regulator.} 
\bea
\lim_{s\rightarrow 0}\partial_s {\cal A}(s,0) &= \frac{1}{2\pi i}\oint_\gamma \frac{{\rm d}s}{s^2} {\cal A}(s,0)\\
&= -C_\infty^{(s)} + \frac{1}{\pi}\int_0^\infty \frac{{\rm d}s}{s} [\sigma(s) - \overline\sigma(s)].
\label{eq:disps}
\eea
Here, $\gamma$ is a small contour around the origin evaluated at sufficiently small $|s|$ that the amplitude can be computed in the EFT, and therefore related to the coefficient of $s$ in the EFT amplitude ${\cal A}(s,0)$, as we have on the first line.
In the second line, we expand the contour to large $|s|$, so that we can rewrite this result in terms of UV quantities, and exploit the optical theorem: $\sigma(s) = s\,{\rm Im}\,{\cal A}(s,0)$ is the cross section associated with the scattering process on the left-hand side, and $\overline\sigma(s) = s\,{\rm Im}\,\bar{\cal A}(s,0)$ is the cross section for the crossed amplitude $\bar {\cal A}$, where particles 1 and 4 have been exchanged under crossing symmetry, e.g., the fully inclusive cross section for $e^+ e^-$ versus $e^- e^-$ (where we have conjugated the charge, but in general any additional internal quantum numbers will be conjugated also).\footnote{While at general kinematics, states with spin transform nontrivially under crossing, in the forward limit helicities behave just like an internal quantum number and are simply conjugated under crossing~\cite{Bellazzini:2016xrt,Remmen:2020uze}.}
The residue at infinity is determined as
\be 
C_\infty^{(s)} = {\rm Res}\left[\frac{{\cal A}(s,0)}{s^2},s=\infty\right]
= -\lim_{|s|\rightarrow \infty}\frac{{\cal A}(s,0)}{s}.
\label{eq:Cs}
\ee
Heuristically, $C_\infty^{(s)}$ is the $\propto s$ coefficient in the expansion of ${\cal A}(s,0)$ at large $s$ and as already noted is commonly referred to as a boundary term.
Observe that in order to go from the first to the second line of \Eq{eq:disps}, we needed to assume that the UV was causal, which implies that the UV amplitude is an analytic function of $s$ except for branch cuts and poles along the real $s$ axis.
Using the sum rule in \Eq{eq:disps}, one can place bounds on Wilson coefficients of dimension-six operators under varying assumptions about the UV, e.g., whether $\sigma(s)$ or $\overline \sigma(s)$ dominates;  this was investigated systematically in \Refc{LianTao}, see also~\Refs{Adams:2008hp,Low:2009di,Falkowski:2012vh,Bellazzini:2014waa}.

From \Eq{eq:Cs}, if the amplitude satisfies $\lim_{|s|\to\infty} |{\cal A}_{\alpha \beta}(s,0)| < O(s)$ (i.e., it is super-Froissart in the language of Ref.~\cite{Davighi:2021osh}), then $C_\infty^{(s)} = 0$.
As discussed in the introduction, while one may expect the UV to soften the energy scaling of the amplitude, it is not required.
The fact that UV states exchanged in the $t$ channel generically introduce a residue at infinity can be readily seen.
Schematically, for an amplitude that generates a dimension-six operator in the IR from UV $t$-channel exchange, we will have ${\cal A}(s,t) \propto p^2/(-t+m^2-i\epsilon)$, where $m$ is the mass of the UV state and $p^2\sim s,t,u$.
As the cross sections in \Eq{eq:disps} are determined by the imaginary part of the amplitudes, for such a completion the integrand of the dispersive integral contains ${\rm Im}\,{\cal A}(s,0) \propto p^2\delta(m^2)$, which vanishes since $m>0$.
Consequently, the EFT results can only be reproduced by $C_\infty^{(s)}$, which is nonzero if the $p^2$ in the amplitude's numerator goes like $s$ in the forward limit.\footnote{If the amplitude scales as $t/(-t+m^2-i\epsilon)$, as can arise from the $t$-channel exchange of a derivatively coupled scalar, then as the amplitude is constant in $s$ it will not generate $C_\infty^{(s)}$.
Nevertheless, such an amplitude can generate a boundary term in the beyond-forward dispersion relation we consider in \Eq{eq:dispt}.
Explicit UV completions demonstrating these points are given in Sec.~\ref{sec:boundary}.}
We will discuss this point, and the contributions from boundary terms more generally, in detail in Sec.~\ref{sec:boundary}.

Applying the optical theorem in the forward limit does not leverage all of the information we can access about the structure of the amplitude.
In particular, in addition to extracting the $\propto s$ part of ${\cal A}(s,0)$ through \Eq{eq:disps}, one can also write a beyond-forward dispersion relation for the $\propto t$ scaling of ${\cal A}(0,t)$.
Setting $s=0$ with $t$ nonzero effectively interchanges an in- and out-state, for instance a swapping of states 1 and 4.\footnote{To be explicit, if we have an amplitude where the in-states (1 and 2) and out-states (3 and 4) have incoming and outgoing momenta, respectively, then we can achieve $s \to 0$ by taking $p_1 \to - p_2$ and $p_4 \to - p_3$.
The minus signs then exchange the roles of the states so that state 4 is incoming and 1 is outgoing.}
While such an exchange is trivial for a bosonic amplitude, it induces an extra minus sign if we are exchanging fermions~\cite{Bellazzini:2016xrt,Remmen:2020uze}.
Requiring $t\rightarrow 0$ to correspond to the forward limit of elastic scattering, if the two initial particles have helicities $h_1$ and $h_2$, then working in a convention with all particles incoming (as we do so throughout), we must take $h_4 = -h_1$ and $h_3 = -h_2$.
Running through a generalization of the contour integral considered in Ref.~\cite{Remmen:2020uze}, we arrive at the following dispersion relation:
\be
\lim_{t\rightarrow 0} \partial_t {\cal A}_{h_1;h_2}(0,t) =-C_\infty^{(t)} + \frac{(-1)^{2h_1}}{\pi}\int_0^\infty {\rm d}s \lim_{t\rightarrow 0}\partial_t \left[\frac{{\rm Im}\, {\cal A}_{h_1;h_2}(s,t)}{s} + \frac{{\rm Im}\, \bar{\cal A}_{-h_1;h_2}(s,t)}{s+t}\right].
\label{eq:dispt}
\ee
This relation introduces a distinct residue at infinity (i.e., a new boundary term), 
\be 
C_\infty^{(t)} = (-1)^{2h_1}\,{\rm Res}\left[\frac{\lim_{t\rightarrow 0} \partial_t {\cal A}_{h_1;h_2}(s,t)}{s},s=\infty\right]\!.
\label{eq:Ct}
\ee

Before continuing, we note that our choice of achieving the $s \to 0$ limit by exchanging the in and out nature of particles 1 and 4 was arbitrary, and we could equally well have exchanged the properties of particles 2 and 3.
The sign of the right-hand sides in Eqs.~\eqref{eq:dispt} and \eqref{eq:Ct} would then be determined by $h_2$ rather than $h_1$.
For purely fermionic amplitudes this exchange is irrelevant, but when scattering a fermion and boson, consistency would require $\lim_{t\rightarrow 0} \partial_t {\cal A}_{h_1;h_2}(0,t) = 0$, and indeed this is the case.
Taking the fermion to correspond to particles 1 and 4, the kinematic dependence of the amplitude can only be supported by expressions such as $\langle 12 \rangle [24]$, $[12] \langle 24 \rangle$, $\langle 13 \rangle [34]$, or $[13] \langle 34 \rangle$, all of which vanish as $s \to 0$.

Moving forward, our strategy is to expand the amplitudes that appear in the integrands of Eqs.~\eqref{eq:disps} and \eqref{eq:dispt} in partial waves of total angular momentum. We then apply unitarity at the level of each partial amplitude $a^{(j)}(s)$ of fixed total angular momentum $j$, requiring that ${\rm Im}\,a^{(j)}(s)>0$.
Doing so will allow us to derive sum rules that access finer details of the UV than \Eq{eq:disps}.
For initial particles with nonzero helicity, recasting the amplitude in partial waves of the total angular momentum requires the Jacob-Wick expansion~\cite{Jacob:1959at} (see also \Refs{Horejsi:1993hz,ItzyksonZuber,Arkani-Hamed:2017jhn}), in which the Legendre polynomials are replaced by the Wigner small $d$-matrices.
We have\footnote{The sum runs over $j \in \mathbb{N}_0 + \lambda$, so that if, e.g., $\lambda=3/2$, $j\in \{3/2,5/2,\ldots\}$.}
\be
{\cal A}_{\alpha,h_{1};\beta,h_{2}}(s,t) =16\pi\sum_{j=\lambda}^{\infty}(2j+1)a_{\alpha,h_{1};\beta,h_{2}}^{(j)}(s)\,d_{\lambda,\lambda}^{(j)}(z),\label{eq:partialwave}
\ee
where we have introduced $\lambda = |h_1-h_2|$, $z=\cos \theta = 1 +2t/s$ with $\theta$ the scattering angle, and $\alpha$ and $\beta$ as labels denoting the internal quantum numbers (for example, flavor) of the two scattered states.
We can similarly expand the crossed amplitude $\bar {\cal A}$, which in the forward limit becomes ${\cal A}_{\bar\alpha,-h_1;\beta,h_2}$.
When implementing this expansion, it is convenient to express the Wigner small $d$-matrices in terms of the Jacobi polynomials,
\be
d_{\lambda,\lambda}^{(j)}(z)=\left(\frac{1+z}{2}\right)^{\lambda}P_{j-\lambda}^{(0,2\lambda)}(z).
\ee
In particular, in the forward limit, where $z=1$, we have $P_{n}^{(0,\beta)}(1)=1$, and further to evaluate the derivative in \Eq{eq:dispt}, we use $\lim_{z\rightarrow 1} {\rm d}P_{n}^{(0,\beta)}(z)/{\rm d}z=n(\beta+n+1)/2$, so that
\be 
\lim_{z\rightarrow 1}\frac{{\rm d}}{{\rm d}z}d_{\lambda,\lambda}^{(j)}(z)=\frac{1}{2}\left[\lambda+(j-\lambda)(j+\lambda+1)\right]\!.
\ee

Combining these results, we can apply the Jacob-Wick expansion to the $s$-channel dispersion relation in \Eq{eq:disps}.
Doing so yields
\bea
&\lim_{s\rightarrow 0}\partial_s {\cal A}_{\alpha,h_{1};\beta,h_{2}}(s,0) + C_{\infty}^{(s)} \\
= &16 \int_0^\infty \frac{{\rm d}s}{s^2} \Biggr\{ \sum_{j=\lambda}^{\infty} (2j+1) {\rm Im}\,a_{\alpha,h_{1};\beta,h_{2}}^{(j)}(s) - \sum_{j=\bar \lambda}^{\infty} (2j+1) {\rm Im}\,a_{\bar \alpha,-h_{1};\beta,h_{2}}^{(j)}(s) \Biggr\},
\label{eq:sums}
\eea
where we have defined $\bar{\lambda} = |h_1 + h_2|$.
Continuing, the partial-wave expansion of \Eq{eq:dispt} yields
\bea
&\lim_{t\rightarrow0}\partial_{t}{\cal A}_{\alpha,h_{1};\beta,h_{2}}(0,t) + C_\infty^{(t)} \\
= & 16(-1)^{2h_1}\!\! \int_0^\infty \frac{{\rm d} s}{s^2}  \Biggl\{ \sum_{j=\lambda}^\infty (2j + 1)\left[\lambda +  \left(j-\lambda\right)\left(j+\lambda+1\right)\right] {\rm Im}\,a_{\alpha,h_{1};\beta,h_{2}}^{(j)}(s)\\
& \hspace{2.9cm}+\sum_{j=\bar{\lambda}}^\infty (2j + 1)\left[\bar{\lambda} + \left( j- \bar{\lambda}\right) \left(j+ \bar{\lambda}+1\right) -1\right] {\rm Im}\,a_{\bar \alpha,-h_{1};\beta,h_{2}}^{(j)}(s)\Biggr\}.
\label{eq:sum1}
\eea
Finally, combining the two, we have an additional sum rule with different weighting of the partial waves,
\bea
&\lim_{t\rightarrow0}\partial_{t}{\cal A}_{\alpha,h_{1};\beta,h_{2}}(0,t) + \lim_{s\rightarrow0}\partial_{s}{\cal A}_{\alpha,h_{1};\beta,h_{2}}(s,0) + C_\infty^{(s)} + C_\infty^{(t)}\\
= &16 (-1)^{2h_1}\! \! \int_0^\infty \frac{{\rm d} s}{s^2}  \Biggl\{ \sum_{j=\lambda}^\infty (2j + 1)\left[\lambda +  \left(j-\lambda\right)\left(j+\lambda+1\right) \! + \! (-1)^{2h_1} \right] {\rm Im}\,a_{\alpha,h_{1};\beta,h_{2}}^{(j)}(s)\\
& \hspace{2.9cm}+\sum_{j=\bar{\lambda}}^\infty (2j + 1)\left[\bar{\lambda} + \left( j- \bar{\lambda}\right) \left(j+ \bar{\lambda}+1\right)\! -\! 1 \! - \! (-1)^{2h_1}\right] {\rm Im}\,a_{\bar \alpha,-h_{1};\beta,h_{2}}^{(j)}(s)\! \Biggr\}.
\label{eq:sum2}
\eea
Depending on the helicities of the incoming states, either \Eq{eq:sum1} or \Eq{eq:sum2} will be most useful in formulating informative sum rules.
For example, if $\bar\lambda \geq 1$, then the right-hand side of \Eq{eq:sum1} is either positive- or negative-definite, with sign $(-1)^{2h_1}$, while if $(-1)^{2h_1} = -1$ and $\lambda\geq 1$, the right-hand side of \Eq{eq:sum2} is negative-definite.
In \Refc{Remmen:2020uze}, this latter observation was exploited by considering the scattering of opposite-helicity fermions, so that $\lambda = 1$ and $\bar{\lambda}=0$.

In Sec.~\ref{sec:sumrules}, we will use the dispersion relations in Eqs.~\eqref{eq:sum1} and \eqref{eq:sum2} to derive the full set of spinning sum rules.
As already mentioned, these relations will connect the sign of the Wilson coefficients in the IR to the physical characteristics of the UV in terms of which angular momentum contributions dominate the sums over $j$ above.
The spinning sum rules can be contrasted with the result in \Eq{eq:sums}, which is the expansion of the conventional forward dispersion relation.
The sign difference that enters in that relation between the sum over $\lambda$ and that over $\bar \lambda$ obstructs similar arguments regarding which angular momentum states dominate in the UV, before even considering boundary terms.
Before deriving the main results of this work, however, let us first review the basis of the SMEFT and establish our conventions for it.

\section{Basis of SMEFT Operators}\label{sec:basis}

\renewcommand{\arraystretch}{1.3}
\begin{table}[htbp]
\begin{center}
\begin{tabular}{C{2.1cm}  L{4.8cm} | C{2.1cm} L{4.8cm}}
\multicolumn{2}{c|}{$F^3$ operators} & \multicolumn{2}{c}{\hspace{-0.8cm}$H^4$ operators} \\ 
&\phantom.{}\\
${\cal O}_{G}$ & $f^{abc} G^{a\;\nu}_{\mu} G^{b\;\rho}_\nu G^{c\;\mu}_\rho$ & ${\cal O}_{H1}$& $(H^\dagger H)\Box(H^\dagger H)$
\\$\widetilde {\cal O}_{G}$ & $f^{abc} G^{a\;\nu}_{\mu} G^{b\;\rho}_\nu \widetilde G^{c\;\mu}_\rho$ & ${\cal O}_{H2}$ & $(H^\dagger D_\mu H)^* (H^\dagger D^\mu H)$
\\
${\cal O}_{W}$ & $\varepsilon^{IJK} W^{I\;\nu}_{\mu} W^{J\;\rho}_\nu W^{K\;\mu}_\rho$ 
\\$\widetilde {\cal O}_{W}$  & $\varepsilon^{IJK} W^{I\;\nu}_{\mu} W^{J\;\rho}_\nu \widetilde W^{K\;\mu}_\rho$ & \cline{1-2}  & \phantom{.}
 \\ \cline{1-2} &\phantom{.}  & \multicolumn{2}{c}{$\psi^4$ operators} 
\\ \multicolumn{2}{c|}{$H^2 F^2$ operators}\\ &\phantom{.} & ${\cal O}_e$ & $-(\bar e_m \gamma_\mu e_n)(\bar e_p \gamma^\mu e_q)$\\
${\cal O}_{HB}$ & $(H^\dagger H) B_{\mu\nu}B^{\mu\nu}$ & ${\cal O}_u$& $-(\bar u_m \gamma_\mu u_n)(\bar u_p \gamma^\mu u_q)$\\
$\widetilde {\cal O}_{HB}$ & $(H^\dagger H) B_{\mu\nu}\widetilde B^{\mu\nu}$ & ${\cal O}_d$ &$-(\bar d_m \gamma_\mu d_n)(\bar d_p \gamma^\mu d_q)$\\
${\cal O}_{HW}$ & $(H^\dagger H) W^I_{\mu\nu}W^{I\mu\nu}$ & ${\cal O}_L$& $-(\bar L_m \gamma_\mu L_n)(\bar L_p \gamma^\mu L_q)$\\
$\widetilde {\cal O}_{HW}$ & $(H^\dagger H) W^I_{\mu\nu}\widetilde W^{I\mu\nu}$ & ${\cal O}_{Q1}$& $-(\bar Q_m \gamma_\mu Q_n)(\bar Q_p \gamma^\mu Q_q)$\\
${\cal O}_{HG}$ & $(H^\dagger H) G^a_{\mu\nu}G^{a\mu\nu}$ & ${\cal O}_{Q2}$ & $-(\bar Q_m \tau^I \gamma_\mu Q_n)(\bar Q_p \tau^I \gamma^\mu Q_q)$\\
$\widetilde {\cal O}_{HG}$ & $(H^\dagger H) G^a_{\mu\nu}\widetilde G^{a\mu\nu}$ & ${\cal O}_{eu}$& $-(\bar e_m \gamma_\mu e_n)(\bar u_p \gamma^\mu u_q)$
\\ &\phantom{.}& ${\cal O}_{ed}$& $-(\bar e_m \gamma_\mu e_n)(\bar d_p \gamma^\mu d_q)$
 \\ \cline{1-2} &\phantom{.}  & ${\cal O}_{ud1}$& $-(\bar u_m \gamma_\mu u_n)(\bar d_p \gamma^\mu d_q)$
 \\ \multicolumn{2}{c|}{$H^2 \psi^2$ operators}& ${\cal O}_{ud2}$& $-(\bar u_m T^a\gamma_\mu u_n)(\bar d_p T^a\gamma^\mu d_q)$
 \\ &\phantom{.}& ${\cal O}_{LQ1}$& $-(\bar L_m \gamma_\mu L_n)(\bar Q_p \gamma^\mu Q_q)$
 \\${\cal O}_{He}$ & $(H^\dagger i \overleftrightarrowalt{D}_\mu H) (\bar e_m \gamma^\mu e_n)$ & ${\cal O}_{LQ2}$& $-(\bar L_m \tau^I \gamma_\mu L_n)(\bar Q_p \tau^I \gamma^\mu Q_q)$
  \\${\cal O}_{Hu}$ & $(H^\dagger i \overleftrightarrowalt{D}_\mu H) (\bar u_m \gamma^\mu u_n)$ & ${\cal O}_{Le}$& $(\bar L_m \gamma_\mu L_n)(\bar e_p \gamma^\mu e_q)$
   \\${\cal O}_{Hd}$ & $(H^\dagger i \overleftrightarrowalt{D}_\mu H) (\bar d_m \gamma^\mu d_n)$ & ${\cal O}_{Lu}$& $(\bar L_m \gamma_\mu L_n)(\bar u_p \gamma^\mu u_q)$
    \\${\cal O}_{HL1}$ & $(H^\dagger i \overleftrightarrowalt{D}_\mu H) (\bar L_m \gamma^\mu L_n)$ & ${\cal O}_{Ld}$& $(\bar L_m \gamma_\mu L_n)(\bar d_p \gamma^\mu d_q)$
     \\${\cal O}_{HL2}$ & $(H^\dagger i \overleftrightarrowalt{D}\vphantom{D}^I_\mu H) (\bar L_m \tau^I\gamma^\mu L_n)$ & ${\cal O}_{Qe}$& $(\bar Q_m \gamma_\mu Q_n)(\bar e_p \gamma^\mu e_q)$
         \\${\cal O}_{HQ1}$ & $(H^\dagger i \overleftrightarrowalt{D}_\mu H) (\bar Q_m \gamma^\mu Q_n)$ & ${\cal O}_{Qu1}$& $(\bar Q_m \gamma_\mu Q_n)(\bar u_p \gamma^\mu u_q)$
     \\${\cal O}_{HQ2}$ & $(H^\dagger i \overleftrightarrowalt{D}\vphantom{D}^I_\mu H) (\bar Q_m \tau^I\gamma^\mu Q_n)$ & ${\cal O}_{Qu2}$& $(\bar Q_m T^a \gamma_\mu Q_n)(\bar u_p T^a\gamma^\mu u_q)$
 \\  &\phantom{.}  & ${\cal O}_{Qd1}$& $(\bar Q_m \gamma_\mu Q_n)(\bar d_p \gamma^\mu d_q)$
  \\  &\phantom{.}  & ${\cal O}_{Qd2}$& $(\bar Q_m T^a \gamma_\mu Q_n)(\bar d_p T^a \gamma^\mu d_q)$
\end{tabular}
\caption{The subset of operators in the dimension-six SMEFT~\cite{Grzadkowski:2010es} considered in this work: those containing quartic interactions allowing for two-to-two elastic scattering of eigenstates of spin and the SM gauge group.}
\label{tab:ops}
\end{center}
\end{table}

We wish to apply the sum rules in Eqs.~\eqref{eq:sum1} and \eqref{eq:sum2} to the dimension-six operators in the SMEFT.
Let us now enumerate the operators in the basis of \Refc{Grzadkowski:2010es} that we will consider.
We write the SM field content as follows.
We have left-handed fermions ($L$, $Q$) and right-handed fermions ($e$, $d$, $u$) comprising the quarks ($Q$, $u$, $d$) and leptons ($L$, $e$).
The multiplet $L$ is a fundamental of ${\rm SU}(2)_L$, $u$ and $d$ are fundamentals of ${\rm SU}(3)_C$, while $Q$ is a fundamental and $e$ a singlet of both.
We will define the hypercharges carried by the fermions under ${\rm U}(1)_Y$ as $(Y_Q,Y_L,Y_u,Y_d,Y_e)= (+1/6,-1/2,+2/3,-1/3,-1)$.
We write the generators of ${\rm SU}(2)_L$ and ${\rm SU}(3)_C$ as $\tau^I = \sigma^I/2$ and $T^a = \lambda^a/2$, where $\sigma^I$ are the Pauli and $\lambda^a$ the Gell-Mann matrices.
We associate each of the fermionic fields with a generation index $m,n,p,q$, running from $1$ to $N_f$ ($=3$ in the SM), parameterizing flavor.
We will write the SM bosons as $B_\mu$, $W_\mu^I$, and $G_\mu^a$ for ${\rm U}(1)_Y$, ${\rm SU}(2)_L$, and ${\rm SU}(3)_C$, with couplings $g'$, $g$, and $g_s$ (and Weinberg angle $\tan \tW = g'/g$) and defining the respective field strengths as $B_{\mu\nu}$, $W_{\mu\nu}^I$, and $G_{\mu\nu}^a$.
Finally, we write the full, complex Higgs doublet---a fundamental of ${\rm SU}(2)_L$ carrying half a unit of hypercharge---as 
\be
H_i = \frac{1}{\sqrt{2}}\left(\begin{array}{c} \phi_1 + i \phi_2 \\ \phi_3 + i \phi_4 \end{array} \right)\!,\label{eq:Higgs}
\ee
with $\phi_{1,2,3,4}\in \mathbb{R}$, where $i$ indexes the fundamental SU(2)$_L$ representation, which we will contract using $\epsilon_{ij}$.

To apply our sum rules, we need to consider forward, two-to-two elastic scattering amplitudes for states of definite helicity.
For simplicity, we will also restrict to the scattering of definite eigenstates of the SM gauge group (i.e., we will not consider processes with superpositions of $e$ and $L$, $L$ and $Q$, $H$ and $u$, $G_{\mu}^a$ and $H$, etc.).
Thus, we need the quartic operators in the dimension-six SMEFT with two copies of one field and two copies of another.
A full, minimal basis at dimension-six was constructed in \Refc{Grzadkowski:2010es}.
The subset that we will need is given in \Tab{tab:ops}, where we make use of the following notation:
\bea
(H^\dagger \overleftrightarrowalt{D}_\mu H) &= H^\dagger (D_\mu H) - (D_\mu H)^\dagger H
\\
(H^\dagger \overleftrightarrowalt{D}\vphantom{D}^I_\mu H) &= H^\dagger \tau^I (D_\mu H) - (D_\mu H)^\dagger \tau^I H.
\eea
All of the operators in \Tab{tab:ops}, except for the four $F^3$ operators, have tree-level UV completions, as shown in \Refc{deBlas:2017xtg}.
The full class of one-loop completions of ${\cal O}_G$ and ${\cal O}_W$ are also known, and were enumerated in \Refc{Quevillon:2018mfl}.

\section{SMEFT Spinning Sum Rules}\label{sec:sumrules}

We now apply the sum rules in Eqs.~\eqref{eq:sum1} and \eqref{eq:sum2} directly to the SMEFT operators of \Sec{sec:basis}, considering each possible combination of helicities from \Tab{tab:ops} in turn.
Spinning sum rules will be established for purely scalar and purely fermionic amplitudes.
The obstruction in the remaining cases will be that the EFT amplitudes we wish to constrain exactly vanish.
Nevertheless, even in these cases we will show that the dispersion relations carry information about the structure of the UV.

At the outset, let us emphasize two points.
Firstly, the sum rules we consider fundamentally exploit the beyond-forward dispersion relation in \Eq{eq:dispt}, which will be how we can obtain information beyond the widely-explored forward relation of \Eq{eq:disps}.
Secondly, to establish a direct connection between the UV spin and the sign of Wilson coefficients, we will require the absence of the boundary terms $C_\infty^{(s)}$ and $C_\infty^{(t)}$: our sharpest relations hold only for the subset of UV completions that do not generate such poles, and we explore this point in detail in \Sec{sec:boundary}.
We will, however, keep track of the boundary contributions throughout our discussion in this section, explicitly stating wherever we require the boundary terms to vanish and otherwise including them in our discussion of what properties of the UV we can infer.

\subsection{$h_1 =h_2 =0$}\label{sec:spin00}

We consider first the only case where the external particles carry no spin, the scattering of Higgs bosons through the two $H^4$ operators in \Tab{tab:ops}.
Using these operators, we wish to compute the amplitude and sum rule associated with scattering an arbitrary superposition of the real $\phi_i$ fields, with the state of the particles given by 
\be
\ket{1} = \alpha_i \ket{\phi_i},\hspace{0.5cm}
\ket{2} = \beta_i \ket{\phi_i},\hspace{0.5cm}
\ket{3} = \gamma_i \ket{\phi_i},\hspace{0.5cm}
\ket{4} = \delta_i \ket{\phi_i}.
\label{eq:scalarstates}
\ee
For elastic scattering, we require $\delta_i = \bar{\alpha}_i$ and $\gamma_i = \bar{\beta}_i$.\footnote{Throughout, we work in a convention where the momenta of states involved in two-to-two scattering are all taken as incoming and labeled cyclicly.
Cyclic labeling implies that, with our metric convention, we have Mandelstam invariants $s = - (p_1+p_2)^2 = - (p_3+p_4)^2$, $t = - (p_1+p_4)^2 = - (p_2+p_3)^2$, and $u = - (p_1+p_3)^2 = - (p_2+p_4)^2$.}
The amplitude is\footnote{Canonical normalization of the external state requires $|\alpha| = |\beta| = 1$.
However, we choose to keep such terms unsimplified in order to leave expressions homogeneous in both $\alpha$ and $\beta$.}
\bea
{\cal A}_{\alpha;\beta} = \;& 2c^{H1}\left(s|\alpha\cdot\beta|^2 + t|\alpha|^2 |\beta|^2 + u |\alpha\cdot\bar{\beta}|^2\right) \\
&+ c^{H2}s\left[(\alpha \gamma^1 \gamma^3 \bar{\alpha})(\beta\gamma^1  \gamma^3 \bar{\beta}) - \frac{1}{2} (\alpha \gamma^1 \bar{\alpha})(\beta  \gamma^1  \bar{\beta}) - \frac{1}{2}(\alpha \gamma^3 \bar{\alpha})(\beta \gamma^3 \bar{\beta})\right] \\
&+ \frac{1}{2} c^{H2} t\left(\varepsilon^{ijkl}\alpha_i \beta_j \bar{\beta}_k \bar{\alpha}_l -|\alpha|^2 |\beta|^2 + |\alpha\cdot\bar{\beta}|^2 +3 |\alpha \gamma^1 \gamma^3 \beta|^2\right)\!,
\label{eq:ampH4gen}
\eea
where $\epsilon^{ijkl}$ is the four-dimensional Levi-Civita tensor, and we write $\gamma^\mu$ for the gamma matrices in the Dirac representation.
(Note the scattering amplitude is purely bosonic, the $\gamma$ matrices are simply being used to capture the matrix structures that appear in the amplitude in a more compact form.)

The amplitude simplifies considerably if we choose a less general $\ket{2}$ state, taking $\beta_i = (1,0,0,0)$, i.e., we analyze the scattering of an arbitrary superposition of the $\phi_i$ off of $\phi_1$.
Doing so, the amplitude in \Eq{eq:ampH4gen} reduces to
\be 
{\cal A}_{\alpha} = t \left[2 c^{H1}\left(|\alpha_2|^2 +|\alpha_3|^2 + |\alpha_4|^2\right)+ \frac{1}{2}c^{H2}\left(2|\alpha_2|^2 -|\alpha_3|^2 - |\alpha_4|^2\right)\right]\!.
\label{eq:ampH4gen-simp}
\ee
This amplitude is fully invariant under crossing the 1 and 4 states: it depends only on $t$ and is invariant under $\alpha_i \leftrightarrow \bar{\alpha}_i$.
If this symmetry persists in the UV, then we would have ${\rm Im}\,a_{\alpha}^{(j)}(s) = {\rm Im}\,a_{\bar\alpha}^{(j)}(s)$, and as $\partial_s {\cal A}_\alpha(s,0) = 0$, \Eq{eq:sums} implies that $C_\infty^{(s)} = 0$.
Alternatively, so long as the UV does not generate a contribution to $C_\infty^{(s)}$, then \Eq{eq:sums} would demand $\sum_j (2j+1){\rm Im}\,a_{\alpha}^{(j)}(s) = \sum_j (2j+1){\rm Im}\,a_{\bar\alpha}^{(j)}(s)$.
In any case, remaining agnostic about the status of this symmetry in the UV, we have the spinning sum rule from \Eq{eq:sum1}:
\bea
&\lim_{t\rightarrow0}\partial_{t}{\cal A}_{\alpha;\beta}(0,t) + C_\infty^{(t)} \\
 & = 16\int_0^\infty \frac{{\rm d} s}{s^2} \sum_{j=0}^\infty (2j + 1) \left\{ \left[j\left(j+1\right)\right] {\rm Im}\,a_{\alpha;\beta}^{(j)}(s) + \left[j\left(j+1\right) {-}1\right] {\rm Im}\,a_{\bar \alpha;\beta}^{(j)}(s)\right\}.
\label{eq:Hsumfull}
\eea
From the sum over partial waves, we see that if the UV completion is dominated by scalars or states with spin, then the above linear combinations are either negative or positive, respectively.
This statement is, of course, modulo the boundary term $C^{(t)}_{\infty}$, and in Sec.~\ref{sec:boundary} we will expand on when we expect such contributions.
We emphasize that the connection between sign and spin is reversed as compared with the result we stated for fermion scattering in \Eq{eq:dim6fermionbound}, and we will see in \Sec{sec:spinonehalf} that the fermionic spinning sum rule is positive for scalars and negative for vectors. 
For specific scattering states, the same conclusion holds for the following combinations of the operators:
\bea
2c^{H1}+ c^{H2} &=  \! 16\! \int_0^\infty \! \frac{{\rm d} s}{s^2}   \sum_{j=0}^\infty(2j{+}1)\left\{ \left[j\left(j{+}1\right)\right] {\rm Im}\,a_{2}^{(j)}(s) + \left[j\left(j{+}1\right) {-}1\right] {\rm Im}\,\bar{a}_{2}^{(j)}(s)\right\} - C^{(t)}_{\infty}\\
2 c^{H1}-\frac{1}{2}c^{H2} &=  \! 16\! \int_0^\infty \! \frac{{\rm d} s}{s^2}   \sum_{j=0}^\infty(2j{+}1)\left\{ \left[j\left(j{+}1\right)\right] {\rm Im}\,a_{3}^{(j)}(s) + \left[j\left(j{+}1\right) {-}1\right] {\rm Im}\,\bar{a}_{3}^{(j)}(s)\right\} - C^{(t)}_{\infty}\\
&=  \! 16\! \int_0^\infty \! \frac{{\rm d} s}{s^2}   \sum_{j=0}^\infty(2j{+}1)\left\{ \left[j\left(j{+}1\right)\right] {\rm Im}\,a_{4}^{(j)}(s) + \left[j\left(j{+}1\right) {-}1\right] {\rm Im}\,\bar{a}_{4}^{(j)}(s)\right\} - C^{(t)}_{\infty},
\label{eq:sumhiggs}
\eea
where $a^{(j)}_{2,3,4}(s)$ are the UV partial-wave amplitudes for $\phi_1 + \phi_{2,3,4} \rightarrow X$, respectively.

\begin{figure}[t]
\begin{center}
\includegraphics[width=6.5cm]{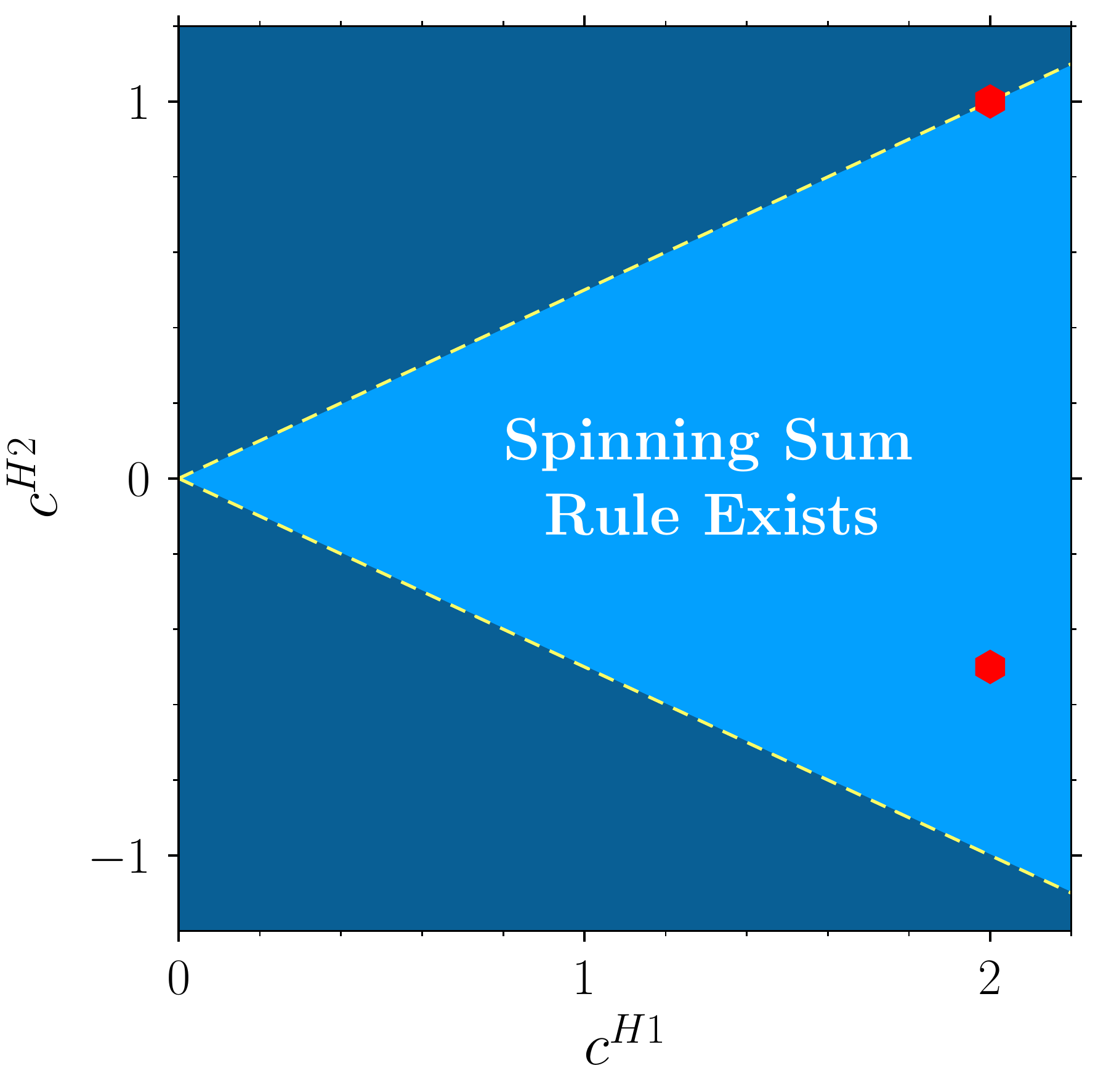}
\end{center}
\vspace{-0.5cm}
\caption{The cone within which the coefficients of the two Higgs operators in \Tab{tab:ops}---${\cal O}_{H1}$ and ${\cal O}_{H2}$---can be chosen such that we can construct a spinning sum rule: a relationship between the sign of the IR coefficients and the dominant spin in the UV completion.
The edges of the cone extend from the origin out through $2c^{H1} \pm c^{H2}$, which are both constructed explicitly in the text. 
The hexagons show the specific sum rules considered in \Eq{eq:sumhiggs}, and the figure highlights that these are just two examples of a broad class of sum rules.
}
\label{fig:scalar-region}
\end{figure}

These results are not exhaustive, however.
Indeed, as ${\cal O}_{H2}$ violates custodial symmetry---the ${\rm O}(4)$ symmetry associated with rotating the $\phi_i$ among each other---we should not expect scattering off a single state to fully determine the physics, that is to say, \Eq{eq:ampH4gen-simp} does not contain the full information of \Eq{eq:ampH4gen}.
As an explicit example, if we take $\alpha = (1,-i,0,0)/\sqrt{2}$ and $\beta = (0,0,i,1)/\sqrt{2}$,\footnote{Note that if the full setup of the scattering experiment is crossing invariant, then $\sigma = \bar\sigma$, the IR amplitude is proportional to $t$, and the sum rule in \Eq{eq:disps} vanishes.
Thus, in the $\phi_i$ basis, the only way to obtain an amplitude that is not identical after crossing---i.e., to obtain a term $\propto s$ in the IR---is to include complex values in the superposition.
As discussed in Ref.~\cite{LianTao}, a relation such as $\sigma = \bar\sigma$ can result in the UV if custodial symmetry is preserved.} the resulting amplitude is ${\cal A}(s,t) = 2 c^{H1} t + c^{H2} u$, which allows us to construct a spinning sum rule for the combination $2c^{H1}- c^{H2}$.
More generally, by allowing arbitrary complex $\alpha_i$ and $\beta_i$, we can construct a spinning sum rule for any choice of coefficients within the region shown in Fig.~\ref{fig:scalar-region}.
In particular, note that a spinning sum rule can be constructed for $c^{H1}$ alone, but not $c^{H2}$.
We emphasize that the sign of the combination of Wilson coefficients in this region is not positive, but instead dictated by the dominant spin in the UV (unless the UV generates a pole at infinity).
We could, of course, obtain additional sum rules by subtracting various spinning sum rules from each other, but doing so would not in general be compatible with this connection between signs and UV spin.

We can compare the spinning sum rules we have obtained to previous sum rules used to study dimension-six Higgs operators.
These were the operators considered in the early work of \Refc{Low:2009di}.
In particular, accounting for sign conventions, \Refc{Low:2009di} finds via dispersion relations for quartic Higgs scattering in the charged-state basis that a negative coefficient of ${\cal O}_{H1}$ can only be generated at tree level via scalar exchange, which agrees with what we find in the real Higgs basis in Eq.~\eqref{eq:Hsumfull} above.
A further notable example those authors construct is the coupling of heavy vectors to the Higgs to generate a positive $c^{H1}$, which is again consistent with Eq.~\eqref{eq:Hsumfull}.

More recently, \Refc{LianTao} considered what sum rules can be determined on the same Higgs operators using only the forward dispersion relation in \Eq{eq:disps}.
The authors of that work adopt a different basis from what we outlined in \Tab{tab:ops}.
Specifically, they consider ${\cal O}_H = - (\partial_{\mu} |H|^2)^2/2$ and ${\cal O}_T = - (H^{\dagger} \overleftrightarrowalt{D}_{\mu} H)^2/2$, where we have introduced an overall sign to account for the opposite signature convention employed in that work.
It is straightforward to map between these bases, and doing so reveals that the two sum rules they establish on $c_H + 3 c_T$ and $-2c_T$ correspond to results for $2c^{H1}+c^{H2}$ and $-c^{H2}$, respectively.
The first of these lies on the boundary of the region shown in Fig.~\ref{fig:scalar-region}, while the second lies outside it.
There is no contradiction, of course, as this is simply a result of the difference between traditional sum rules as established in, e.g., \Refc{LianTao} and the spinning sum rules we consider here.
For the combination $2c^{H1}+c^{H2}$, it so happens that both types of sum rules can be established.

\subsection{$h_1 = \pm 1/2$, $h_2 =0$}\label{sec:spinhalf0}

Next, we consider the case of fermion-scalar scattering (e.g., elastic scattering of a lepton or quark with a Higgs), with $h_1 = \pm 1/2$ and $h_2 = 0$.
Accordingly, we have $\lambda=\bar \lambda=1/2$, and then further introducing a new summation index $n = j-1/2$, we can write \Eq{eq:sum1} as\footnote{Where our results are independent of the helicity of the external states, as is the case here, we will not explicitly flag the helicity in the amplitude as a subscript.
When it is, as is the case for $|h_1| = |h_2| = 1/2$ or $1$, we will.}
\bea
&\lim_{t\rightarrow0}\partial_{t}{\cal A}_{\alpha;\beta}(0,t) + C_{\infty}^{(t)} \\
&= -32 \int_0^\infty  \frac{{\rm d} s}{s^2}  \sum_{n=0}^\infty (n+1)\Biggl\{\left[n(n+2)+\frac{1}{2}\right] {\rm Im}\,a_{\alpha;\beta}^{(n)}(s)+\left[n(n+2)-\frac{1}{2}\right] {\rm Im}\,a_{\bar \alpha;\beta}^{(n)}(s)\Biggr\},
\label{eq:disptfermionscalar}
\eea
where $a_{\bar\alpha;\beta}$ denotes the partial waves for the helicity-flipped process with $h_1 = \mp 1/2$ and with any internal quantum numbers of the fermion complex-conjugated.

As discussed in Sec.~\ref{sec:dispersion}, when scattering a boson and fermion we expect $\lim_{t\rightarrow0}\partial_{t}{\cal A}_{\alpha;\beta}(0,t) = 0$, and indeed we will confirm this in explicit examples below.
However, before any analysis this observation implies that a spinning sum rule, which makes central use of this EFT amplitude, cannot be constructed for fermion-scalar scattering.
Nevertheless, we show that even in this case the dispersion relations still carry considerable information.

Let us now apply \Eq{eq:disptfermionscalar} to the $H^2 \psi^2$ operators in \Tab{tab:ops}.
To begin with, note that the Wilson coefficient tensors $c_{mn}$ for each of these operators must be hermitian, $c_{mn}=\bar{c}_{nm}$, which follows from the hermiticity of the operators themselves. 
Moreover, the operators conserve CP if and only if all $c_{mn}$ are real.
Expanding the Higgs in real scalars, as in \Eq{eq:Higgs}, we have
\be
H^{\dagger}i\overleftrightarrowalt{\partial}_{\mu}H= \phi_{2}\partial_{\mu}\phi_{1}-\phi_{1}\partial_{\mu}\phi_{2}+\phi_{4}\partial_{\mu}\phi_{3}-\phi_{3}\partial_{\mu}\phi_{4}.
\ee
Computing $\bar{e}^{-}H\rightarrow \bar{e}^- H$ scattering (where the sign labels the helicity, not electric charge), with $\alpha_{m}$ and $\delta_m$ denoting the flavor of the $e_{m}$ and $\beta_{i}$ and $\gamma_i$ denoting the superposition of $\phi_{i}$, we have
\be 
{\cal A} =-i\alpha_{m}\delta_{n}c_{mn}^{He}(\beta_{1}\gamma_{2}-\beta_{2}\gamma_{1}+\beta_{3}\gamma_{4}-\beta_{4}\gamma_{3})\left([12] \langle 24 \rangle - [13] \langle 34 \rangle \right)\!.
\ee
For the case of elastic scattering, where we take $\alpha=\bar{\delta}$ and $\beta=\bar{\gamma}$, we have
\be 
{\cal A}_{\alpha;\beta}=2\alpha_{m}\bar{\alpha}_{n}c_{mn}^{He}\left[{\rm Im}(\beta_{1}\bar{\beta}_{2})+{\rm Im}(\beta_{3}\bar{\beta}_{4})\right]\left([12] \langle 24 \rangle -[13] \langle 34 \rangle \right)\!.
\label{eq:eHscattering}
\ee
At special kinematics,
\bea
{\cal A}_{\alpha;\beta}(s,0) & =- 4\alpha_{m}\bar{\alpha}_{n}c_{mn}^{He}\left[{\rm Im}(\beta_{1}\bar{\beta}_{2})+{\rm Im}(\beta_{3}\bar{\beta}_{4})\right]s\\
{\cal A}_{\alpha;\beta}(0,t) & =0.
\label{eq:specialkin}
\eea
Similar expressions hold for all of the other $H^2 \psi^2$ operators.
For example, for the scattering of $L^{-}H$, we must specify both a flavor superposition and an ${\rm SU}(2)_L$ charge vector for the lepton.
Using the completeness relation for the Pauli matrices, we find:
\be 
{\cal O}_{HL2,mn}=\frac{i}{2}(H^{\dagger j}\overleftrightarrowalt{D}_{\mu}H_{i})(\bar{L}_{m}^{i}\gamma^{\mu}L_{nj})-\frac{i}{4}(H^{\dagger j}\overleftrightarrowalt{D}_{\mu}H_{j})(\bar{L}_{m}^{i}\gamma^{\mu}L_{ni}).
\ee
Without loss of generality, we can use the fact that the overall operator ${\cal O}_{HL2}$ is ${\rm SU}(2)$-neutral to rotate the $L$ into ``up'' isospin.
Doing so, we obtain:
\be 
{\cal O}_{HL2,mn}\supset \frac{1}{4}(\phi_{2}\partial_{\mu}\phi_{1}-\phi_{1}\partial_{\mu}\phi_{2}-\phi_{4}\partial_{\mu}\phi_{3}+\phi_{3}\partial_{\mu}\phi_{4})(\bar{L}_{m}^{1}\gamma^{\mu}L_{n1}).
\ee
The amplitude for $HL$ scattering is thus given by
\bea
{\cal A}_{\alpha;\beta}=2\alpha_{m}\bar{\alpha}_{n} &\left[\left(c_{mn}^{HL1}+\frac{1}{4}c_{mn}^{HL2}\right){\rm Im}(\beta_{1}\bar{\beta}_{2})+\left(c_{mn}^{HL1}-\frac{1}{4}c_{mn}^{HL2}\right){\rm Im}(\beta_{3}\bar{\beta}_{4})\right] \\
\times &\left( \langle 12 \rangle [24] - \langle 13 \rangle [34] \right)\!.
\label{eq:HLamp}
\eea
Again this amplitude vanishes for $s \to 0$.

The vanishing of the amplitudes as $s \to 0$ implies that according to \Eq{eq:sum1} the UV partial waves for $H\psi \rightarrow X$, where $\psi$ corresponds to any (superposition of) SM fermion(s), satisfy a special sum rule:
\be 
C_{\infty}^{(t)}=32\int_0^\infty \frac{{\rm d} s}{s^2}  \sum_{n=0}^\infty (n+1)\left\{\left[n(n+2)+\frac{1}{2}\right] {\rm Im}\,a_{\psi,H}^{(n)}(s)  +\left[n(n+2)-\frac{1}{2}\right] {\rm Im}\,a_{\bar\psi,H}^{(n)}(s)\right\}\!.\label{eq:special}
\ee
Since each of the partial waves has positive imaginary part, \Eq{eq:special} implies that either the boundary term is nonzero or that the scattering of $H\psi\rightarrow X$ must be dominated by the $n=0$ mode (i.e., $X$ of spin one-half), since this is the only contribution within the sum that can be negative.
If there is a symmetry in the UV such that $a_{\alpha;\beta} = a_{\bar\alpha;\beta}$ (for examples, see Refs.~\cite{Agashe:2006at,LianTao}), then \Eq{eq:special} would imply that, in theories where $C_\infty^{(t)}$ vanishes, only the $n=0$ term in the sum could be nonzero: the UV must be strictly composed of only the lowest partial wave, a fermionic intermediate state.

Regardless, given that ${\cal A}_{\alpha;\beta}(0,t)=0$, the remaining information will be conveyed by the expanded $s$-dispersion relation,
\be 
\lim_{s\rightarrow 0} \partial_s {\cal A}_{\alpha;\beta}(s,0) = 32 \int_0^\infty \frac{{\rm d} s}{s^2} \sum_{n=0}^\infty (n+1) \left[{\rm Im}\,a_{\alpha;\beta}^{(n)}(s) - {\rm Im}\,a_{\bar \alpha;\beta}^{(n)}(s) \right] - C_{\infty}^{(s)},
\label{eq:sumrulesimp}
\ee 
which is a simple rewriting of \Eq{eq:sums}, and therefore not a spinning sum rule, as discussed in Sec.~\ref{sec:dispersion}.
If $a_{\alpha;\beta} = a_{\bar\alpha;\beta}$, then we should find that the EFT amplitude satisfies $\lim_{s\rightarrow 0} \partial_s {\cal A}_{\alpha;\beta}(s,0) = - C_{\infty}^{(s)}$ and therefore can only be nontrivial if there is a boundary term.
As an explicit example, we can take the $\bar{e}^-H$ scattering considered above and imagine that the SMEFT continues to describe physics in the UV, such that \Eq{eq:eHscattering} describes the scatterings at high energies.
We can achieve $a_{\alpha;\beta} = a_{\bar\alpha;\beta}$ by taking CP-conserving coefficients (i.e., $c_{mn}^{He} = c_{nm}^{He}$), and then from Eqs.~\eqref{eq:Cs} and \eqref{eq:specialkin}, we have $C_{\infty}^{(s)} = 4 \alpha_m \bar{\alpha}_n c_{mn}^{He} \left[{\rm Im}(\beta_{1}\bar{\beta}_{2})+{\rm Im}(\beta_{3}\bar{\beta}_{4})\right]$, as required to reproduce the IR result.

Explicitly, we can write the Wilson coefficients of the SMEFT operators in terms of the sum rule in \Eq{eq:sumrulesimp}. For the $He$, $Hu$, and $Hd$ operators,
\bea
\alpha_{m}\bar{\alpha}_{n}c_{mn}^{He}\left[{\rm Im}(\beta_{1}\bar{\beta}_{2})+{\rm Im}(\beta_{3}\bar{\beta}_{4})\right] &=8 \int_0^\infty \frac{{\rm d} s}{s^2} \sum_{n=0}^\infty (n+1) \left[{\rm Im}\,a_{\bar e,H}^{(n)}(s) - {\rm Im}\,a_{e,H}^{(n)}(s) \right] - \frac{C_{\infty}^{(s)}}{4} \\
\alpha_{m}\bar{\alpha}_{n}c_{mn}^{Hu}\left[{\rm Im}(\beta_{1}\bar{\beta}_{2})+{\rm Im}(\beta_{3}\bar{\beta}_{4})\right] &=8 \int_0^\infty \frac{{\rm d} s}{s^2} \sum_{n=0}^\infty (n+1) \left[{\rm Im}\,a_{\bar u, H}^{(n)}(s) - {\rm Im}\,a_{u, H}^{(n)}(s) \right] - \frac{C_{\infty}^{(s)}}{4} \\
\alpha_{m}\bar{\alpha}_{n}c_{mn}^{Hd}\left[{\rm Im}(\beta_{1}\bar{\beta}_{2})+{\rm Im}(\beta_{3}\bar{\beta}_{4})\right] &=8 \int_0^\infty \frac{{\rm d} s}{s^2} \sum_{n=0}^\infty (n+1) \left[{\rm Im}\,a_{\bar d, H}^{(n)}(s) - {\rm Im}\,a_{d, H}^{(n)}(s) \right] - \frac{C_{\infty}^{(s)}}{4},
\label{eq:He}
\eea
while for $HL$ and $HQ$,
\bea
&\alpha_{m}\bar{\alpha}_{n}\left[\left(c_{mn}^{HL1}+\frac{1}{4}c_{mn}^{HL2}\right){\rm Im}(\beta_{1}\bar{\beta}_{2})+\left(c_{mn}^{HL1}-\frac{1}{4}c_{mn}^{HL2}\right){\rm Im}(\beta_{3}\bar{\beta}_{4})\right] \\ &\qquad\qquad\qquad = 8 \int_0^\infty \frac{{\rm d} s}{s^2} \sum_{n=0}^\infty (n+1) \left[{\rm Im}\,a_{\bar L, H}^{(n)}(s) - {\rm Im}\,a_{L, H}^{(n)}(s) \right] - \frac{C_{\infty}^{(s)}}{4} \\
&\alpha_{m}\bar{\alpha}_{n}\left[\left(c_{mn}^{HQ1}+\frac{1}{4}c_{mn}^{HQ2}\right){\rm Im}(\beta_{1}\bar{\beta}_{2})+\left(c_{mn}^{HQ1}-\frac{1}{4}c_{mn}^{HQ2}\right){\rm Im}(\beta_{3}\bar{\beta}_{4})\right] \\ &\qquad\qquad\qquad = 8 \int_0^\infty \frac{{\rm d} s}{s^2} \sum_{n=0}^\infty (n+1) \left[{\rm Im}\,a_{\bar Q, H}^{(n)}(s) - {\rm Im}\,a_{Q, H}^{(n)}(s) \right] - \frac{C_{\infty}^{(s)}}{4}.
\label{eq:HL}
\eea

As the above results are based on the conventional $s$-channel dispersion relation, there should be a clear connection between our results and those of \Refc{LianTao}.
After converting between bases, the authors of that work established $s$-channel sum rules proportional to the following combinations of Wilson coefficients (where we suppress flavor indices as that work considered only a single generation): $c^{He}$, $c^{Hu}$, $c^{Hd}$, $c^{HL1}\pm c^{HL2}/4$, and $c^{HQ1}\pm c^{HQ2}/4$.
Each of these combinations can be isolated from Eqs.~\eqref{eq:He} and \eqref{eq:HL} by appropriate choices of $\beta_i$.

\subsection{$|h_1| = |h_2| = 1/2$}\label{sec:spinonehalf}

For the next set of external helicities, we turn to four-fermion scattering.
This case was considered in \Refc{Remmen:2020uze}, and as there we consider scattering fermions of opposite helicity, which fixes $\lambda=1$ and $\bar \lambda=0$, so that Eqs.~\eqref{eq:sum1} and \eqref{eq:sum2} become
\bea
&\lim_{t\rightarrow0}\partial_{t}{\cal A}_{\alpha-;\beta+}(0,t) + C_\infty^{(t)} \\
= & 16 \int_0^\infty \frac{{\rm d} s}{s^2}  \left\{{\rm Im}\,a_{\bar \alpha+;\beta+}^{(0)}(s)- \sum_{j=1}^\infty (2j + 1)\left[j(j+1)-1\right] \left[{\rm Im}\,a_{\alpha-;\beta+}^{(j)}(s) + {\rm Im}\,a_{\bar \alpha+;\beta+}^{(j)}(s) \right]\right\}
\label{eq:sumfermions0}
\eea
and
\bea
&\lim_{t\rightarrow0}\partial_{t}{\cal A}_{\alpha-;\beta+}(0,t) + \lim_{s\rightarrow0}\partial_{s}{\cal A}_{\alpha-;\beta+}(s,0) + C_\infty^{(s)} + C_\infty^{(t)}\\
= &-16  \int_0^\infty \frac{{\rm d} s}{s^2}  \left\{ \sum_{j=1}^\infty (2j \!+\! 1) (j\!-\!1)(j\!+\!2) {\rm Im}\,a_{\alpha-;\beta+}^{(j)}(s)+\sum_{j=0}^\infty (2j \!+\! 1)j(j\!+\!1) {\rm Im}\,a_{\bar \alpha+;\beta+}^{(j)}(s) \right\}\!.
\label{eq:sumfermions}
\eea
The right-hand side of the first relation is strictly negative except for the contribution from ${\rm Im}\,a_{\bar \alpha +;\beta +}^{(0)}(s)$, and so this relation allows for the construction of a spinning sum rule.\footnote{For identical-helicity scattering, so that $\lambda = 0$ and $\bar \lambda = 1$, the sum rules we obtain are analogous to those above, but with the left-hand sides in Eqs.~\eqref{eq:sumfermions0} and \eqref{eq:sumfermions} exchanged and with $a_{\alpha;\beta}$ and $a_{\bar\alpha;\beta}$ (and the appropriate helicities) exchanged on the right-hand side.
Thus, for identical-helicity scattering, we instead arrive at a sum rule in which the only positive contribution arises from the lowest partial wave of the uncrossed amplitude.} 
By contrast, the right-hand side of the second relation in \Eq{eq:sumfermions} is of fixed sign.
The definite sign suggests the possibility of a positivity bound at a dimension-six (modulo boundary terms), however, we will see next why this is not possible.

Let us consider a minimal example from \Tab{tab:ops}, ${\cal O}_e$, and use this operator to mediate elastic scattering between a positron and electron.
We take the positron and electron to be flavor superpositions denoted by $\alpha_m$ and $\beta_m$ respectively.
Under these conditions, the amplitude at special kinematics takes the form (see \Refc{Remmen:2020uze} for details):
\bea
{\cal A}_{\alpha-;\beta+}(s,0) &= - 8 \alpha_m \beta_n \bar{\beta}_p \bar{\alpha}_q c_{mnpq}^e s \\
{\cal A}_{\alpha-;\beta+}(0,t) &= + 8 \alpha_m \beta_n \bar{\beta}_p \bar{\alpha}_q c_{mnpq}^e t.
\eea
Accordingly, in this case our relation~\eqref{eq:sumfermions} becomes
\bea
&C_\infty^{(s)} + C_\infty^{(t)}\\
&= -16  \int_0^\infty \frac{{\rm d} s}{s^2}  \left\{ \sum_{j=1}^\infty (2j\!+\!1) (j\!-\!1)(j\!+\!2) {\rm Im}\,a_{\alpha-;\beta+}^{(j)}(s)+\sum_{j=0}^\infty (2j\!+\!1)j(j\!+\!1) {\rm Im}\,a_{\bar \alpha+;\beta+}^{(j)}(s) \right\}\!.
\label{eq:e4sumpre}
\eea
The Wilson coefficients have vanished, removing the possibility of a positivity bound.
Indeed, as noted in \Refc{Remmen:2020uze}, this is a property of the kinematic structure of any four-fermion scattering amplitude at dimension six and not specific to ${\cal O}_e$.
Instead, the dispersion relation constrains the poles at infinity in terms of the partial waves of the massive states in the UV completion.
If the UV completion is such that the boundary contributions vanish, then as noted in \Refc{Remmen:2020uze}, positivity of the partial waves appearing in Eq.~\eqref{eq:e4sumpre} implies that only the $j=1$ partial wave can contribute to $a_{\alpha -;\beta +}^{(j)}$ and only the $j=0$ counterpart can contribute to $a_{\bar\alpha +;\beta +}^{(j)}$; that is, if the boundary terms vanish, the completion must be at tree level (at leading order in a weakly coupled theory) and contain only massive scalars and vectors.
In that case, Eq.~\eqref{eq:sumfermions0} becomes
\be 
\alpha_m \beta_n \bar{\beta}_p \bar{\alpha}_q c_{mnpq}^e  = 2 \int_0^\infty \frac{{\rm d} s}{s^2}  \left[{\rm Im}\,a_{\bar \alpha+;\beta+}^{(0)}(s)- 3\, {\rm Im}\,a_{\alpha-;\beta+}^{(1)}(s) \right]\!,\label{eq:ceesimp}
\ee
the sign of which is determined by whether scalars or vectors dominate the UV.
This result is the detailed version of the qualitative relation we gave in \Eq{eq:dim6fermionbound}, and again we emphasize that the sign is reversed as compared to the scalar spinning sum rules of \Sec{sec:spin00}. 
Similar spinning sum rules and positivity bounds apply to the other four-fermion operators listed in Table~\ref{tab:ops}.
See Ref.~\cite{Remmen:2020uze} for details, as well as a discussion of the importance of these sum rules for flavor and CP violation.

We can again derive a sum rule using the forward dispersion relation in \Eq{eq:sums}.
This is exactly the case considered in \Refc{LianTao}, in which a sum rule is established on $2c^e$, where again flavor indices are suppressed as it was not considered in that work.
Our approach for \Eq{eq:sums} would establish identical relations, but with $c^e \to \alpha_m \beta_n \bar{\beta}_p \bar{\alpha}_q c_{mnpq}^e$, so that relations between different flavor structures can be derived.

\subsection{$h_1 = \pm 1$, $h_2 = 0$}\label{sec:spin10}

Turning to the scattering of a gauge boson in a definite-helicity state off of a Higgs, we can mechanically derive the appropriate relations from Eqs.~\eqref{eq:sum1} and \eqref{eq:sum2}.
Doing so results in a positivity bound and a mixed-sign sum rule (up to boundary terms), namely, 
\bea
&\lim_{t\rightarrow0}\partial_{t}{\cal A}_{\alpha;\beta}(0,t) + C_\infty^{(t)}\\
&=16\int_{0}^{\infty}\frac{{\rm d}s}{s^{2}}\sum_{j=1}^{\infty}(2j+1)\left\{[j(j+1)-1]{\rm Im}a_{\alpha;\beta}^{(j)}(s)+[j(j+1)-2]\,{\rm Im}a_{\bar{\alpha};\beta}^{(j)}(s)\right\}\\&>0
\label{eq:sum101}
\eea
and
\bea
&\lim_{t\rightarrow0}\partial_{t}{\cal A}_{\alpha;\beta}(0,t)+\lim_{s\rightarrow0} \partial_{s} {\cal A}_{\alpha;\beta}(s,0) + C_\infty^{(s)} + C_\infty^{(t)}\\
&=16\int_{0}^{\infty}\frac{{\rm d}s}{s^{2}}\sum_{j=1}^{\infty}(2j+1)\left\{j(j+1)\,{\rm Im}a_{\alpha\beta}^{(j)}(s)+[j(j+1)-3]{\rm Im}a_{\bar{\alpha}\beta}^{(j)}(s)\right\}\!.
\label{eq:sum102}
\eea

In order to make use of these relations, however, we need to compute the relevant SMEFT amplitudes.
By direct calculation, the amplitude for the hypercharge $B$ gauge field and a general superposition of the Higgs $\phi_i$ via the operators in \Tab{tab:ops} is given by
\be
{\cal A}=2\beta_{i}\gamma_{i}\left\{ c^{HB}\left[(e_{1}\cdot e_{4})t+2(e_{1}\cdot p_{4})(e_{4}\cdot p_{1})\right]+2\widetilde{c}^{HB}e_{1}^{\mu}e_{4}^{\nu}p_{1}^{\rho}p_{4}^{\sigma}\epsilon_{\mu\nu\rho\sigma}\right\}\!,
\label{eq:amp01}
\ee
where the gauge boson corresponds to states 1 and 4, $e_i$ and $p_i$ are the polarization and momentum four-vectors for particle $i$, and $\beta_i$ and $\gamma_i$ dictate the superpositions of the initial and final Higgs.
Choosing definite-helicity states for the gauge field to correspond to elastic scattering, so that $h_1 = -h_4 = \pm 1$,  little-group scaling arguments imply that the contact amplitude for dimension-six scalar-vector scattering must vanish~\cite{LianTao}.
It is therefore not possible to constrain the $H^2 B^2$ (and similarly the $H^2 W^2$ and $H^2 G^2$) dimension-six Wilson coefficients with either conventional or spinning sum rules.

As the elastic, fixed-helicity dimension-six amplitude vanishes, \Eq{eq:sum101} implies that
\bea
C_\infty^{(t)}
=\,16\int_{0}^{\infty}\frac{{\rm d}s}{s^{2}}\sum_{j=1}^{\infty}(2j+1)\left\{[j(j+1)-1]{\rm Im}a_{\alpha;\beta}^{(j)}(s)+[j(j+1)-2]\,{\rm Im}a_{\bar{\alpha};\beta}^{(j)}(s)\right\}
>\,0.
\eea
If the pole at infinity is present, it must therefore be positive.
Instead, if it vanishes, then all partial cross sections of the amplitude must vanish except ${\rm Im}\,a_{\bar\alpha;\beta}^{(1)}$.
If we further assume $C_\infty^{(s)} = 0$, so that the UV generates neither boundary term, then Eqs.~\eqref{eq:sum101} and \eqref{eq:sum102} imply that all imaginary parts of partial waves must vanish.
As an explicit example, a simple UV extension of ${\cal O}_{HB}$ is given by $y_1 \varphi H^\dagger H + y_2 \varphi B_{\mu\nu}B^{\mu\nu}$ for a real massive scalar $\varphi$.
Integrating out $\varphi$ from the action, one generates $(H^{\dagger} H) B_{\mu \nu} B^{\mu \nu}$ with Wilson coefficient $c^{HB} = y_1 y_2/m_{\varphi}^2$.
The indefinite sign of the coefficient is representative of the fact that there is no spinning sum rule for these operators.

\subsection{$|h_1| = |h_2| = 1$}

Finally, we consider scattering of gauge bosons via the four $F^3$ operators in \Tab{tab:ops}.
For same-helicity scattering, $h_1 = h_2 = \pm 1$, we have the spinning sum rules\footnote{Here we use $\pm$ subscripts to indicate the helicities of the two initial bosons.}
\bea
\lim_{t\rightarrow0}\partial_{t}{\cal A}_{\alpha,\pm;\beta,\pm}(0,t) + C_\infty^{(t)} &=  16 \int_0^\infty  \frac{{\rm d} s}{s^2}  \Biggl\{ \sum_{j=0}^\infty(2j +1)j(j+1) {\rm Im}\,a_{\alpha,\pm;\beta,\pm}^{FF(j)}(s)\\
&\hspace{2.2cm} + \sum_{j=2}^\infty (2j + 1)\left[j(j+1)-5\right] {\rm Im}\,a_{\bar \alpha,\mp;\beta,\pm}^{FF(j)}(s)\Biggr\}\\&\,>0
\label{eq:sum11}
\eea
and
\bea
&\lim_{t\rightarrow0}\partial_{t}{\cal A}_{\alpha,\pm;\beta,\pm}(0,t) + \lim_{s\rightarrow0}\partial_{s}{\cal A}_{\alpha,\pm;\beta,\pm}(s,0) + C_\infty^{(s)} + C_\infty^{(t)} \\
= \,&16 \int_0^\infty  \frac{{\rm d} s}{s^2}  \Biggl\{ \sum_{j=0}^\infty (2j + 1)\left[ j(j+1)+1\right] {\rm Im}\,a_{\alpha,\pm;\beta,\pm}^{FF(j)}(s)\\
&\hspace{1.63cm} +  \sum_{j=2}^\infty(2j + 1)\left[j(j+1)-6\right] {\rm Im}\,a_{\bar \alpha,\mp;\beta,\pm}^{FF(j)}(s)\Biggr\}\\&\,>0.
\label{eq:sum12}
\eea
For opposite-helicity scattering, $h_1 = -h_2 = \pm 1$, we find the sum rules
\bea
\lim_{t\rightarrow0}\partial_{t}{\cal A}_{\alpha,\pm;\beta,\mp}(0,t) + C_\infty^{(t)} 
&\,=  16 \int_0^\infty  \frac{{\rm d} s}{s^2}  \Biggl\{ \sum_{j=2}^\infty (2j + 1)\left[j(j+1)-4\right] {\rm Im}\,a_{\alpha,\pm;\beta,\mp}^{FF(j)}(s)\\
&\hspace{2.26cm} + \sum_{j=0}^\infty (2j +1)\left[ j(j+1)-1\right] {\rm Im}\,a_{\bar \alpha,\mp;\beta,\mp}^{FF(j)}(s)\Biggr\}
\label{eq:sum21}
\eea
and 
\bea
&\lim_{t\rightarrow0}\partial_{t}{\cal A}_{\alpha,\pm;\beta,\mp}(0,t) + \lim_{s\rightarrow0}\partial_{s}{\cal A}_{\alpha,\pm;\beta,\mp}(s,0) + C_\infty^{(s)} + C_\infty^{(t)} \\
=\,& 16 \int_0^\infty \frac{{\rm d} s}{s^2}  \Biggl\{ \sum_{j=2}^\infty(2j+ 1)\left[j(j+1)-3\right] {\rm Im}\,a_{\alpha,\pm;\beta,\mp}^{FF(j)}(s)\\
&\hspace{1.63cm} +  \sum_{j=0}^\infty (2j + 1)\left[j(j+1)-2\right] {\rm Im}\,a_{\bar \alpha,\mp;\beta,\mp}^{FF(j)}(s)\Biggr\}.
\label{eq:sum22}
\eea
We note that while Eqs.~\eqref{eq:sum11} and \eqref{eq:sum12} establish relations of definite sign and suggest a possible positivity relation (up to boundary terms), Eqs.~\eqref{eq:sum21} and \eqref{eq:sum22} are of indefinite sign.

The $F^3$ operators in the SMEFT generate both cubic and quartic vertices, with the cubic vertices contributing to four-point scattering via tree diagrams involving single-gluon exchange.
Analogous diagrams exist in the pure Yang-Mills sectors of the SM itself.
Let us choose the color vectors $t_{1,2,3,4}$ to be in an elastic configuration, $t_1^a = \bar t_4^a$ and $t_2^a = \bar t_3^a$.
We avoid kinematic singularities from the cubic exchange diagrams when the $s\rightarrow 0$ or $t\rightarrow 0$ limits are taken by choosing colors that commute, i.e., $f^{abc} t_1^a t_2^b = 0$, in which case both the SM amplitude and corrected SMEFT amplitudes vanish identically (cf. \Refc{Remmen:2019cyz}).
As the EFT amplitude vanishes, the dispersion relation then implies that either the poles at infinity must be non-vanishing or else by Eqs.~\eqref{eq:sum11} and \eqref{eq:sum12} all imaginary parts of partial waves---i.e., the contributions to the amplitude from the UV states---are zero.
This is in keeping with the observation---made in Ref.~\cite{LianTao} via little-group scaling arguments and the on-shell formalism---that elastic vector-vector scattering at dimension six must vanish.

\section{Boundary Terms and UV Completions}\label{sec:boundary}

Boundary terms are a fundamental aspect of dispersion relations at dimension six and a fundamental obstruction to the clean UV-IR connection spinning sum rules would otherwise provide.
The complete classification of the conditions required for a UV theory to generate a pole at infinity is not unknown.
We will not attempt an exhaustive analysis in this work.
Instead, in this section we will study various weakly coupled UV completions of our four-scalar and four-fermion spinning sum rules, focusing exclusively on tree-level completions; loop-level completions are left as a particularly interesting open direction.
Our goal is to build intuition for the role the boundary terms that appear in Eqs.~\eqref{eq:sum1} and \eqref{eq:sum2} play in the dispersion relations and the properties of theories in which they can be neglected.

There have been a number of discussions of the boundary term associated with the forward dispersion relation, $C_\infty^{(s)}$.
For instance, \Refs{Falkowski:2012vh,Bellazzini:2014waa} have argued that the generic expectation is that such a pole does not arise from strongly coupled UV theories, whereas for weakly coupled theories, massive vectors exchanged in the $t$ channel have been identified as a canonical source of a pole at infinity (see, e.g., \Refs{Bellazzini:2014waa,LianTao}).
We have already discussed the connection between $t$-channel exchanges and boundary contributions in \Sec{sec:dispersion}: if such terms contribute to the EFT, then as they cannot be reproduced from the dispersive integrals, they must appear as poles at infinity.
This connection will extend to the beyond-forward boundary term $C_\infty^{(t)}$.
Indeed, we will see that a scalar exchanged in the $t$ channel can generate $C_\infty^{(t)}$, while leaving $C_\infty^{(s)} = 0$, whereas only a massive vector generates a non-vanishing $C_\infty^{(s)}$.

\subsection{UV completions of $H^4$ operators}

We begin by considering obstructions to the spinning sum rules introduced for ${\cal O}_{H1}$ and ${\cal O}_{H2}$ in \Sec{sec:spin00}.
We do so by considering two tree-level completions of these operators, drawing from the exhaustive list enumerated in \Refc{deBlas:2017xtg}.
In particular, we will study a real singlet scalar $S$ and a complex vector $V$ that is a singlet under ${\rm SU}(3)_C$ and ${\rm SU}(2)_L$, but carries unit hypercharge.\footnote{In the notation of \Refc{deBlas:2017xtg}, $S$ and $V$ correspond to ${\cal S}$ and ${\cal B}_1$, respectively.}
The first of these will violate our spinning sum rules, whereas the latter satisfies them, and we will isolate the violation as explicitly associated with a $t$-channel exchange and boundary terms.
More generally, we will use these completions as an opportunity to demonstrate in detail how the dispersion relations are satisfied and where the boundary terms play a key role.

\subsubsection{Scalar Completion}

The singlet scalar can couple to the Higgs at the renormalizable and dimension-five level as\footnote{The theory in \Eq{eq:LscalarHiggs} is manifestly non-renormalizable and therefore should be considered as a UV extension---a theory that improves the momentum scaling of amplitudes in the UV and therefore raises the cutoff---rather than a UV completion.
Yet the theory is causal and unitary, and therefore must be---and in fact is---consistent with our dispersion relations.}
\be
{\cal L} \supset - \mu\, S H^{\dagger} H + y\, S D_{\mu} H^{\dagger} D^{\mu} H.
\label{eq:LscalarHiggs}
\ee
Observe that $\mu$ and $y$ have unit and inverse mass dimensions, respectively, and further that hermiticity requires that they both be real.
Other couplings, even at the renormalizable level, are allowed, but play no role in the generation of the dimension-six Higgs operators we are considering, so they are irrelevant to our discussion.
Integrating out $S$ at the level of the action generates ${\cal O}_{H1}$ and ${\cal O}_{H2}$ with the Wilson coefficients
\be
c^{H1} = \frac{\mu^2}{2\mS^4} - \frac{\mu y}{2\mS^2}
\qquad {\rm and} \qquad 
c^{H2} = 0.
\label{eq:scalarHiggscoeffs}
\ee
As $\mu$ and $y$ are of indefinite sign, so too is $c^{H1}$, and therefore this completion cannot satisfy the expectations of the spinning sum rule, which from Fig.~\ref{fig:scalar-region} would require $c^{H1} < 0$ for a scalar completion.

To study this case further, we require the UV amplitude for four-Higgs scattering.
The amplitude receives contributions from $s$-, $t$-, and $u$-channel exchanges, where the vertices can be associated with either the $\mu$ or $y$ couplings.
Explicitly, choosing the scattering states in \Eq{eq:scalarstates}, we obtain the following elastic amplitude:
\be
{\cal A}(s,t) 
= \frac{(s\,\mu y-\mu^2)|\alpha \cdot \beta|^2}{s-\mS^2+i\epsilon}
+ \frac{(t\,\mu y-\mu^2)|\alpha|^2 |\beta|^2}{t-\mS^2+i\epsilon}
+ \frac{(u\,\mu y-\mu^2)|\alpha \cdot \bar{\beta}|}{u-\mS^2+i\epsilon}.
\label{eq:scalarHiggsamp}
\ee
Expanding this result for $\mS^2 \gg p^2$ and comparing with the EFT amplitude in \Eq{eq:ampH4gen} readily confirms the matching in \Eq{eq:scalarHiggscoeffs}.

As an aside, note that in \Eq{eq:scalarHiggsamp} we have not included terms at ${\cal O}(y^2)$.
These contributions are present, and they can be trivially included by modifying the $s$-channel numerator with $(s\,\mu y-\mu^2) \to -(s\,y/2-\mu)^2$ and similarly for other channels.
We have neglected the correction, however, as it only generates Higgs EFT operators at dimension eight.
Nevertheless, it is only after this contribution is added that one would obtain a positive ${\rm Im}\, {\cal A}(s,0)$.

From \Eq{eq:scalarHiggsamp}, we can study the IR and UV sides of the dispersion relations.
First, consider the forward relation in \Eq{eq:disps}.
The IR contribution is given by
\be
\lim_{s \to 0} \partial_s {\cal A}(s,0) = \left( \frac{\mu^2}{\mS^4} - \frac{\mu y}{\mS^2} \right) ( |\alpha \cdot \beta|^2 - |\alpha \cdot \bar{\beta}|^2).
\ee
Comparing this with the UV amplitude, it is clear that the $t$-channel contribution has no low-energy remnant.
As such, we would expect $C_{\infty}^{(s)}=0$, which is confirmed by direct calculation from \Eq{eq:Cs}.
The UV then only needs to account for the $s$ and $u$ channels, which it does so as
\bea
{\rm Im}\, {\cal A}(s,0) &= \pi (\mu^2-\mS^2 \mu y) |\alpha \cdot \beta|^2 \delta(s-\mS^2) + \cdots \\
{\rm Im}\, \bar{\cal A}(s,0) &= \pi (\mu^2-\mS^2 \mu y) |\alpha \cdot \bar{\beta}|^2 \delta(s-\mS^2) + \cdots,
\label{eq:scalarImHiggs}
\eea
where the ellipses contain only terms that do not have support in the dispersive integrals over the positive $s$-axis.

We next consider the beyond-forward dispersion relation of \Eq{eq:dispt}, on which the spinning sum rules were based.
The EFT now records an explicit contribution from the $t$ channel $\propto |\alpha|^2 |\beta|^2$,
\be
\lim_{t \to 0} \partial_t {\cal A}(0,t) = \left( \frac{\mu^2}{\mS^4} - \frac{\mu y}{\mS^2} \right) ( |\alpha|^2 |\beta|^2 - |\alpha \cdot \bar{\beta}|^2).
\ee
This contribution is---and can only be---reconstructed by a boundary term, which from \Eq{eq:Ct} is given by
\be
C_{\infty}^{(t)} = - \left( \frac{\mu^2}{\mS^4} - \frac{\mu y}{\mS^2} \right) |\alpha|^2 |\beta|^2.
\ee
To confirm the full dispersion relation, note that the integral can be rewritten in terms of three terms,
\bea
&\frac{1}{\pi}\int_0^\infty {\rm d}s \lim_{t\rightarrow 0}\partial_t \left[\frac{{\rm Im}\, {\cal A}(s,t)}{s} + \frac{{\rm Im}\, \bar{\cal A}(s,t)}{s+t}\right] \\
= &\frac{1}{\pi} \int^{\infty}_0 {\rm d}s \left[ \frac{\lim_{t \to 0} \partial_t {\rm Im}\, {\cal A}(s,t)}{s}
+ \frac{\lim_{t \to 0} \partial_t {\rm Im}\, \bar{\cal A}(s,t)}{s}
- \frac{{\rm Im}\, \bar{\cal A}(s,0)}{s^2} \right]\!.
\eea
Given the range of integration, only terms proportional to $\delta(s-\mS^2)$ will contribute, but as in the amplitude such terms are independent of $t$, the $t$ derivatives will cause them to vanish.
Accordingly, only the final term contributes, and using \Eq{eq:scalarImHiggs} we see that this reproduces the $u$-channel contribution, so \Eq{eq:dispt} is indeed satisfied.

In summary, a scalar exchanged in the $t$ channel can generate the beyond-forward boundary term, even when $C_{\infty}^{(s)}=0$, as claimed.
The presence of this boundary term, which is of indefinite sign, violates the UV scaling assumptions needed for spinning sum rules, explaining why this completion can generate a positive $c^{H1}$, which would otherwise be suggestive of a vector in the UV.

\subsubsection{Vector Completion}

We next study a UV completion that satisfies the spinning sum rules: a vector $V$ with unit hypercharge.
The relevant interactions take the form
\be
{\cal L} \supset \lambda\, V^{\mu \dagger} (i \partial_{\mu} H^T) i \sigma_2 H + {\rm h.c.},
\ee
such that when the vector is integrated out, we obtain
\be 
c^{H1} = \frac{|\lambda|^2}{2\mV^2}
\qquad {\rm and} \qquad
c^{H2} = -\frac{|\lambda|^2}{\mV^2}.
\label{eq:vectorHiggscoeffs}
\ee 
These coefficients lead to a positive value for all the spinning sum rules considered in \Sec{sec:spin00}, which is consistent with the prediction for the signal a UV vector would leave.

Nonetheless, it is still illuminating to study how the dispersion relations operate in detail, and we will see that this completion generates both boundary terms and yet can still satisfy the conditions for the spinning sum rules.
Our starting point is the four-Higgs UV amplitude, which is now given by
\bea
{\cal A}(s,t) = \frac{|\lambda|^2}{4} 
\left[ 
\frac{(u-t) {\cal C}_s^{\alpha \beta}}{s-\mV^2+i\epsilon}
+ \frac{(u-s) {\cal C}_t^{\alpha \beta}}{t-\mV^2+i\epsilon}
+ \frac{(s-t) {\cal C}_u^{\alpha \beta}}{t-\mV^2+i\epsilon}
\right]\!.
\label{eq:vectorHiggsamp}
\eea
We have introduced a shorthand notation for the Higgs flavor structure, obtained by first defining
\bea
{\cal C}(\alpha,\beta,\gamma,\delta)
=&\,\Big\{[(\alpha_3-i\alpha_4)(\beta_1-i\beta_2)-(\alpha_1-i\alpha_2)(\beta_3-i\beta_4)] \\
&\;\;\;\;\;\;\times[(\delta_3+i\delta_4)(\gamma_1+i\gamma_2)-(\delta_1+i\delta_2)(\gamma_3+i\gamma_4)]\Big\} \\
&+\Big\{(\alpha,\beta) \leftrightarrow (\delta,\gamma)\Big\},
\eea
in terms of which we can write
\bea
{\cal C}_s^{\alpha \beta} = {\cal C}(\alpha,\beta,\bar{\beta},\bar{\alpha}),
\hspace{0.5cm}
{\cal C}_t^{\alpha \beta} = {\cal C}(\alpha,\bar{\alpha},\bar{\beta},\beta),
\hspace{0.5cm}
{\cal C}_u^{\alpha \beta} = {\cal C}(\alpha,\bar{\beta},\beta,\bar{\alpha}).
\eea
These flavor structures satisfy a number of relations.
First, all three combinations are real.
Under crossing, they transform as ${\cal C}_s^{\bar{\alpha} \beta} = {\cal C}_u^{\alpha \beta}$ and ${\cal C}_t^{\bar{\alpha} \beta} = - {\cal C}_t^{\alpha \beta}$, so that the full amplitude is crossing symmetric.
Additionally, while ${\cal C}_t^{\alpha \beta}$ is of indefinite sign, both ${\cal C}_s^{\alpha \beta}$ and ${\cal C}_u^{\alpha \beta}$ are positive for all $\alpha$ and $\beta$, which ensures that \Eq{eq:vectorHiggsamp} satisfies ${\rm Im}\, {\cal A}(s,0) > 0$.
Finally, one can confirm that taking $\mV^2 \gg p^2$ in \Eq{eq:vectorHiggsamp} and matching onto \Eq{eq:ampH4gen}, the flavors are such that the Wilson coefficients in \Eq{eq:vectorHiggscoeffs} are reproduced.

Turning to the dispersion relations, for the forward expression, in the IR we have
\be
\lim_{s \to 0} \partial_s {\cal A}(s,0) = \frac{|\lambda|^2}{4\mV^2} \left( 
{\cal C}_s^{\alpha \beta} + 2 {\cal C}_t^{\alpha \beta} - {\cal C}_u^{\alpha \beta} \right)\!,
\ee
which receives contributions from all three channels.
In the UV, the $s$ and $u$ channels are recreated from the dispersive integral using
\bea
{\rm Im}\, {\cal A}(s,0) &= \frac{\pi |\lambda|^2 \mV^2}{4} {\cal C}_s^{\alpha \beta} \delta(s-\mV^2) + \cdots \\
{\rm Im}\, \bar{\cal A}(s,0) &= \frac{\pi |\lambda|^2 \mV^2}{4} {\cal C}_u^{\alpha \beta} \delta(s-\mV^2) + \cdots,
\eea
whereas the $t$ channel is reconstructed from the boundary term,
\be
C_{\infty}^{(s)} = - \frac{|\lambda|^2}{2\mV^2} {\cal C}_t^{\alpha \beta}.
\ee
Consistency of the beyond-forward calculation requires a contribution from all parts of the dispersion relation.
Again the EFT result receives contributions from all channels,
\be
\lim_{t \to 0} \partial_t {\cal A}(0,t) = \frac{|\lambda|^2}{4\mV^2} \left(
2\, {\cal C}_s^{\alpha \beta} + {\cal C}_t^{\alpha \beta} + {\cal C}_u^{\alpha \beta} \right)\!.
\ee
The integral recovers the non-$t$-channel contributions, 
\bea
\frac{1}{\pi} \int^{\infty}_0 {\rm d}s \left[ \frac{\lim_{t \to 0} \partial_t {\rm Im}\, {\cal A}(s,t)}{s}
+ \frac{\lim_{t \to 0} \partial_t {\rm Im}\, \bar{\cal A}(s,t)}{s} \right]
&= \frac{|\lambda|^2}{2\mV^2} \left( {\cal C}_s^{\alpha \beta} + {\cal C}_u^{\alpha \beta} \right) \\
-\frac{1}{\pi} \int^{\infty}_0 \frac{{\rm d}s}{s^2} {\rm Im}\, \bar{\cal A}(s,0)
&= - \frac{|\lambda|^2}{4\mV^2} {\cal C}_u^{\alpha \beta},\hspace{4cm}
\eea
leaving consistency hinging on the pole at infinity, which is correctly reproduced by
\bea
C_{\infty}^{(t)} = - \frac{|\lambda|^2}{4\mV^2} {\cal C}_t^{\alpha \beta}.
\eea

Thus, the vector completion satisfies both dispersion relations and the spinning sum rules, and yet does so while generating both boundary terms.
This naively seems inconsistent with the obstruction a pole at infinity provides to the arguments in \Sec{sec:spin00}.
The resolution is that the calculation above was performed for general $\alpha$ and $\beta$, whereas the explicit spinning sum rules derived always took specific values of the flavor coefficients.
Indeed, the calculation leading to Eq.~\eqref{eq:sumhiggs} took $\beta = (1,0,0,0)$, which gives ${\cal C}_t^{\alpha \beta}=0$ and therefore vanishing boundary terms.
The choice $\alpha = (1,-i,0,0)/\sqrt{2}$ and $\beta = (0,0,i,1)/\sqrt{2}$, which was used to generate a spinning sum rule for $2c^{H1}-c^{H2}$, also leads to ${\cal C}_t^{\alpha \beta}=0$.
More generally, it is possible to choose flavors such that the boundary terms do not vanish.
For such cases, there are no guarantees that $\lim_{t \to 0} \partial_t {\cal A}(0,t)$ will be positive.
Nevertheless, it turns out that $2\, {\cal C}_s^{\alpha \beta} + {\cal C}_t^{\alpha \beta} + {\cal C}_u^{\alpha \beta}>0$ for all possible flavors, so it is simply the case that for this UV completion, while the boundary terms could impose an obstruction, in practice they do not.

\subsection{UV completions of $\psi^4$ operators}

We now turn to UV completions of dimension-six four-fermion operators, thereby exploring the applicability of the spinning sum rules derived in \Sec{sec:spinonehalf}.
In \Refc{Remmen:2020uze}, where these fermionic spinning sum rules were first introduced, a number of explicit UV completions were considered, and the expected result was demonstrated that a violation of the UV-IR connection arises for $t$-channel UV completions.
Rather than consider explicit UV completions, in this section we will be more general.
In particular, we will consider how our formalism applies to tree-level completions in the $s$, $t$, and $u$ channels, without specifying the particular UV Lagrangian generating such completions.
As we restrict our attention to opposite-helicity scattering, there is an explicit connection between the channel and the spin of the exchanged state; the $s$ and $t$ channels are associated with vectors, and the $u$ channel scalars.
Nevertheless, beyond what can be inferred from conservation of angular momentum, we remain UV-agnostic.
Doing so will allow us to disentangle the UV behavior and boundary terms as general features of particular classes of completions, independent of their details, and will also allow the discussion to sidestep a number of sign-related subtleties associated with fermionic scattering.
Again, the purpose of the exercise is not only to demonstrate where boundary terms can arise at tree level, but also to furnish several explicit examples of a complete implementation of the sum rules, where both sides of the expressions can be manifestly evaluated.

\subsubsection{$s$-channel completion}

Our focus in this section is tree-level completions of dimension-six four-fermion operators. 
As appropriate for the sum rules considered in Sec.~\ref{sec:spinonehalf}, we focus on opposite-helicity scattering.
To be explicit, let us consider the scattering of a right-handed chiral fermion $\psi$ (an example in the SM would be the right-handed electron).
An amplitude to consider that satisfies all our requirements is ${\cal A}(\bar{\psi}^-_1 \psi^+_2 \bar{\psi}^-_3 \psi^+_4)$.\footnote{Again, we are working in the all-incoming convention.}
The helicity structure of the external fermions ensures that any amplitude for this process will be proportional to $[13] \langle 24 \rangle$.
Accordingly, we can write a generic tree-level completion in the $s$ channel as
\be
{\cal A}(s,t)=\frac{c_s\,[13] \langle 24 \rangle}{s-m^2+i\epsilon}.
\label{eq:schannel}
\ee
Here, $c_s$ is an arbitrary constant that could depend on the additional quantum numbers carried by the fermions (an explicit example can be found in \Refc{Remmen:2020uze}).
Nevertheless, we can fix the sign of $c_s$ as positive from the optical theorem.
Taking the forward limit $t \to 0$ ($p_{4,3} \to -p_{1,2}$), we have
\be
\lim_{t \to 0} {\cal A}(s,t) = \frac{-c_s s}{s-m^2+i\epsilon},
\label{eq:schannelt20}
\ee
so that the imaginary part of the forward amplitude is $\pi c_s s\,\delta(s-m^2)$, and we require $c_s > 0$, as claimed.

Let us now show explicitly the consistency of the dispersion relations used to derive our sum rules for this completion, as we did for the Higgs completions above.
First, our forward dispersion relation in Eq.~\eqref{eq:disps} gives
\be
\lim_{s\rightarrow0}\partial_{s}{\cal A}(s,0)=\frac{1}{\pi}\int_{s_{0}}^{\infty}\frac{{\rm d}s}{s^{2}}\left[{\rm Im}\, {\cal A}(s,0)-{\rm Im}\, \bar{\cal A}(s,0)\right]-C_{\infty}^{(s)},
\label{eq:Disp1}
\ee
where again $C_{\infty}^{(s)}=-\lim_{|s|\rightarrow\infty}{\cal A}(s,0)/s$.
The left-hand side of this expression is understood to be computed at low energies, where $p^2 \ll m^2$, so that the EFT description is appropriate.
In this simple example, our EFT amplitude is given by ${\cal A}(s,t)=-(c_s/m^2)[13] \langle 24 \rangle$, so that ${\cal A}(s,0)=(c_s/m^2)s$, and the low-energy contribution to the dispersion relation is
\be
\lim_{s\rightarrow0}\partial_{s}{\cal A}(s,0) = \frac{c_s}{m^2}.
\label{eq:schannelexampleLHS}
\ee
Turning to the contributions that should be evaluated in the UV completion, we see immediately from Eq.~\eqref{eq:schannelt20} that $C_{\infty}^{(s)} = 0$.
In order to evaluate the integral, we require the amplitude with particles 2 and 3 crossed, given by
\be
\bar{\cal A}(s,t) = \frac{c_s\,[12] \langle 34 \rangle}{u-m^2+i\epsilon}.
\ee
Note that we have ${\rm Im}\,\bar{\cal A}(s,0)=-\pi c_ss\,\delta(s+m^{2})>0$, as required.
We can further use this result to evaluate the dispersive integral,
\be
\frac{1}{\pi}\int_{s_{0}}^{\infty}\frac{{\rm d}s}{s^{2}}\left[{\rm Im}\, {\cal A}(s,0)-{\rm Im}\, \bar{\cal A}(s,0)\right]
= c_s \int_{s_{0}}^{\infty}\frac{{\rm d}s}{s} \left[ \delta(s-m^2) + \delta(s+m^2) \right]
= \frac{c_s}{m^2},
\ee
in agreement with Eq.~\eqref{eq:schannelexampleLHS}.

We now move on to confirm the beyond-forward dispersion relation we exploited, Eq.~\eqref{eq:dispt}, which we restate for convenience, specializing to the fermion scattering we are considering,
\be
\lim_{t\rightarrow 0} \partial_t {\cal A}(0,t) =-C_\infty^{(t)} - \frac{1}{\pi}\int_0^\infty {\rm d}s \lim_{t\rightarrow 0}\partial_t \left[\frac{{\rm Im}\, {\cal A}(s,t)}{s} + \frac{{\rm Im}\, \bar{\cal A}(s,t)}{s+t}\right]\!,
\label{eq:Disp2}
\ee
where the boundary contribution is now given by $C_\infty^{(t)} = -{\rm Res}\left[\lim_{t\rightarrow 0} \partial_t {\cal A}(s,t)/s,\,s=\infty\right]$.
From Eq.~\eqref{eq:schannel}, and taking $m^2 \gg p^2$, we can evaluate the EFT contribution immediately as $\lim_{t\rightarrow 0} \partial_t {\cal A}(0,t) = -c_s/m^2$.
For the expressions evaluated in the UV completion, we need an explicit form for the spinor-helicity expressions in terms of Mandelstam invariants.
These are provided by $[13] \langle 24 \rangle = u \Phi_u$ and $[12] \langle 34 \rangle = s \Phi_s$.
We caution that the first of these results cannot be evaluated for $s \to 0$ at finite $t$, as this corresponds to a crossing that the scalar expression $u$ cannot account for.
The little-group phases of the fermions are encoded in $\Phi_u = e^{i(\varphi_1-\varphi_2+\varphi_3-\varphi_4)}$ and $\Phi_s = e^{i(\varphi_1+\varphi_2-\varphi_3-\varphi_4)}$, where $\varphi_i$ is the little-group phase associated with particle $i$.
In the elastic forward limit, these reduce to $\Phi_u = \Phi_s = 1$.
Accordingly, we have
\bea
&- \frac{1}{\pi}\int_0^\infty {\rm d}s \lim_{t\rightarrow 0}\partial_t \left[\frac{{\rm Im}\, {\cal A}(s,t)}{s} + \frac{{\rm Im}\, \bar{\cal A}(s,t)}{s+t}\right] \\
&= - c_s\int_0^\infty {\rm d}s \lim_{t\rightarrow 0}\partial_t \left[\frac{(s+t)\delta(s-m^2)\Phi_u}{s} - \frac{s\delta(s+t+m^2) \Phi_s}{s+t}\right] \\
&= - \frac{c_s}{m^2}.
\eea
Consistency then requires that the boundary condition cannot contribute, and indeed for the UV amplitude $C_\infty^{(t)}=0$.
Finally, as the boundary terms both vanish, the negativity of $\lim_{t\rightarrow 0} \partial_t {\cal A}(0,t)$ is indicative of the UV theory being dominated by vector exchange.
This is consistent with the fact that for opposite-helicity scattering, the state exchanged in the $s$ channel must have unit angular momentum.

\subsubsection{$u$-channel completion}

The analogous case for a $u$-channel UV completion is almost identical.
The relevant amplitudes are now
\bea
{\cal A}(s,t) &= \frac{-c_u\,[13]\langle 24 \rangle}{u-m^2+i\epsilon} \\
\bar{\cal A}(s,t) &= \frac{-c_u\,[12] \langle 34 \rangle}{s-m^2+i\epsilon},
\eea
where the sign of the constant $c_u > 0$ can again be fixed by the optical theorem.
Once more, both boundary terms vanish, and an identical computation reveals that the dispersion relations are satisfied, although this time the dispersive integral receives a contribution from $\bar{\cal A}(s,t)$, rather than ${\cal A}(s,t)$.
In the EFT, we can compute $\lim_{t\rightarrow 0} \partial_t {\cal A}(0,t) = c_u/m^2 > 0$, consistent with a scalar exchange, as expected for this channel.

\subsubsection{$t$-channel completion}

Finally, let us consider a $t$-channel completion, which will furnish an explicit example that generates boundary contributions for a fermionic UV completion.
The amplitude is now given by
\be
{\cal A}(s,t) = \frac{-c_t\,[13] \langle 24 \rangle}{t-m^2+i\epsilon}.
\ee
The imaginary part of the associated forward amplitude vanishes, as ${\rm Im}\, {\cal A}(s,0) = - \pi c_t s\,\delta(m^2) = 0$ (since $m^2 \neq 0$).
Consequently, the optical theorem has no bearing on $c_t$, indicating that the constant is of indefinite sign.
The effective theory contributions to the dispersion relation are given by 
\be
\lim_{s\rightarrow0}\partial_{s}{\cal A}(s,0) = - \lim_{t\rightarrow 0} \partial_t {\cal A}(0,t) = -\frac{c_t}{m^2}.
\label{eq:tchannelEFT}
\ee
That these results are of indefinite sign already suggests the need for boundary terms.
This is confirmed by the fact that the integrals in both Eqs.~\eqref{eq:Disp1} and \eqref{eq:Disp2} vanish: ${\rm Im}\, {\cal A}(s,t)$ and ${\rm Im}\, \bar{\cal A}(s,t)$ are proportional to $\delta(t-m^2)$, which is only non-vanishing for $t=m^2 > 0$, while for real momenta, $t < 0$.
The two relations then reduce to
\bea
\lim_{s\rightarrow0}\partial_{s}{\cal A}(s,0) &=-C_{\infty}^{(s)} \\
\lim_{t\rightarrow 0} \partial_t {\cal A}(0,t) &=-C_\infty^{(t)},
\eea
and explicit computation shows $C_{\infty}^{(s)} = -C_{\infty}^{(t)} = c_t/m^2$, consistent with Eq.~\eqref{eq:tchannelEFT}.

The above calculation demonstrates that any $t$-channel tree-level completion will induce boundary terms.
The presence of such terms renders the arguments leading to Eq.~\eqref{eq:ceesimp} inapplicable and would seemingly imply that such completions could give rise to CP- or flavor-violating contributions that are parametrically larger than their symmetry-conserving analogues.
Nevertheless, in practice engineering a violation of these constraints is less straightforward.
For instance, we could imagine a UV completion of ${\cal O}_e$ that couples the UV state to a flavor-conserving electron vertex $c_{11}$ and a flavor-violating electron-muon vertex $c_{12}$.
An amplitude involving both new couplings could mediate flavor-violating decays such as $\mu \to 3e$, with a scaling in the UV couplings of $c_{11} c_{12}$.
However, the theory also allows flavor-conserving interactions proportional to $c_{11}^2$ and $c_{12}^2$, which are of the same parametric order of the flavor-violating amplitude.
Achieving a separation, where flavor violation dominates, requires a more complex completion containing both scalars and vectors with masses and couplings tuned to suppress the conserving interactions (exploiting the different signs they contribute to \Eq{eq:ceesimp}).
We leave a detailed exploration of this point to future work; however, we note that this argument suggests that the observation of flavor violation in a low-energy experiment without concomitant flavor-conserving operators appearing in the near-term at the LHC would signify the presence of both new scalars and vectors.

\section{Oblique Parameters and Universal Theories}\label{sec:oblique}

An alternative to characterizing deviations from the SM with Wilson coefficients is to instead consider parameters that are more directly connected to precision measurements.
In particular, the electroweak oblique parameters $S$, $T$, $U$ of Peskin and Takeuchi~\cite{Peskin:1991sw} (see also \Refs{Maksymyk:1993zm,Barbieri:2004qk}), which parameterize deviation of vacuum polarization functions of the SM bosons, are a traditional benchmark for comparing signatures of new physics.
The anomalous measurement of the $W$-boson mass reported by CDF in \Refc{CDF:2022hxs} provides a recent reminder of their preeminence for electroweak precision measurements, as the result can be interpreted as deviations in the $S$ and $T$ parameters away from their SM values of zero.
For our purposes, however, the important detail is that both $S$ and $T$ correspond to deviations induced by dimension-six operators in the SMEFT, so it is interesting to consider the consequences of our dispersion relations for the oblique parameters.

In general, as discussed in \Refc{Wells:2015uba}, applying the oblique parameters is appropriate only for particular beyond-SM  scenarios called ``universal theories,'' which are the subset of all theories for which the higher-dimension operators in the SMEFT can, via field redefinitions, be rewritten entirely in terms of bosonic operators.
As described in \Refc{Wells:2015uba}, the set of universal theories at dimension six is completely characterized by a subset of sixteen Wilson coefficients: five oblique parameters, four anomalous triple gauge couplings (TGC), three parameters describing the rescaling of the SM $h^3$ coupling, three non-SM Higgs-vector-vector $hVV$ couplings, and one additional four-fermion coupling.
Universal theories manifestly do not capture the most general UV physics that could be seen at dimension six; rather, that is what the full set of Wilson coefficients in the SMEFT describe.

To define the oblique parameters on which we will focus, we review the essential notation.
For a vector boson $V$, we can extract the transverse part of the self-energy $\Pi_{VV}(p^2)$ from
\be 
\Pi^{\mu\nu}_{VV} = -\left(g^{\mu\nu} - \frac{p^\mu p^\nu}{p^2}  \right)\Pi_{VV}(p^2) + \frac{p^\mu p^\nu}{p^2}(\cdots),
\ee 
where we expand the self-energy function in a series in $p^2$, implicitly defining $\Pi_{VV}(0)$, $\Pi'_{VV}(0)$, and $\Pi''_{VV}(0)$:
\be
\Pi_{VV}(p^2) = \Pi_{VV}(0) - \Pi'_{VV}(0)p^2 + \frac{1}{2}\Pi_{VV}''(0)(p^2)^2 + \cdots,
\ee
with primes denoting derivatives with respect to $p^2$.\footnote{Relative to Ref.~\cite{Wells:2015uba}, we have introduced minus signs to account for the difference in metric convention (as we use mostly-plus signature), so that our $\Pi_{VV}(0)$, $\Pi'_{VV}(0)$, and $\Pi''_{VV}(0)$ numerically agree with theirs.
That is, the notation $\Pi'_{VV}(0)$ is here defined to be $-\lim_{p^2\rightarrow 0} {\rm d}\Pi(p^2)/{\rm d}p^2$.}
In terms of these expanded self-energy functions, we can define the oblique parameters as
\bea
\hat S &= \frac{g^2}{16\pi} S = - \bar\Pi'(0)_{W^3 B}\cot\tW \\
\hat T &= \frac{g^2}{4\pi} T \sin^2 \tW = \frac{1}{\mW^2} \left[\bar\Pi_{W^+ W^-} (0) - \bar\Pi_{W^3 W^3}(0) \right]\\
W &= -\frac{\mW^2}{2}\bar\Pi_{W^3 W^3}''(0) \\
Y &= -\frac{\mW^2}{2} \bar\Pi_{BB}''(0)\\
Z &= -\frac{\mW^2}{2}\bar\Pi_{GG}''(0).
\label{eq:STWYZ}
\eea
The bars indicate that we have chosen a definition of the SM fields and parameters such that, as discussed in \Refc{Wells:2015uba}, only bosonic fields are present, $B$ and $W^\pm$ have canonical kinetic terms, and $\Pi_{W^+ W^-}(0)$ vanishes, thus pinning $g$, $g'$, and the Higgs vev to fixed values.
For a particular, concrete beyond-SM theory (i.e., a UV completion of SMEFT operators by some massive degrees of freedom), the oblique parameters are well-defined observables and thus invariant under whichever SMEFT basis we choose; indeed, \Refc{Wells:2015uba} demonstrates translations between the Warsaw basis~\cite{Grzadkowski:2010es} and various other bases.
For the present analysis, we will use this freedom to work in whichever basis is most convenient.

In what follows, we will exploit the relation between the parameters in \Eq{eq:STWYZ} and the dimension-six SMEFT to draw connections to our spinning sum rules.
For all parameters except $\hat{S}$, a relation can be established with no assumptions beyond those with which the spinning sum rules were derived.
For $\hat{S}$, however, a clean connection does not exist, and only after invoking several additional assumptions about the UV will we be able to draw any relation between the UV and IR.
Given the widespread use of the oblique parameters, we find it interesting to consider what it would take to draw such a connection, however, particularly in the case of $\hat{S}$ a detailed exploration of whether the relation can even be brought to bear on a realistic UV completion is an interesting question that we leave to future work.

\subsection{$W,Y,Z$}

We first consider the second-derivative oblique parameters $W$, $Y$, and $Z$.
We will see how the sign of $Y$ can be explicitly derived, depending on the spins of states in the UV, from the sum rule for $e^+ e^-$ scattering and unitarity of the partial-wave expansion.\footnote{In \Refc{Cacciapaglia:2006pk}, the K\"all\'en-Lehmann representation of the two-point function was used to advocate for positivity of $\Pi''(0)$, fixing the sign of $Y$ and $W$, assuming that there are no extra gauge fields in the UV completion (see also \Refc{Englert:2019zmt}).
Note that Ref.~\cite{Cacciapaglia:2006pk} uses the opposite sign convention to that of Eq.~\eqref{eq:STWYZ} and Ref.~\cite{Wells:2015uba} in defining the oblique parameters relative to $\Pi(0)$, $\Pi'(0)$, and $\Pi''(0)$, but also that Refs.~\cite{Wells:2015uba} and \cite{Cacciapaglia:2006pk} (which both use mostly-minus metric signature) differ from each other in the definition of the overall sign of the self-energy $\Pi(p^2)$.}
For $W$ and $Z$, we use other sum rules to illustrate how the magnitudes of these oblique parameters may be upper bounded in terms of $|Y|$.

In a universal theory in the Warsaw basis, four-fermion self-quartics (i.e., $e^4$, $L^4$, etc.) appear only through the combinations in the current-squared operators, given in our mostly-plus metric signature by
\bea
{\cal O}_{2JB} &= - J_{B\mu}J^{B\mu} \\
{\cal O}_{2JW} &= - J^I_{W\mu}J_W^{I\mu} \\
{\cal O}_{2JG} &= - J^a_{G\mu}J_G^{a\mu},
\label{eq:universal-selfquartic}
\eea
where $J_B^\mu$, $J_W^{I\mu}$ and $J_G^{a\mu}$ are the currents coupling to the respective SM gauge bosons, for example, $J_G^{a\mu} = g_s \sum_\psi \bar{\psi} \gamma^{\mu} T^a \psi$, where $\psi$ runs over $u$, $d$, and $Q$. 
(The operator formed from the square of scalar Yukawa currents contributes only to fermion cross-quartics, not self-quartics~\cite{Wells:2015uba}.
The Wilson coefficient of this operator is one of those characterizing a universal theory, although we will not need to consider it here.) 
We write the Wilson coefficients of the operators in Eq.~\eqref{eq:universal-selfquartic} as $c_{2JB}$, $c_{2JW}$, and $c_{2JG}$.
Then as shown in \Refc{Wells:2015uba}, the oblique parameters are
\bea
W &= - \frac{1}{2}g^2 c_{2JW} \\
Y &= - \frac{1}{2} g^2 c_{2JB} \\
Z &= - \frac{1}{2} g^2 c_{2JG}.
\label{eq:WYZdef}
\eea

To emphasize, in a universal theory there is a single Wilson coefficient for each of the current-squared operators in \Eq{eq:universal-selfquartic}.
Accordingly, we can determine the constraints on these coefficients by considering different fermion scattering experiments, knowing that the coefficients of each are related.
To begin with, the $e^4$ four-fermi operator appears only in ${\cal O}_{2JB} \supset -g'^2 Y_{e}^2 (\bar{e}_{m}\gamma_{\mu}e_{m})(\bar{e}_{n}\gamma^{\mu}e_{n})$.
As in Sec.~\ref{sec:spinonehalf} and \Refc{Remmen:2020uze}, we will consider opposite incoming helicities in deriving sum rules for four-fermion scattering.
Taking a superposition of flavors, the amplitude for $\bar e^- e^+$ scattering via ${\cal O}_{2JB}$ is ${\cal A} = 8 \alpha_m \beta_m \gamma_n \delta_n g'^2 Y_e^2 c_{2JB}[13] \langle 24 \rangle$.
Taking the flavor superposition coefficients to satisfy $\alpha_m = \bar{\delta}_m$ and $\beta_m = \bar{\gamma}_m$, we can apply our sum rules.
We find that, if the boundary terms vanish (which is true if we consider tree-level completions without a vector exchanged in the $t$ channel, as we showed in \Sec{sec:boundary}), the oblique parameter $Y$ has sign dictated by the spin of the UV states as in \Eq{eq:ceesimp}:
\be
Y  = \frac{\cot^2\tW}{|\alpha\cdot\beta|^2 Y_e^2 } \int_0^\infty \frac{{\rm d} s}{s^2}  \left[3\, {\rm Im}\,a_{\alpha_-;\beta_+}^{ee(1)}(s)-{\rm Im}\,a_{\bar \alpha_+;\beta_+}^{ee(0)}(s) \right]\!.\label{eq:Y}
\ee

Looking to the universality of the Wilson coefficients, we can similarly consider $d^4$ scattering, which contributes to the following operators:
\bea
{\cal O}_{2JB} & \supset - g'^{2}Y_{d}^{2}(\bar{d}_{m}\gamma_{\mu}d_{m})(\bar{d}_{n}\gamma^{\mu}d_{n})\\
{\cal O}_{2JG} & \supset - g_{s}^{2}\left[-\frac{1}{6}(\bar{d}_{m}\gamma_{\mu}d_{m})(\bar{d}_{n}\gamma^{\mu}d_{n})+\frac{1}{2}(\bar{d}_{m}\gamma_{\mu}d_{n})(\bar{d}_{n}\gamma^{\mu}d_{m})\right]\!.
\eea
Writing the Wilson coefficient of $(\bar{d}_{m}\gamma_{\mu}d_{n})(\bar{d}_{p}\gamma^{\mu}d_{q})$ as $-c_{mnpq}^d$, we have
\be
c_{mnpq}^d =\left(g'^{2}Y_{d}^{2}c_{2JB}-\frac{1}{6}g_{s}^{2}c_{2JG}\right)\delta_{mn}\delta_{pq}+\frac{1}{2}g_{s}^{2}c_{2JG}\delta_{mq}\delta_{np}.
\ee
Computing the $d^4$ amplitude and applying the analogous sum rule from \Eq{eq:ceesimp}, derived in \Refc{Remmen:2020uze}, we have the bounds 
\bea
c_{mnpq}^{d}(\alpha_{m}\beta_{n}\bar{\beta}_{p}\bar{\alpha}_{q}+\alpha_{m}\bar{\alpha}_{n}\bar{\beta}_{p}\beta_{q}) & =4 \int_0^\infty \frac{{\rm d} s}{s^2}  \left[{\rm Im}\,a_{\bar \alpha_+;\beta_+}^{dd(0),\parallel}(s)- 3\, {\rm Im}\,a_{\alpha_-;\beta_+}^{dd(1),\parallel}(s) \right]\\
c_{mnpq}^{d}\alpha_{m}\bar{\alpha}_{n} \bar{\beta}_{p}\beta_{q} & =4 \int_0^\infty \frac{{\rm d} s}{s^2}  \left[{\rm Im}\,a_{\bar \alpha_+;\beta_+}^{dd(0),\perp}(s)- 3\, {\rm Im}\,a_{\alpha_-;\beta_+}^{dd(1),\perp}(s) \right]\!,
\eea
where the superscripts in parentheses indicate spin of the intermediate state as always, and we have further denoted whether we are scattering incoming states of the same ($\parallel$) or orthogonal ($\perp$) ${\rm SU}(3)_C$ charges.
That is,
\bea
&\left(g'^{2}Y_{d}^{2}Y+\frac{1}{3}g_{s}^{2}Z\right)\left(|\alpha|^2|\beta|^2 +|\alpha\cdot\beta|^2\right) 
\\& =2g^2 \int_0^\infty \frac{{\rm d} s}{s^2}  \left[3\, {\rm Im}\,a_{\alpha_-;\beta_+}^{dd(1),\parallel}(s) -{\rm Im}\,a_{\bar \alpha_+;\beta_+}^{dd(0),\parallel}(s)\right]
\eea
and
\bea
&\left(g'^{2}Y_{d}^{2}Y-\frac{1}{6}g_{s}^{2}Z\right)|\alpha|^2|\beta|^2 + \frac{1}{2}g_{s}^{2}Z|\alpha\cdot\beta|^2 \\& =2 g^2 \int_0^\infty \frac{{\rm d} s}{s^2}  \left[3\, {\rm Im}\,a_{\alpha_-;\beta_+}^{dd(1),\perp}(s)-{\rm Im}\,a_{\bar \alpha_+;\beta_+}^{dd(0),\perp}(s) \right]\!,
\eea
so in particular, considering $\alpha \parallel \beta$, we find that 
\be
g'^{2}Y_{d}^{2}Y+\frac{1}{3}g_{s}^{2}Z  \gtrless 0,\label{eq:Zbd1}
\ee
so that the combination is $< 0$ ($> 0$) for a completion dominated by scalars (vectors). Similarly, considering $\alpha\perp\beta$, we find that
\be 
g'^{2}Y_{d}^{2}Y-\frac{1}{6}g_{s}^{2}Z  \gtrless 0.\label{eq:Zbd2}
\ee
Thus, if the completion is globally dominated exclusively by scalars or by vectors regardless of flavor, then if the UV amplitudes do not generate a pole at infinity, one finds that $|Z|$ must be bounded from above by ${\cal O}(1) \times (g'^2 Y_d^2/g_s^2) \times |Y|$ as a consequence of unitarity.
Moreover, while requiring vanishing boundary integrals, it is not possible in a universal theory to realize a scenario where the amplitude for $e^+ e^-$ scattering is dominated by scalar exchange (which implies $Y<0$) while requiring the amplitudes for $d^+ d^-$ scattering to be dominated by vector exchange, since then Eqs.~\eqref{eq:Zbd1} and \eqref{eq:Zbd2}, combined with \Eq{eq:Y}, would lead to contradictory requirements on the sign of $Z$.
The same statement also holds with signs and scalars/vectors swapped.

Analogously, $L^4$ scattering appears in a universal theory through the operators
\bea
{\cal O}_{2JW} & \supset - g^{2}\left[-\frac{1}{4}(\bar{L}_{m}\gamma_{\mu}L_{m})(\bar{L}_{n}\gamma^{\mu}L_{n})+\frac{1}{2}(\bar{L}_{m}\gamma_{\mu}L_{n})(\bar{L}_{n}\gamma^{\mu}L_{m})\right]\\
{\cal O}_{2JB} & \supset - g'^{2}Y_{L}^{2} (\bar{L}_{m}\gamma_{\mu}L_{m})(\bar{L}_{n}\gamma^{\mu}L_{n}).
\eea
Our effective Wilson coefficient tensor is $c_{mnpq}^{L}=\left(g'^{2}Y_{L}^{2}c_{2JB}-\frac{1}{4}g^{2}c_{2JW}\right)\delta_{mn}\delta_{pq}+\frac{1}{2}g^{2}c_{2JW}\delta_{mq}\delta_{np}$.
Computing the $L^4$ amplitude and applying the sum rule, we find
\bea
c_{mnpq}^{L}(\alpha_{m}\beta_{n}\bar{\beta}_{p}\bar{\alpha}_{q}+\alpha_{m}\bar{\alpha}_{n}\bar{\beta}_{p}\beta_{q}) & =4 \int_0^\infty \frac{{\rm d} s}{s^2}  \left[{\rm Im}\,a_{\bar \alpha_+;\beta_+}^{LL(0),\parallel}(s)- 3\, {\rm Im}\,a_{\alpha_-;\beta_+}^{LL(1),\parallel}(s) \right]\\
c_{mnpq}^{L}\alpha_{m}\bar{\alpha}_{n} \bar{\beta}_{p}\beta_{q} & =4 \int_0^\infty \frac{{\rm d} s}{s^2}  \left[{\rm Im}\,a_{\bar \alpha_+;\beta_+}^{LL(0),\perp}(s)- 3\, {\rm Im}\,a_{\alpha_-;\beta_+}^{LL(1),\perp}(s) \right]\!, 
\eea
that is,
\bea
&\left(2Y_{L}^{2}Y\tan^2 \tW +\frac{1}{2}W\right)\left(|\alpha|^2|\beta|^2 +|\alpha\cdot\beta|^2\right)  \\& =4 \int_0^\infty \frac{{\rm d} s}{s^2}  \left[3\, {\rm Im}\,a_{\alpha_-;\beta_+}^{LL(1),\parallel}(s)  - {\rm Im}\,a_{\bar \alpha_+;\beta_+}^{LL(0),\parallel}(s)\right]
\eea
and
\bea
&\left(2Y_{L}^{2}Y\tan^2 \tW  -\frac{1}{2}W\right)|\alpha|^2|\beta|^2+ W |\alpha\cdot\beta|^2 \\& =4 \int_0^\infty \frac{{\rm d} s}{s^2}  \left[3\, {\rm Im}\,a_{\alpha_-;\beta_+}^{LL(1),\perp}(s)  - {\rm Im}\,a_{\bar \alpha_+;\beta_+}^{LL(0),\perp}(s)\right]\!.
\eea
In particular, for $\alpha$ and $\beta$ parallel, we find
\be 
2Y_{L}^{2}Y\tan^2 \tW +\frac{1}{2}W \gtrless 0,\label{eq:Wbd1}
\ee
with $<$ ($>$) for a scalar- (vector-)dominated completion, and similarly, for $\alpha$ and $\beta$ perpendicular,
\be 
2Y_{L}^{2}Y\tan^2 \tW -\frac{1}{2}W \gtrless 0.\label{eq:Wbd2}
\ee
Hence, for UV completions that are globally scalar- or vector-dominated, we have an upper bound on the size of the oblique parameter $W$, $|W|<4 Y_L^2 |Y| \tan^2 \tW$.
Similarly, the spinning sum rules imply that it is not possible, in a universal theory with vanishing boundary integrals, for $e^+ e^-$ scattering to be dominated by scalar (vector) exchange while $L^+ L^-$ scattering is dominated by vector (respectively, scalar) exchange, since Eqs.~\eqref{eq:Wbd1} and \eqref{eq:Wbd2}, combined with \Eq{eq:Y}, would lead to contradictory predictions for the sign of $W$.

\subsection{$\hat T$}

Next, let us consider the $\hat T$ parameter.
In terms of the basis used by Elias-Mir\'{o}, Grojean, Gupta, and Marzocca (EGGM) in \Refc{EGGM}, we define the operators, in our mostly-plus metric, as
\bea
{\cal O}_{T} & = - \frac{1}{2}(H^{\dagger}\overleftrightarrowalt{D}_{\mu}H)^{2}\\
{\cal O}_{H} & = - \frac{1}{2}(\partial_{\mu}|H|^{2})^{2},
\label{eq:EGGMHiggs}
\eea
and write $E_{T,H}$ for the Wilson coefficients of ${\cal O}_{T,H}$, respectively.\footnote{\Eq{eq:EGGMHiggs} contains the same operators used by \Refc{LianTao} and discussed in \Sec{sec:spin00}.
Nevertheless, we adopt a different symbol for the Wilson coefficients here for consistency with the notation of \Refc{Wells:2015uba}. }
No other Higgs quartics appear in the EGGM basis.
Relating these operators to the oblique parameters, one has $\hat T = E_T$.
Let us further define the rescaling in the TGC for the $hWW$ vertex (in the field basis where the vectors have been rescaled to have canonical kinetic terms) as $1+\Delta\bar\kappa_V$.
Then in the EGGM basis, we have $\Delta \bar\kappa_V = -E_H/2$.

Expanding the operators in \Eq{eq:EGGMHiggs} in the basis of real scalars via \Eq{eq:Higgs}, the amplitudes are the same as those computed in \Eq{eq:ampH4gen}, with the identifications $E_T = c^{H2}/2$ and $E_H = (4c^{H1} - c^{H2})/2$.
In particular, from \Eq{eq:ampH4gen-simp}, we have the amplitudes for $\phi_1 + \phi_{2,3,4} \rightarrow \phi_1 + \phi_{2,3,4}$:
\bea
{\cal A}_{1,2} &= (3E_T + E_H)t \\
{\cal A}_{1,3} = {\cal A}_{1,4} &= E_H t.
\eea
Applying the sum rules from \Sec{sec:spin00}, we thus have:
\bea
3\hat T -2\Delta\bar\kappa_V &=  \! 16\! \int_0^\infty \! \frac{{\rm d} s}{s^2}   \sum_{j=0}^\infty(2j{+}1)\left\{ \left[j\left(j{+}1\right)\right] {\rm Im}\,a_{2}^{(j)}(s) + \left[j\left(j{+}1\right) {-}1\right] {\rm Im}\,\bar{a}_{2}^{(j)}(s)\right\} - C^{(t)}_{\infty}\\
\Delta\bar\kappa_V &=  \! -8\! \int_0^\infty \! \frac{{\rm d} s}{s^2}   \sum_{j=0}^\infty(2j{+}1)\left\{ \left[j\left(j{+}1\right)\right] {\rm Im}\,a_{3}^{(j)}(s) + \left[j\left(j{+}1\right) {-}1\right] {\rm Im}\,\bar{a}_{3}^{(j)}(s)\right\} + C^{(t)}_{\infty}\\
&=  \! -8\! \int_0^\infty \! \frac{{\rm d} s}{s^2}   \sum_{j=0}^\infty(2j{+}1)\left\{ \left[j\left(j{+}1\right)\right] {\rm Im}\,a_{4}^{(j)}(s) + \left[j\left(j{+}1\right) {-}1\right] {\rm Im}\,\bar{a}_{4}^{(j)}(s)\right\} + C^{(t)}_{\infty}.
\eea
As a result, for completions where $C^{(t)}_{\infty} = 0$, if the UV is dominated by scalar exchange, our dispersion relations predict $\Delta \bar\kappa_V >0$ and $\hat T < 2\Delta\bar\kappa_V/3$.
On the other hand, if the UV completion is dominated by higher-spin currents, then $\Delta \bar\kappa_V < 0$ and $\hat T > 2\Delta\bar\kappa_V/3$.

\subsection{$\hat S$}

Finally, we consider what is required in order to draw a connection between spinning sum rules and the first oblique parameter $\hat S$.\footnote{We note that a dispersive representation of $S$ was in fact constructed by Peskin and Takeuchi in \Refc{Peskin:1991sw}, where a sum rule for $S$ was established in terms of the vector and axial isospin analogues of the $R$-ratio, in the spirit of the Weinberg sum rules~\cite{Weinberg:1967kj}.
Their sum rule connected $S$ to the imaginary part of self-energy functions, whereas our focus is a connection to two-to-two scatterings that we can relate to dimension-six SMEFT coefficients.}
At the outset, we note that we will not be able to establish a clean connection.
Additional assumptions will be required, and the study below should be considered as in the spirit of determining what is needed in order to connect $\hat{S}$ to the UV.

The first step is to relate $\hat{S}$ to Wilson coefficients in the SMEFT.
To do so, we again follow \Refc{Wells:2015uba}, but with an accounting for our differing conventions, and define the following operators: 
\bea
{\cal O}_{HJB} &= \frac{ig'}{2} (H^\dagger \overleftrightarrowalt{D}_\mu H)J_B^\mu \\&=  \frac{1}{2} g'^2 \delta_{mn} (Y_Q {\cal O}_{HQ1} + Y_L {\cal O}_{HL1} + Y_u {\cal O}_{Hu} + Y_d {\cal O}_{Hd} + Y_e {\cal O}_{He})\\
{\cal O}_{HJW} &= ig (H^\dagger \overleftrightarrowalt{D}\vphantom{D}^I_\mu H)J_W^{I\mu} \\&= g^2 \delta_{mn} ({\cal O}_{HL2} + {\cal O}_{HQ2}) \\
{\cal O}_{HWB} &= 2 (H^\dagger \tau^I H)W^I_{\mu\nu} B^{\mu\nu},
\label{eq:universalquartics}
\eea
where, when possible, we have related the operators to those we listed in \Tab{tab:ops}.
We can relate the Wilson coefficients of these operators, together with those in \Eq{eq:universal-selfquartic}, to specify
\be
\hat S = g^2 \left(\frac{1}{gg'} c_{HWB} + \frac{1}{4} c_{HJB} + \frac{1}{4} c_{HJW} - \frac{1}{2}c_{2JB} - \frac{1}{2}c_{2JW}\right)\!.
\label{eq:hatS}
\ee
This relation immediately presents a challenge. 
As we saw in \Sec{sec:spin10}, our dispersive arguments cannot constrain the Wilson coefficients of $H^2 B^2$ and $H^2 W^2$, and the same conclusion will hold for ${\cal O}_{HWB}$ via the analogous sum rule one obtains from scattering a Higgs and a superposition of $B$ and $W$ gauge bosonn: the obstruction is that the elastic, fixed-helicity EFT amplitude vanishes by little-group scaling~\cite{LianTao}.
Accordingly, we should expect that $c_{HWB}$ can take on any sign independent of any properties of the UV, exactly as we saw for the simple scalar completion considered in \Sec{sec:spin10}.
To make progress, we will simply consider the case in which the UV does not generate $c_{HWB}$.

From here, studying the form within Eq.~\eqref{eq:universalquartics} in which the fermion and Higgs cross-quartics can appear in universal theories, we can then write $\hat S$ in terms of the lepton and Higgs Wilson coefficients, along with $W$ and $Y$ from \Eq{eq:WYZdef}:
\be
\hat S - Y - W = - \frac{\cot^2 \tW}{2Y_e} c^{He}_{mm} - \frac{1}{4} c^{HL2}_{mm}.
\ee
Once more we have exploited the universality of the coefficients to replace $c_{HJB}$ and $c_{HJW}$ solely with the coefficients $c^{He}_{mm}$ and $c^{HL2}_{mm}$.
We can now use our dispersion relations to study these SMEFT coefficients.
For the Higgs modes we are scattering parameterized by the $\beta_i$, suppose we specifically choose a superposition such that $H=H^\dagger\in \mathbb{R}$, with negative isospin, e.g., $\beta_1 = 1/2$, $\beta_2 = i/2$, $\beta_3 = 1/2$, $\beta_4 = -i/2$; call this choice $\boldsymbol{\beta}_{\downarrow}$.
Then by \Eq{eq:HL} we have the sum rule
\be 
c_{mm}^{HL2}  = -64 \sum_{\alpha}\int_0^\infty \frac{{\rm d} s}{s^2} \sum_{n=0}^\infty (n+1) \left[{\rm Im}\,a_{\alpha;\boldsymbol{\beta}_{\downarrow}}^{HL(n)}(s) - {\rm Im}\,a_{\bar \alpha;\boldsymbol{\beta}_{\downarrow}}^{HL(n)}(s) \right]\!,
\ee
where we have assumed $C_{\infty}^{(s)}=0$ and the sum on $\alpha$ is over an orthonormal basis of the generation index vector for $L$.
Similarly, we can define another choice of Higgs superposition $\boldsymbol{\beta}_\uparrow$ as $\beta_1 = 1/2$, $\beta_2 = i/2$, $\beta_3 = 1/2$, $\beta_4 = i/2$, in which $H=H^\dagger \in \mathbb{R}$, but with positive isospin.
We then find by \Eq{eq:He} that when the pole at infinity can be neglected,
\be
c_{mm}^{He} = -16 \sum_\alpha \int_0^\infty \frac{{\rm d} s}{s^2} \sum_{n=0}^\infty \frac{{\rm d} s}{s^2} (n+1) \left[{\rm Im}\,a_{\alpha;\boldsymbol{\beta}_\uparrow}^{He(n)}(s) - {\rm Im}\,a_{\bar \alpha;\boldsymbol{\beta
}_\uparrow}^{He(n)}(s) \right]\!,
\ee
so that we have the sum rule
\bea
\hat S - Y - W =&\;   \frac{8\cot^2 \tW}{Y_e}\sum_{\alpha}\int_0^\infty \frac{{\rm d} s}{s^2} \sum_{n=0}^\infty (n+1) \left[{\rm Im}\,a_{\alpha;\boldsymbol{\beta}_{\uparrow}}^{HL(n)}(s) - {\rm Im}\,a_{\bar \alpha;\boldsymbol{\beta}_{\uparrow}}^{HL(n)}(s) \right]\\
&\;+16 \sum_{\alpha}\int_0^\infty \frac{{\rm d} s}{s^2} \sum_{n=0}^\infty (n+1) \left[{\rm Im}\,a_{\alpha;\boldsymbol{\beta}_{\downarrow}}^{HL(n)}(s) - {\rm Im}\,a_{\bar \alpha;\boldsymbol{\beta}_{\downarrow}}^{HL(n)}(s) \right]\!.
\label{eq:SYW}
\eea

As discussed in \Sec{sec:spinhalf0}, the forward dispersion relations we have used to obtain the results for $c_{mm}^{HL2}$ and $c_{mm}^{He}$ do not provide clear-cut sign constraints on the coefficients.
Given this fact, and following the discussion in that section, we invoke our last assumption: suppose there is a symmetry in the UV such that in each case $a_{\alpha;\beta} = a_{\bar{\alpha};\beta}$. 
Should this be true, then $\hat S = Y + W$.
Moreover, given the SM values of the Weinberg angle and hypercharges, Eqs.~\eqref{eq:Wbd1}, and \eqref{eq:Wbd2} would then imply that for a theory whose fermion scattering in the UV is globally dominated by scalars or vectors, $|W|\lesssim 0.3\,|Y|$, so $Y+W$ has the same sign as $Y$, which by Eq.~\eqref{eq:Y} is negative (positive) in the scalar (respectively, vector) case. In such a theory, we would then have a definite sign for $\hat S$:
\be
\hat S \gtrless 0,
\ee
again with $<$ ($>$) obtaining for a scalar (vector) completion.
Nevertheless, we emphasize once more that such a result only holds for UV completions that do not generate ${\cal O}_{HWB}$, that have vanishing boundary terms, and for which the amplitude of $He$ and $HL$ scattering is crossing invariant.

\section{Conclusions}\label{sec:conclusions}

The three-thousand-dimensional space describing the dimension-six SMEFT is going to be probed extensively in the coming decades by an array of experiments operating at the energy, intensity, and precision frontiers.
Deviations and discoveries could well be on our horizon.
The focus of this work has been to further chart where in this enormous space the UV may have left distinctive relics.

The key findings of our exploration are the spinning sum rules: a connection between the sign of the IR Wilson coefficients and the dominant UV spin.
That these relations reveal information beyond what has been considered so far at dimension six can be traced to the use of the beyond-forward dispersion relation we outlined in \Sec{sec:dispersion}, which isolated the $t$ scaling of the EFT amplitude, as opposed to the traditional focus on the $s$ coefficient.
As demonstrated in \Sec{sec:sumrules}, spinning sum rules can be established for purely bosonic and fermionic scattering.
The obstruction to establishing such relations more generally is that the required EFT amplitudes vanish, either entirely or in the $s \to 0$ limit.
Nevertheless, as shown in \Sec{sec:oblique}, the relations we obtained are sufficient to establish connections to the oblique parameters that define universal theories.
In particular, while we were able to draw a connection to the well-known Peskin-Takeuchi $S$ and $T$ parameters, our statements for $S$ hinged on a number of strong assumptions about the UV beyond what we required to establish the spinning sum rules themselves.

As we have emphasized throughout this work, the fundamental obstruction to a sharper UV-IR connection at dimension six is the presence of poles at infinity, which enter with arbitrary sign.
The challenges for spinning sum rules are only enhanced over previous studies, given the use of two dispersion relations with distinct boundary terms.
Section~\ref{sec:boundary} focused on furthering our understanding of the boundary contributions (in addition to providing explicit demonstrations of how the dispersion relations are satisfied).
As we showed, at the level of tree-level completions, the boundary terms are associated with states exchanged in the $t$ channel.
The connection is straightforward: the imaginary part of $t$-channel propagators vanishes in the $t \to 0$ limit, so that these results cannot contribute to the dispersive integral and will only be reproduced by the pole at infinity.
The vanishing of the imaginary part of the forward amplitude for $t$-channel propagators also mandates that the sign of their coefficient is unconstrained by the optical theorem, and accordingly we should expect boundary terms to take on arbitrary sign, exactly as they do in explicit completions.
Ultimately, while the spinning sum rules all carry the boundary term caveat, the results remain one of the cleanest connections between mass dimension six and the UV.

This paper opens many promising directions for future study.
Our results extend what is known at dimension six, but will not be the final word.
Additional operators could be constrained by scattering states in superpositions of SM representations, and further by applying the generalized optical theorem as considered in \Refc{Zhang:2020jyn} to move beyond elastic scattering altogether.
Boundary terms are a topic that require further study.
The examples we considered were purely tree-level completions, a subset of possible weakly coupled UV theories.
Extending our analysis to consider completions where loops play an important role and establishing the expected behavior of the boundary terms in these scenarios would further help elucidate the class of theories for which the spinning sum rules are expected to hold.\footnote{We note that loop completions for conventional sum rules have already been considered in \Refc{LianTao}.}
Finally, the ultimate goal of these relations is to open a direct connection to experiment.
Determining exactly where spinning sum rules will be most impactful---particularly in the near-term---remains a critical next step.
In our current era of particle physics, where the first hints of the laws of nature beyond the SM may come from detection of operators in the SMEFT, pursuing a full understanding of how to leverage the axioms of quantum field theory to turn observables like Wilson coefficients into concrete statements about properties of the UV is a crucial enterprise.

\vspace{5mm}
 
\begin{center} 
{\bf Acknowledgments}
\end{center}
\noindent 
G.N.R. is supported at the Kavli Institute for Theoretical Physics by the Simons Foundation (Grant~No.~216179) and the National Science Foundation (Grant~No.~NSF PHY-1748958) and at the University of California, Santa Barbara by the Fundamental Physics Fellowship.

\bibliographystyle{utphys-modified}
\bibliography{General_dim_6}

\providecommand{\href}[2]{#2}\begingroup\raggedright\begin{thebibliography}{100}

\bibitem{Grzadkowski:2010es}
B.~Grzadkowski, M.~Iskrzynski, M.~Misiak, and J.~Rosiek, ``{Dimension-Six Terms
  in the Standard Model Lagrangian},''
  \href{http://dx.doi.org/10.1007/JHEP10(2010)085}{{\em JHEP} {\bfseries 10}
  (2010) 085}, \href{http://arxiv.org/abs/1008.4884}{{\ttfamily arXiv:1008.4884
  [hep-ph]}}.

\bibitem{Henning:2015alf}
B.~Henning, X.~Lu, T.~Melia, and H.~Murayama, ``{2, 84, 30, 993, 560, 15456,
  11962, 261485, ...: Higher dimension operators in the SM EFT},''
  \href{http://dx.doi.org/10.1007/JHEP08(2017)016}{{\em JHEP} {\bfseries 08}
  (2017) 016}, \href{http://arxiv.org/abs/1512.03433}{{\ttfamily
  arXiv:1512.03433 [hep-ph]}}.
  \href{https://doi.org/10.1007/JHEP09(2019)019}{[Erratum: {\it JHEP} {\bf 09}
  (2019) 019]}.

\bibitem{Pham:1985cr}
T.~N. Pham and T.~N. Truong, ``{Evaluation of the Derivative Quartic Terms of
  the Meson Chiral Lagrangian From Forward Dispersion Relation},''
  \href{http://dx.doi.org/10.1103/PhysRevD.31.3027}{{\em Phys. Rev. D}
  {\bfseries 31} (1985) 3027}.

\bibitem{Ananthanarayan:1994hf}
B.~Ananthanarayan, D.~Toublan, and G.~Wanders, ``{Consistency of the chiral
  pion-pion scattering amplitudes with axiomatic constraints},''
  \href{http://dx.doi.org/10.1103/PhysRevD.51.1093}{{\em Phys. Rev. D}
  {\bfseries 51} (1995) 1093},
  \href{http://arxiv.org/abs/hep-ph/9410302}{{\ttfamily arXiv:hep-ph/9410302}}.

\bibitem{Pennington:1994kc}
M.~R. Pennington and J.~Portoles, ``{The chiral lagrangian parameters,
  $\ell_1$, $\ell_2$, are determined by the $\rho$-resonance},''
  \href{http://dx.doi.org/10.1016/0370-2693(94)01551-M}{{\em Phys. Lett. B}
  {\bfseries 344} (1995) 399},
  \href{http://arxiv.org/abs/hep-ph/9409426}{{\ttfamily arXiv:hep-ph/9409426}}.

\bibitem{Adams:2006sv}
A.~Adams, N.~Arkani-Hamed, S.~Dubovsky, A.~Nicolis, and R.~Rattazzi,
  ``{Causality, analyticity and an IR obstruction to UV completion},''
  \href{http://dx.doi.org/10.1088/1126-6708/2006/10/014}{{\em JHEP} {\bfseries
  10} (2006) 014}, \href{http://arxiv.org/abs/hep-th/0602178}{{\ttfamily
  arXiv:hep-th/0602178}}.

\bibitem{Distler:2006if}
J.~Distler, B.~Grinstein, R.~A. Porto, and I.~Z. Rothstein, ``{Falsifying
  Models of New Physics via $WW$ Scattering},''
  \href{http://dx.doi.org/10.1103/PhysRevLett.98.041601}{{\em Phys. Rev. Lett.}
  {\bfseries 98} (2007) 041601},
\href{http://arxiv.org/abs/hep-ph/0604255}{{\ttfamily arXiv:hep-ph/0604255
  [hep-ph]}}.

\bibitem{Jenkins:2006ia}
A.~Jenkins and D.~O'Connell, ``{The Story of ${\cal O}$: Positivity constraints
  in effective field theories},''
\href{http://arxiv.org/abs/hep-th/0609159}{{\ttfamily arXiv:hep-th/0609159
  [hep-th]}}.

\bibitem{Vecchi:2007na}
L.~Vecchi, ``{Causal versus analytic constraints on anomalous quartic gauge
  couplings},'' \href{http://dx.doi.org/10.1088/1126-6708/2007/11/054}{{\em
  JHEP} {\bfseries 11} (2007) 054},
\href{http://arxiv.org/abs/0704.1900}{{\ttfamily arXiv:0704.1900 [hep-ph]}}.

\bibitem{Manohar:2008tc}
A.~V. Manohar and V.~Mateu, ``{Dispersion Relation Bounds for pi pi
  Scattering},'' \href{http://dx.doi.org/10.1103/PhysRevD.77.094019}{{\em Phys.
  Rev. D} {\bfseries 77} (2008) 094019},
  \href{http://arxiv.org/abs/0801.3222}{{\ttfamily arXiv:0801.3222 [hep-ph]}}.

\bibitem{Wang:2020jxr}
Y.-J. Wang, F.-K. Guo, C.~Zhang, and S.-Y. Zhou, ``{Generalized positivity
  bounds on chiral perturbation theory},''
  \href{http://dx.doi.org/10.1007/JHEP07(2020)214}{{\em JHEP} {\bfseries 07}
  (2020) 214}, \href{http://arxiv.org/abs/2004.03992}{{\ttfamily
  arXiv:2004.03992 [hep-ph]}}.

\bibitem{Bellazzini:2018paj}
B.~Bellazzini and F.~Riva, ``{New phenomenological and theoretical perspective
  on anomalous $ZZ$ and $Z\gamma$ processes},''
  \href{http://dx.doi.org/10.1103/PhysRevD.98.095021}{{\em Phys. Rev. D}
  {\bfseries 98} (2018) 095021},
  \href{http://arxiv.org/abs/1806.09640}{{\ttfamily arXiv:1806.09640
  [hep-ph]}}.

\bibitem{Zhang:2018shp}
C.~Zhang and S.-Y. Zhou, ``{Positivity bounds on vector boson scattering at the
  LHC},'' \href{http://dx.doi.org/10.1103/PhysRevD.100.095003}{{\em Phys. Rev.
  D} {\bfseries 100} (2019) 095003},
  \href{http://arxiv.org/abs/1808.00010}{{\ttfamily arXiv:1808.00010
  [hep-ph]}}.

\bibitem{Remmen:2019cyz}
G.~N. Remmen and N.~L. Rodd, ``{Consistency of the Standard Model Effective
  Field Theory},'' \href{http://dx.doi.org/10.1007/JHEP12(2019)032}{{\em JHEP}
  {\bfseries 12} (2019) 032}, \href{http://arxiv.org/abs/1908.09845}{{\ttfamily
  arXiv:1908.09845 [hep-ph]}}.

\bibitem{Bi:2019phv}
Q.~Bi, C.~Zhang, and S.-Y. Zhou, ``{Positivity constraints on aQGC: carving out
  the physical parameter space},''
  \href{http://dx.doi.org/10.1007/JHEP06(2019)137}{{\em JHEP} {\bfseries 06}
  (2019) 137}, \href{http://arxiv.org/abs/1902.08977}{{\ttfamily
  arXiv:1902.08977 [hep-ph]}}.

\bibitem{Nicolis:2009qm}
A.~Nicolis, R.~Rattazzi, and E.~Trincherini, ``{Energy's and amplitudes'
  positivity},'' \href{http://dx.doi.org/10.1007/JHEP05(2010)095}{{\em JHEP}
  {\bfseries 05} (2010) 095}, \href{http://arxiv.org/abs/0912.4258}{{\ttfamily
  arXiv:0912.4258 [hep-th]}}.
\href{https://doi.org/10.1007/JHEP11(2011)128}{[Erratum: {\it JHEP} {\bf 11}
  (2011) 128]}.

\bibitem{Dvali:2012zc}
G.~Dvali, A.~Franca, and C.~Gomez, ``{Road Signs for UV-Completion},''
\href{http://arxiv.org/abs/1204.6388}{{\ttfamily arXiv:1204.6388 [hep-th]}}.

\bibitem{deRham:2017imi}
C.~de~Rham, S.~Melville, A.~J. Tolley, and S.-Y. Zhou, ``{Massive Galileon
  Positivity Bounds},'' \href{http://dx.doi.org/10.1007/JHEP09(2017)072}{{\em
  JHEP} {\bfseries 09} (2017) 072},
\href{http://arxiv.org/abs/1702.08577}{{\ttfamily arXiv:1702.08577 [hep-th]}}.

\bibitem{Chandrasekaran:2018qmx}
V.~Chandrasekaran, G.~N. Remmen, and A.~Shahbazi-Moghaddam, ``{Higher-Point
  Positivity},'' \href{http://dx.doi.org/10.1007/JHEP11(2018)015}{{\em JHEP}
  {\bfseries 11} (2018) 015},
\href{http://arxiv.org/abs/1804.03153}{{\ttfamily arXiv:1804.03153 [hep-th]}}.

\bibitem{Herrero-Valea:2019hde}
M.~Herrero-Valea, I.~Timiryasov, and A.~Tokareva, ``{To positivity and beyond,
  where Higgs-Dilaton Inflation has never gone before},''
  \href{http://dx.doi.org/10.1088/1475-7516/2019/11/042}{{\em JCAP} {\bfseries
  11} (2019) 042}, \href{http://arxiv.org/abs/1905.08816}{{\ttfamily
  arXiv:1905.08816 [hep-ph]}}.

\bibitem{Li:2021cjv}
X.~Li, H.~Xu, C.~Yang, C.~Zhang, and S.-Y. Zhou, ``{Positivity in Multifield
  Effective Field Theories},''
  \href{http://dx.doi.org/10.1103/PhysRevLett.127.121601}{{\em Phys. Rev.
  Lett.} {\bfseries 127} (2021) 121601},
  \href{http://arxiv.org/abs/2101.01191}{{\ttfamily arXiv:2101.01191
  [hep-ph]}}.

\bibitem{Remmen:2021zmc}
G.~N. Remmen, ``{Amplitudes and the Riemann Zeta Function},''
  \href{http://dx.doi.org/10.1103/PhysRevLett.127.241602}{{\em Phys. Rev.
  Lett.} {\bfseries 127} (2021) 241602},
  \href{http://arxiv.org/abs/2108.07820}{{\ttfamily arXiv:2108.07820
  [hep-th]}}.

\bibitem{Arkani-Hamed:2020blm}
N.~Arkani-Hamed, T.-C. Huang, and Y.-t. Huang, ``{The EFT-Hedron},''
  \href{http://dx.doi.org/10.1007/JHEP05(2021)259}{{\em JHEP} {\bfseries 05}
  (2021) 259}, \href{http://arxiv.org/abs/2012.15849}{{\ttfamily
  arXiv:2012.15849 [hep-th]}}.

\bibitem{Bellazzini:2020cot}
B.~Bellazzini, J.~Elias~Mir\'o, R.~Rattazzi, M.~Riembau, and F.~Riva,
  ``{Positive moments for scattering amplitudes},''
  \href{http://dx.doi.org/10.1103/PhysRevD.104.036006}{{\em Phys. Rev. D}
  {\bfseries 104} (2021) 036006},
  \href{http://arxiv.org/abs/2011.00037}{{\ttfamily arXiv:2011.00037
  [hep-th]}}.

\bibitem{Tolley:2020gtv}
A.~J. Tolley, Z.-Y. Wang, and S.-Y. Zhou, ``{New positivity bounds from full
  crossing symmetry},'' \href{http://dx.doi.org/10.1007/JHEP05(2021)255}{{\em
  JHEP} {\bfseries 05} (2021) 255},
  \href{http://arxiv.org/abs/2011.02400}{{\ttfamily arXiv:2011.02400
  [hep-th]}}.

\bibitem{Caron-Huot:2020cmc}
S.~Caron-Huot and V.~Van~Duong, ``{Extremal Effective Field Theories},''
  \href{http://dx.doi.org/10.1007/JHEP05(2021)280}{{\em JHEP} {\bfseries 05}
  (2021) 280}, \href{http://arxiv.org/abs/2011.02957}{{\ttfamily
  arXiv:2011.02957 [hep-th]}}.

\bibitem{Bellazzini:2021oaj}
B.~Bellazzini, M.~Riembau, and F.~Riva, ``{The IR-Side of Positivity Bounds},''
  \href{http://arxiv.org/abs/2112.12561}{{\ttfamily arXiv:2112.12561
  [hep-th]}}.

\bibitem{Caron-Huot:2021rmr}
S.~Caron-Huot, D.~Maz\'a{\v c}, L.~Rastelli, and D.~Simmons-Duffin, ``{Sharp
  boundaries for the swampland},''
  \href{http://dx.doi.org/10.1007/JHEP07(2021)110}{{\em JHEP} {\bfseries 07}
  (2021) 110}, \href{http://arxiv.org/abs/2102.08951}{{\ttfamily
  arXiv:2102.08951 [hep-th]}}.

\bibitem{Komargodski:2011vj}
Z.~Komargodski and A.~Schwimmer, ``{On Renormalization Group Flows in Four
  Dimensions},'' \href{http://dx.doi.org/10.1007/JHEP12(2011)099}{{\em JHEP}
  {\bfseries 12} (2011) 099},
\href{http://arxiv.org/abs/1107.3987}{{\ttfamily arXiv:1107.3987 [hep-th]}}.

\bibitem{Elvang:2012st}
H.~Elvang, D.~Z. Freedman, L.-Y. Hung, M.~Kiermaier, R.~C. Myers, and
  S.~Theisen, ``{On renormalization group flows and the $a$-theorem in 6d},''
  \href{http://dx.doi.org/10.1007/JHEP10(2012)011}{{\em JHEP} {\bfseries 10}
  (2012) 011},
\href{http://arxiv.org/abs/1205.3994}{{\ttfamily arXiv:1205.3994 [hep-th]}}.

\bibitem{Adams:2008hp}
A.~Adams, A.~Jenkins, and D.~O'Connell, ``{Signs of analyticity in fermion
  scattering},'' \href{http://arxiv.org/abs/0802.4081}{{\ttfamily
  arXiv:0802.4081 [hep-ph]}}.

\bibitem{Bellazzini:2016xrt}
B.~Bellazzini, ``{Softness and amplitudes\textquoteright{} positivity for
  spinning particles},'' \href{http://dx.doi.org/10.1007/JHEP02(2017)034}{{\em
  JHEP} {\bfseries 02} (2017) 034},
  \href{http://arxiv.org/abs/1605.06111}{{\ttfamily arXiv:1605.06111
  [hep-th]}}.

\bibitem{Bellazzini:2017bkb}
B.~Bellazzini, F.~Riva, J.~Serra, and F.~Sgarlata, ``{The other effective
  fermion compositeness},''
  \href{http://dx.doi.org/10.1007/JHEP11(2017)020}{{\em JHEP} {\bfseries 11}
  (2017) 020},
\href{http://arxiv.org/abs/1706.03070}{{\ttfamily arXiv:1706.03070 [hep-ph]}}.

\bibitem{Remmen:2020vts}
G.~N. Remmen and N.~L. Rodd, ``{Flavor Constraints from Unitarity and
  Analyticity},'' \href{http://dx.doi.org/10.1103/PhysRevLett.125.081601}{{\em
  Phys. Rev. Lett.} {\bfseries 125} (2020) 081601},
  \href{http://arxiv.org/abs/2004.02885}{{\ttfamily arXiv:2004.02885
  [hep-ph]}}.

\bibitem{Remmen:2020uze}
G.~N. Remmen and N.~L. Rodd, ``{Signs, spin, SMEFT: Sum rules at dimension
  six},'' \href{http://dx.doi.org/10.1103/PhysRevD.105.036006}{{\em Phys. Rev.
  D} {\bfseries 105} (2022) 036006},
  \href{http://arxiv.org/abs/2010.04723}{{\ttfamily arXiv:2010.04723
  [hep-ph]}}.

\bibitem{Cheung:2016yqr}
C.~Cheung and G.~N. Remmen, ``{Positive Signs in Massive Gravity},''
  \href{http://dx.doi.org/10.1007/JHEP04(2016)002}{{\em JHEP} {\bfseries 04}
  (2016) 002},
\href{http://arxiv.org/abs/1601.04068}{{\ttfamily arXiv:1601.04068 [hep-th]}}.

\bibitem{deRham:2017xox}
C.~de~Rham, S.~Melville, and A.~J. Tolley, ``{Improved Positivity Bounds and
  Massive Gravity},'' \href{http://dx.doi.org/10.1007/JHEP04(2018)083}{{\em
  JHEP} {\bfseries 04} (2018) 083},
\href{http://arxiv.org/abs/1710.09611}{{\ttfamily arXiv:1710.09611 [hep-th]}}.

\bibitem{Camanho:2014apa}
X.~O. Camanho, J.~D. Edelstein, J.~Maldacena, and A.~Zhiboedov, ``{Causality
  Constraints on Corrections to the Graviton Three-Point Coupling},''
  \href{http://dx.doi.org/10.1007/JHEP02(2016)020}{{\em JHEP} {\bfseries 02}
  (2016) 020},
\href{http://arxiv.org/abs/1407.5597}{{\ttfamily arXiv:1407.5597 [hep-th]}}.

\bibitem{Camanho:2016opx}
X.~O. Camanho, G.~Lucena~G\'omez, and R.~Rahman, ``{Causality Constraints on
  Massive Gravity},'' \href{http://dx.doi.org/10.1103/PhysRevD.96.084007}{{\em
  Phys. Rev. D} {\bfseries 96} (2017) 084007},
\href{http://arxiv.org/abs/1610.02033}{{\ttfamily arXiv:1610.02033 [hep-th]}}.

\bibitem{Bellazzini:2017fep}
B.~Bellazzini, F.~Riva, J.~Serra, and F.~Sgarlata, ``{Beyond Positivity Bounds
  and the Fate of Massive Gravity},''
  \href{http://dx.doi.org/10.1103/PhysRevLett.120.161101}{{\em Phys. Rev.
  Lett.} {\bfseries 120} (2018) 161101},
\href{http://arxiv.org/abs/1710.02539}{{\ttfamily arXiv:1710.02539 [hep-th]}}.

\bibitem{Bonifacio:2018vzv}
J.~Bonifacio and K.~Hinterbichler, ``{Bounds on Amplitudes in Effective
  Theories with Massive Spinning Particles},''
  \href{http://dx.doi.org/10.1103/PhysRevD.98.045003}{{\em Phys. Rev. D}
  {\bfseries 98} (2018) 045003},
\href{http://arxiv.org/abs/1804.08686}{{\ttfamily arXiv:1804.08686 [hep-th]}}.

\bibitem{Bonifacio:2016wcb}
J.~Bonifacio, K.~Hinterbichler, and R.~A. Rosen, ``{Positivity constraints for
  pseudolinear massive spin-2 and vector Galileons},''
  \href{http://dx.doi.org/10.1103/PhysRevD.94.104001}{{\em Phys. Rev. D}
  {\bfseries 94} (2016) 104001},
\href{http://arxiv.org/abs/1607.06084}{{\ttfamily arXiv:1607.06084 [hep-th]}}.

\bibitem{deRham:2017zjm}
C.~de~Rham, S.~Melville, A.~J. Tolley, and S.-Y. Zhou, ``{UV complete me:
  positivity bounds for particles with spin},''
  \href{http://dx.doi.org/10.1007/JHEP03(2018)011}{{\em JHEP} {\bfseries 03}
  (2018) 011},
\href{http://arxiv.org/abs/1706.02712}{{\ttfamily arXiv:1706.02712 [hep-th]}}.

\bibitem{Hinterbichler:2017qyt}
K.~Hinterbichler, A.~Joyce, and R.~A. Rosen, ``{Massive Spin-2 Scattering and
  Asymptotic Superluminality},''
  \href{http://dx.doi.org/10.1007/JHEP03(2018)051}{{\em JHEP} {\bfseries 03}
  (2018) 051},
\href{http://arxiv.org/abs/1708.05716}{{\ttfamily arXiv:1708.05716 [hep-th]}}.

\bibitem{deRham:2018qqo}
C.~de~Rham, S.~Melville, A.~J. Tolley, and S.-Y. Zhou, ``{Positivity Bounds for
  Massive Spin-1 and Spin-2 Fields},''
  \href{http://dx.doi.org/10.1007/JHEP03(2019)182}{{\em JHEP} {\bfseries 03}
  (2019) 182},
\href{http://arxiv.org/abs/1804.10624}{{\ttfamily arXiv:1804.10624 [hep-th]}}.

\bibitem{Bellazzini:2019bzh}
B.~Bellazzini, F.~Riva, J.~Serra, and F.~Sgarlata, ``{Massive Higher Spins:
  Effective Theory and Consistency},''
  \href{http://dx.doi.org/10.1007/JHEP10(2019)189}{{\em JHEP} {\bfseries 10}
  (2019) 189}, \href{http://arxiv.org/abs/1903.08664}{{\ttfamily
  arXiv:1903.08664 [hep-th]}}.

\bibitem{Alberte:2019xfh}
L.~Alberte, C.~de~Rham, A.~Momeni, J.~Rumbutis, and A.~J. Tolley, ``{Positivity
  Constraints on Interacting Spin-2 Fields},''
  \href{http://dx.doi.org/10.1007/JHEP03(2020)097}{{\em JHEP} {\bfseries 03}
  (2020) 097}, \href{http://arxiv.org/abs/1910.11799}{{\ttfamily
  arXiv:1910.11799 [hep-th]}}.

\bibitem{Alberte:2019zhd}
L.~Alberte, C.~de~Rham, A.~Momeni, J.~Rumbutis, and A.~J. Tolley, ``{Positivity
  Constraints on Interacting Pseudo-Linear Spin-2 Fields},''
  \href{http://dx.doi.org/10.1007/JHEP07(2020)121}{{\em JHEP} {\bfseries 07}
  (2020) 121}, \href{http://arxiv.org/abs/1912.10018}{{\ttfamily
  arXiv:1912.10018 [hep-th]}}.

\bibitem{Wang:2020xlt}
Z.-Y. Wang, C.~Zhang, and S.-Y. Zhou, ``{Generalized elastic positivity bounds
  on interacting massive spin-2 theories},''
  \href{http://dx.doi.org/10.1007/JHEP04(2021)217}{{\em JHEP} {\bfseries 04}
  (2021) 217}, \href{http://arxiv.org/abs/2011.05190}{{\ttfamily
  arXiv:2011.05190 [hep-th]}}.

\bibitem{Alberte:2020bdz}
L.~Alberte, C.~de~Rham, S.~Jaitly, and A.~J. Tolley, ``{QED positivity
  bounds},'' \href{http://dx.doi.org/10.1103/PhysRevD.103.125020}{{\em Phys.
  Rev. D} {\bfseries 103} (2021) 125020},
  \href{http://arxiv.org/abs/2012.05798}{{\ttfamily arXiv:2012.05798
  [hep-th]}}.

\bibitem{Gorghetto:2021luj}
M.~Gorghetto, G.~Perez, I.~Savoray, and Y.~Soreq, ``{Probing CP violation in
  photon self-interactions with cavities},''
  \href{http://dx.doi.org/10.1007/JHEP10(2021)056}{{\em JHEP} {\bfseries 10}
  (2021) 056}, \href{http://arxiv.org/abs/2103.06298}{{\ttfamily
  arXiv:2103.06298 [hep-ph]}}.

\bibitem{Baumann:2015nta}
D.~Baumann, D.~Green, H.~Lee, and R.~A. Porto, ``{Signs of Analyticity in
  Single-Field Inflation},''
  \href{http://dx.doi.org/10.1103/PhysRevD.93.023523}{{\em Phys. Rev. D}
  {\bfseries 93} (2016) 023523},
  \href{http://arxiv.org/abs/1502.07304}{{\ttfamily arXiv:1502.07304
  [hep-th]}}.

\bibitem{Baumann:2019ghk}
D.~Baumann, D.~Green, and T.~Hartman, ``{Dynamical Constraints on RG Flows and
  Cosmology},'' \href{http://dx.doi.org/10.1007/JHEP12(2019)134}{{\em JHEP}
  {\bfseries 12} (2019) 134}, \href{http://arxiv.org/abs/1906.10226}{{\ttfamily
  arXiv:1906.10226 [hep-th]}}.

\bibitem{Grall:2021xxm}
T.~Grall and S.~Melville, ``{Positivity Bounds without Boosts},''
  \href{http://arxiv.org/abs/2102.05683}{{\ttfamily arXiv:2102.05683
  [hep-th]}}.

\bibitem{deRham:2021fpu}
C.~de~Rham, S.~Melville, and J.~Noller, ``{Positivity bounds on dark energy:
  when matter matters},''
  \href{http://dx.doi.org/10.1088/1475-7516/2021/08/018}{{\em JCAP} {\bfseries
  08} (2021) 018}, \href{http://arxiv.org/abs/2103.06855}{{\ttfamily
  arXiv:2103.06855 [astro-ph.CO]}}.

\bibitem{Grall:2020tqc}
T.~Grall and S.~Melville, ``{Inflation in motion: unitarity constraints in
  effective field theories with (spontaneously) broken Lorentz symmetry},''
  \href{http://dx.doi.org/10.1088/1475-7516/2020/09/017}{{\em JCAP} {\bfseries
  09} (2020) 017}, \href{http://arxiv.org/abs/2005.02366}{{\ttfamily
  arXiv:2005.02366 [gr-qc]}}.

\bibitem{Melville:2021lst}
S.~Melville and E.~Pajer, ``{Cosmological Cutting Rules},''
  \href{http://dx.doi.org/10.1007/JHEP05(2021)249}{{\em JHEP} {\bfseries 05}
  (2021) 249}, \href{http://arxiv.org/abs/2103.09832}{{\ttfamily
  arXiv:2103.09832 [hep-th]}}.

\bibitem{Bellazzini:2015cra}
B.~Bellazzini, C.~Cheung, and G.~N. Remmen, ``{Quantum Gravity Constraints from
  Unitarity and Analyticity},''
  \href{http://dx.doi.org/10.1103/PhysRevD.93.064076}{{\em Phys. Rev. D}
  {\bfseries 93} (2016) 064076},
\href{http://arxiv.org/abs/1509.00851}{{\ttfamily arXiv:1509.00851 [hep-th]}}.

\bibitem{Cheung:2016wjt}
C.~Cheung and G.~N. Remmen, ``{Positivity of Curvature-Squared Corrections in
  Gravity},'' \href{http://dx.doi.org/10.1103/PhysRevLett.118.051601}{{\em
  Phys. Rev. Lett.} {\bfseries 118} (2017) 051601},
\href{http://arxiv.org/abs/1608.02942}{{\ttfamily arXiv:1608.02942 [hep-th]}}.

\bibitem{Gruzinov:2006ie}
A.~Gruzinov and M.~Kleban, ``{A note on causality constraining higher curvature
  corrections to gravity},''
  \href{http://dx.doi.org/10.1088/0264-9381/24/13/N02}{{\em Class. Quant.
  Grav.} {\bfseries 24} (2007) 3521},
\href{http://arxiv.org/abs/hep-th/0612015}{{\ttfamily arXiv:hep-th/0612015
  [hep-th]}}.

\bibitem{Guerrieri:2021ivu}
A.~Guerrieri, J.~Penedones, and P.~Vieira, ``{Where Is String Theory in the
  Space of Scattering Amplitudes?},''
  \href{http://dx.doi.org/10.1103/PhysRevLett.127.081601}{{\em Phys. Rev.
  Lett.} {\bfseries 127} (2021) 081601},
  \href{http://arxiv.org/abs/2102.02847}{{\ttfamily arXiv:2102.02847
  [hep-th]}}.

\bibitem{deRham:2021bll}
C.~de~Rham, A.~J. Tolley, and J.~Zhang, ``{Causality Constraints on
  Gravitational Effective Field Theories},''
  \href{http://dx.doi.org/10.1103/PhysRevLett.128.131102}{{\em Phys. Rev.
  Lett.} {\bfseries 128} (2022) 131102},
  \href{http://arxiv.org/abs/2112.05054}{{\ttfamily arXiv:2112.05054 [gr-qc]}}.

\bibitem{Caron-Huot:2022ugt}
S.~Caron-Huot, Y.-Z. Li, J.~Parra-Martinez, and D.~Simmons-Duffin, ``{Causality
  constraints on corrections to Einstein gravity},''
  \href{http://arxiv.org/abs/2201.06602}{{\ttfamily arXiv:2201.06602
  [hep-th]}}.

\bibitem{Chiang:2022jep}
L.-Y. Chiang, Y.-t. Huang, W.~Li, L.~Rodina, and H.-C. Weng,
  ``{(Non)-projective bounds on gravitational EFT},''
  \href{http://arxiv.org/abs/2201.07177}{{\ttfamily arXiv:2201.07177
  [hep-th]}}.

\bibitem{Cheung:2014ega}
C.~Cheung and G.~N. Remmen, ``{Infrared Consistency and the Weak Gravity
  Conjecture},'' \href{http://dx.doi.org/10.1007/JHEP12(2014)087}{{\em JHEP}
  {\bfseries 12} (2014) 087},
\href{http://arxiv.org/abs/1407.7865}{{\ttfamily arXiv:1407.7865 [hep-th]}}.

\bibitem{Bellazzini:2019xts}
B.~Bellazzini, M.~Lewandowski, and J.~Serra, ``{Positivity of Amplitudes, Weak
  Gravity Conjecture, and Modified Gravity},''
  \href{http://dx.doi.org/10.1103/PhysRevLett.123.251103}{{\em Phys. Rev.
  Lett.} {\bfseries 123} (2019) 251103},
  \href{http://arxiv.org/abs/1902.03250}{{\ttfamily arXiv:1902.03250
  [hep-th]}}.

\bibitem{Cheung:2019cwi}
C.~Cheung, J.~Liu, and G.~N. Remmen, ``{Entropy Bounds on Effective Field
  Theory from Rotating Dyonic Black Holes},''
  \href{http://dx.doi.org/10.1103/PhysRevD.100.046003}{{\em Phys. Rev. D}
  {\bfseries 100} (2019) 046003},
\href{http://arxiv.org/abs/1903.09156}{{\ttfamily arXiv:1903.09156 [hep-th]}}.

\bibitem{Cheung:2018cwt}
C.~Cheung, J.~Liu, and G.~N. Remmen, ``{Proof of the Weak Gravity Conjecture
  from Black Hole Entropy},''
  \href{http://dx.doi.org/10.1007/JHEP10(2018)004}{{\em JHEP} {\bfseries 10}
  (2018) 004},
\href{http://arxiv.org/abs/1801.08546}{{\ttfamily arXiv:1801.08546 [hep-th]}}.

\bibitem{Charles:2019qqt}
A.~M. Charles, ``{The Weak Gravity Conjecture, RG Flows, and Supersymmetry},''
\href{http://arxiv.org/abs/1906.07734}{{\ttfamily arXiv:1906.07734 [hep-th]}}.

\bibitem{deRham:2018dqm}
C.~De~Rham, L.~Heisenberg, and A.~J. Tolley, ``{Spin-2 fields and the weak
  gravity conjecture},''
  \href{http://dx.doi.org/10.1103/PhysRevD.100.104033}{{\em Phys. Rev. D}
  {\bfseries 100} (2019) 104033},
  \href{http://arxiv.org/abs/1812.01012}{{\ttfamily arXiv:1812.01012
  [hep-th]}}.

\bibitem{Andriolo:2020lul}
S.~Andriolo, T.-C. Huang, T.~Noumi, H.~Ooguri, and G.~Shiu, ``{Duality and
  axionic weak gravity},''
  \href{http://dx.doi.org/10.1103/PhysRevD.102.046008}{{\em Phys. Rev. D}
  {\bfseries 102} (2020) 046008},
  \href{http://arxiv.org/abs/2004.13721}{{\ttfamily arXiv:2004.13721
  [hep-th]}}.

\bibitem{Alberte:2020jsk}
L.~Alberte, C.~de~Rham, S.~Jaitly, and A.~J. Tolley, ``{Positivity Bounds and
  the Massless Spin-2 Pole},''
  \href{http://dx.doi.org/10.1103/PhysRevD.102.125023}{{\em Phys. Rev. D}
  {\bfseries 102} (2020) 125023},
  \href{http://arxiv.org/abs/2007.12667}{{\ttfamily arXiv:2007.12667
  [hep-th]}}.

\bibitem{Arkani-Hamed:2021ajd}
N.~Arkani-Hamed, Y.-t. Huang, J.-Y. Liu, and G.~N. Remmen, ``{Causality,
  unitarity, and the weak gravity conjecture},''
  \href{http://dx.doi.org/10.1007/JHEP03(2022)083}{{\em JHEP} {\bfseries 03}
  (2022) 083}, \href{http://arxiv.org/abs/2109.13937}{{\ttfamily
  arXiv:2109.13937 [hep-th]}}.

\bibitem{Low:2009di}
I.~Low, R.~Rattazzi, and A.~Vichi, ``{Theoretical Constraints on the Higgs
  Effective Couplings},'' \href{http://dx.doi.org/10.1007/JHEP04(2010)126}{{\em
  JHEP} {\bfseries 04} (2010) 126},
  \href{http://arxiv.org/abs/0907.5413}{{\ttfamily arXiv:0907.5413 [hep-ph]}}.

\bibitem{Falkowski:2012vh}
A.~Falkowski, S.~Rychkov, and A.~Urbano, ``{What if the Higgs couplings to W
  and Z bosons are larger than in the Standard Model?},''
  \href{http://dx.doi.org/10.1007/JHEP04(2012)073}{{\em JHEP} {\bfseries 04}
  (2012) 073}, \href{http://arxiv.org/abs/1202.1532}{{\ttfamily arXiv:1202.1532
  [hep-ph]}}.

\bibitem{Bellazzini:2014waa}
B.~Bellazzini, L.~Martucci, and R.~Torre, ``{Symmetries, Sum Rules and
  Constraints on Effective Field Theories},''
  \href{http://dx.doi.org/10.1007/JHEP09(2014)100}{{\em JHEP} {\bfseries 09}
  (2014) 100}, \href{http://arxiv.org/abs/1405.2960}{{\ttfamily arXiv:1405.2960
  [hep-th]}}.

\bibitem{Ema:2018jgc}
Y.~Ema, R.~Kitano, and T.~Terada, ``{Unitarity constraint on the K\"ahler
  curvature},'' \href{http://dx.doi.org/10.1007/JHEP09(2018)075}{{\em JHEP}
  {\bfseries 09} (2018) 075}, \href{http://arxiv.org/abs/1807.06940}{{\ttfamily
  arXiv:1807.06940 [hep-th]}}.

\bibitem{LianTao}
J.~Gu and L.-T. Wang, ``{Sum Rules in the Standard Model Effective Field Theory
  from Helicity Amplitudes},''
  \href{http://dx.doi.org/10.1007/JHEP03(2021)149}{{\em JHEP} {\bfseries 03}
  (2021) 149}, \href{http://arxiv.org/abs/2008.07551}{{\ttfamily
  arXiv:2008.07551 [hep-ph]}}.

\bibitem{Zhang:2020jyn}
C.~Zhang and S.-Y. Zhou, ``{Convex Geometry Perspective on the (Standard Model)
  Effective Field Theory Space},''
  \href{http://dx.doi.org/10.1103/PhysRevLett.125.201601}{{\em Phys. Rev.
  Lett.} {\bfseries 125} (2020) 201601},
  \href{http://arxiv.org/abs/2005.03047}{{\ttfamily arXiv:2005.03047
  [hep-ph]}}.

\bibitem{Fuks:2020ujk}
B.~Fuks, Y.~Liu, C.~Zhang, and S.-Y. Zhou, ``{Positivity in electron-positron
  scattering: testing the axiomatic quantum field theory principles and probing
  the existence of UV states},''
  \href{http://dx.doi.org/10.1088/1674-1137/abcd8c}{{\em Chin. Phys. C}
  {\bfseries 45} (2021) 023108},
  \href{http://arxiv.org/abs/2009.02212}{{\ttfamily arXiv:2009.02212
  [hep-ph]}}.

\bibitem{Yamashita:2020gtt}
K.~Yamashita, C.~Zhang, and S.-Y. Zhou, ``{Elastic positivity vs extremal
  positivity bounds in SMEFT: a case study in transversal electroweak
  gauge-boson scatterings},''
  \href{http://dx.doi.org/10.1007/JHEP01(2021)095}{{\em JHEP} {\bfseries 01}
  (2021) 095}, \href{http://arxiv.org/abs/2009.04490}{{\ttfamily
  arXiv:2009.04490 [hep-ph]}}.

\bibitem{Gu:2020ldn}
J.~Gu, L.-T. Wang, and C.~Zhang, ``{An unambiguous test of positivity at lepton
  colliders},'' \href{http://arxiv.org/abs/2011.03055}{{\ttfamily
  arXiv:2011.03055 [hep-ph]}}.

\bibitem{Trott:2020ebl}
T.~Trott, ``{Causality, unitarity and symmetry in effective field theory},''
  \href{http://dx.doi.org/10.1007/JHEP07(2021)143}{{\em JHEP} {\bfseries 07}
  (2021) 143}, \href{http://arxiv.org/abs/2011.10058}{{\ttfamily
  arXiv:2011.10058 [hep-ph]}}.

\bibitem{Bonnefoy:2020yee}
Q.~Bonnefoy, E.~Gendy, and C.~Grojean, ``{Positivity bounds on Minimal Flavor
  Violation},'' \href{http://dx.doi.org/10.1007/JHEP04(2021)115}{{\em JHEP}
  {\bfseries 04} (2021) 115}, \href{http://arxiv.org/abs/2011.12855}{{\ttfamily
  arXiv:2011.12855 [hep-ph]}}.

\bibitem{Davighi:2021osh}
J.~Davighi, S.~Melville, and T.~You, ``{Natural selection rules: new positivity
  bounds for massive spinning particles},''
  \href{http://dx.doi.org/10.1007/JHEP02(2022)167}{{\em JHEP} {\bfseries 02}
  (2022) 167}, \href{http://arxiv.org/abs/2108.06334}{{\ttfamily
  arXiv:2108.06334 [hep-th]}}.

\bibitem{Zhang:2021eeo}
C.~Zhang, ``{SMEFTs living on the edge: determining the UV theories from
  positivity and extremality},''
  \href{http://arxiv.org/abs/2112.11665}{{\ttfamily arXiv:2112.11665
  [hep-ph]}}.

\bibitem{Li:2022tcz}
X.~Li and S.~Zhou, ``{Origin of Neutrino Masses on the Convex Cone of
  Positivity Bounds},'' \href{http://arxiv.org/abs/2202.12907}{{\ttfamily
  arXiv:2202.12907 [hep-ph]}}.

\bibitem{Chala:2021wpj}
M.~Chala and J.~Santiago, ``{Positivity bounds in the standard model effective
  field theory beyond tree level},''
  \href{http://dx.doi.org/10.1103/PhysRevD.105.L111901}{{\em Phys. Rev. D}
  {\bfseries 105} (2022) L111901},
  \href{http://arxiv.org/abs/2110.01624}{{\ttfamily arXiv:2110.01624
  [hep-ph]}}.

\bibitem{Azatov:2021ygj}
A.~Azatov, D.~Ghosh, and A.~H. Singh, ``{Four-fermion operators at dimension 6:
  Dispersion relations and UV completions},''
  \href{http://dx.doi.org/10.1103/PhysRevD.105.115019}{{\em Phys. Rev. D}
  {\bfseries 105} (2022) 115019},
  \href{http://arxiv.org/abs/2112.02302}{{\ttfamily arXiv:2112.02302
  [hep-ph]}}.

\bibitem{Murphy:2020rsh}
C.~W. Murphy, ``{Dimension-8 operators in the Standard Model Eective Field
  Theory},'' \href{http://dx.doi.org/10.1007/JHEP10(2020)174}{{\em JHEP}
  {\bfseries 10} (2020) 174}, \href{http://arxiv.org/abs/2005.00059}{{\ttfamily
  arXiv:2005.00059 [hep-ph]}}.

\bibitem{Li:2020gnx}
H.-L. Li, Z.~Ren, J.~Shu, M.-L. Xiao, J.-H. Yu, and Y.-H. Zheng, ``{Complete
  set of dimension-eight operators in the standard model effective field
  theory},'' \href{http://dx.doi.org/10.1103/PhysRevD.104.015026}{{\em Phys.
  Rev. D} {\bfseries 104} (2021) 015026},
  \href{http://arxiv.org/abs/2005.00008}{{\ttfamily arXiv:2005.00008
  [hep-ph]}}.

\bibitem{deRham:2017avq}
C.~de~Rham, S.~Melville, A.~J. Tolley, and S.-Y. Zhou, ``{Positivity bounds for
  scalar field theories},''
  \href{http://dx.doi.org/10.1103/PhysRevD.96.081702}{{\em Phys. Rev. D}
  {\bfseries 96} (2017) 081702},
  \href{http://arxiv.org/abs/1702.06134}{{\ttfamily arXiv:1702.06134
  [hep-th]}}.

\bibitem{Bern:2021ppb}
Z.~Bern, D.~Kosmopoulos, and A.~Zhiboedov, ``{Gravitational effective field
  theory islands, low-spin dominance, and the four-graviton amplitude},''
  \href{http://dx.doi.org/10.1088/1751-8121/ac0e51}{{\em J. Phys. A} {\bfseries
  54} (2021) 344002}, \href{http://arxiv.org/abs/2103.12728}{{\ttfamily
  arXiv:2103.12728 [hep-th]}}.

\bibitem{Jacob:1959at}
M.~Jacob and G.~Wick, ``{On the General Theory of Collisions for Particles with
  Spin},'' \href{http://dx.doi.org/10.1016/0003-4916(59)90051-X}{{\em Annals
  Phys.} {\bfseries 7} (1959) 404}.

\bibitem{Chala:2021pll}
M.~Chala, G.~Guedes, M.~Ramos, and J.~Santiago, ``{Towards the renormalisation
  of the Standard Model effective field theory to dimension eight: Bosonic
  interactions I},''
  \href{http://dx.doi.org/10.21468/SciPostPhys.11.3.065}{{\em SciPost Phys.}
  {\bfseries 11} (2021) 065}, \href{http://arxiv.org/abs/2106.05291}{{\ttfamily
  arXiv:2106.05291 [hep-ph]}}.

\bibitem{DasBakshi:2022mwk}
S.~Das~Bakshi, M.~Chala, A.~D\'\i{}az-Carmona, and G.~Guedes, ``{Towards the
  renormalisation of the Standard Model effective field theory to dimension
  eight: Bosonic interactions II},''
  \href{http://arxiv.org/abs/2205.03301}{{\ttfamily arXiv:2205.03301
  [hep-ph]}}.

\bibitem{Horejsi:1993hz}
J.~Ho\v{r}ej\v{s}\'i, \href{http://dx.doi.org/10.1142/2445}{{\em {Introduction
  to Electroweak Unification: Standard Model from Tree Unitarity}}}.
\newblock World Scientific, 1993.

\bibitem{ItzyksonZuber}
C.~Itzykson and J.-B. Zuber, {\em Quantum Field Theory}.
\newblock McGraw-Hill, 1980.

\bibitem{Arkani-Hamed:2017jhn}
N.~Arkani-Hamed, T.-C. Huang, and Y.-t. Huang, ``{Scattering amplitudes for all
  masses and spins},'' \href{http://dx.doi.org/10.1007/JHEP11(2021)070}{{\em
  JHEP} {\bfseries 11} (2021) 070},
  \href{http://arxiv.org/abs/1709.04891}{{\ttfamily arXiv:1709.04891
  [hep-th]}}.

\bibitem{deBlas:2017xtg}
J.~de~Blas, J.~Criado, M.~Perez-Victoria, and J.~Santiago, ``{Effective
  description of general extensions of the Standard Model: the complete
  tree-level dictionary},''
  \href{http://dx.doi.org/10.1007/JHEP03(2018)109}{{\em JHEP} {\bfseries 03}
  (2018) 109}, \href{http://arxiv.org/abs/1711.10391}{{\ttfamily
  arXiv:1711.10391 [hep-ph]}}.

\bibitem{Quevillon:2018mfl}
J.~Quevillon, C.~Smith, and S.~Touati, ``{Effective action for gauge bosons},''
  \href{http://dx.doi.org/10.1103/PhysRevD.99.013003}{{\em Phys. Rev. D}
  {\bfseries 99} (2019) 013003},
  \href{http://arxiv.org/abs/1810.06994}{{\ttfamily arXiv:1810.06994
  [hep-ph]}}.

\bibitem{Agashe:2006at}
K.~Agashe, R.~Contino, L.~Da~Rold, and A.~Pomarol, ``{A custodial symmetry for
  $Zb \bar b$},'' \href{http://dx.doi.org/10.1016/j.physletb.2006.08.005}{{\em
  Phys. Lett. B} {\bfseries 641} (2006) 62},
  \href{http://arxiv.org/abs/hep-ph/0605341}{{\ttfamily arXiv:hep-ph/0605341}}.

\bibitem{Peskin:1991sw}
M.~E. Peskin and T.~Takeuchi, ``{Estimation of oblique electroweak
  corrections},'' \href{http://dx.doi.org/10.1103/PhysRevD.46.381}{{\em Phys.
  Rev. D} {\bfseries 46} (1992) 381}.

\bibitem{Maksymyk:1993zm}
I.~Maksymyk, C.~Burgess, and D.~London, ``{Beyond $S$, $T$, and $U$},''
  \href{http://dx.doi.org/10.1103/PhysRevD.50.529}{{\em Phys. Rev. D}
  {\bfseries 50} (1994) 529},
  \href{http://arxiv.org/abs/hep-ph/9306267}{{\ttfamily arXiv:hep-ph/9306267}}.

\bibitem{Barbieri:2004qk}
R.~Barbieri, A.~Pomarol, R.~Rattazzi, and A.~Strumia, ``{Electroweak symmetry
  breaking after LEP-1 and LEP-2},''
  \href{http://dx.doi.org/10.1016/j.nuclphysb.2004.10.014}{{\em Nucl. Phys. B}
  {\bfseries 703} (2004) 127},
  \href{http://arxiv.org/abs/hep-ph/0405040}{{\ttfamily arXiv:hep-ph/0405040}}.

\bibitem{CDF:2022hxs}
{\bfseries CDF} {\bfseries Collaboration}, T.~Aaltonen { et~al.},
  ``{High-precision measurement of the $W$ boson mass with the CDF II
  detector},'' \href{http://dx.doi.org/10.1126/science.abk1781}{{\em Science}
  {\bfseries 376} (2022) 170}.

\bibitem{Wells:2015uba}
J.~D. Wells and Z.~Zhang, ``{Effective theories of universal theories},''
  \href{http://dx.doi.org/10.1007/JHEP01(2016)123}{{\em JHEP} {\bfseries 01}
  (2016) 123}, \href{http://arxiv.org/abs/1510.08462}{{\ttfamily
  arXiv:1510.08462 [hep-ph]}}.

\bibitem{Cacciapaglia:2006pk}
G.~Cacciapaglia, C.~Cs\'aki, G.~Marandella, and A.~Strumia, ``{The Minimal Set
  of Electroweak Precision Parameters},''
  \href{http://dx.doi.org/10.1103/PhysRevD.74.033011}{{\em Phys. Rev. D}
  {\bfseries 74} (2006) 033011},
  \href{http://arxiv.org/abs/hep-ph/0604111}{{\ttfamily arXiv:hep-ph/0604111}}.

\bibitem{Englert:2019zmt}
C.~Englert, G.~F. Giudice, A.~Greljo, and M.~McCullough, ``{The
  $\hat{H}$-Parameter: An Oblique Higgs View},''
  \href{http://dx.doi.org/10.1007/JHEP09(2019)041}{{\em JHEP} {\bfseries 09}
  (2019) 041}, \href{http://arxiv.org/abs/1903.07725}{{\ttfamily
  arXiv:1903.07725 [hep-ph]}}.

\bibitem{EGGM}
J.~Elias-Mir\'o, C.~Grojean, R.~S. Gupta, and D.~Marzocca, ``{Scaling and
  tuning of EW and Higgs observables},''
  \href{http://dx.doi.org/10.1007/JHEP05(2014)019}{{\em JHEP} {\bfseries 05}
  (2014) 019}, \href{http://arxiv.org/abs/1312.2928}{{\ttfamily arXiv:1312.2928
  [hep-ph]}}.

\bibitem{Weinberg:1967kj}
S.~Weinberg, ``{Precise relations between the spectra of vector and axial
  vector mesons},'' \href{http://dx.doi.org/10.1103/PhysRevLett.18.507}{{\em
  Phys. Rev. Lett.} {\bfseries 18} (1967) 507}.

\end{thebibliography}\endgroup

\end{document}